\documentclass{article}
\usepackage{jheppub} 
\usepackage[utf8x]{inputenc}	
\usepackage[english]{babel}

\usepackage[ddmmyyyy,hhmmss]{datetime}
\usepackage{subfig}
\usepackage{multirow}

\catcode`@=11
\allowdisplaybreaks
\usepackage{amsxtra}
\usepackage{mathtools}
\usepackage{placeins}
\usepackage{float}
\usepackage{xcolor}
\usepackage{tikz}
\usepackage{macros}
\usepackage{hyperref}
\usepackage{amsmath}
\usepackage{graphicx}
\usepackage{rotating}  

\newcommand{\Figref}[1]{Figure~\ref{#1}}

\usepackage{placeins}
\usetikzlibrary{calc}
\usetikzlibrary{arrows.meta, bending, patterns}
\usepackage{soul}

\usepackage{placeins}
\usetikzlibrary{calc}
\usetikzlibrary{arrows.meta, bending, patterns}
\usepackage{soul}
\usetikzlibrary{hobby,intersections} 
\usetikzlibrary{shapes.geometric}
\usetikzlibrary{shapes.misc}

\tikzset{flavour/.style={draw=none,minimum size=0.3mm,fill=white, regular polygon,regular polygon sides=4,draw}}
\tikzset{flavourr/.style={draw=none,minimum size=0.3mm,fill=red, regular polygon,regular polygon sides=4,draw}}
\tikzset{flavourb/.style={draw=none,minimum size=0.3mm,fill=blue, regular polygon,regular polygon sides=4,draw}}
\tikzset{gaugeBig/.style={inner sep=1mm,draw=none,fill=white,minimum size=2mm,circle, draw}}
\tikzset{bd/.style={circle, draw=black, inner sep=0pt, fill=black, minimum size=2mm}}
\tikzset{wd/.style={circle, draw=black, inner sep=0pt, fill=white, minimum size=2mm}}
\tikzset{Dynkin/.style={circle, draw=black, inner sep=0pt, fill=white, minimum size=2mm}}
\tikzstyle{ligne}=[draw, very thick] 
\tikzstyle{gridline}=[draw, gray] 
\tikzset{gauge/.style={circle, draw,inner sep=2.5pt}}
\tikzset{gaugeo/.style={circle, draw,inner sep=2.5pt,fill=orange}}
\tikzset{gauger/.style={circle, draw,inner sep=2.5pt,fill=red}}
\tikzset{gaugeb/.style={circle, draw,inner sep=2.5pt,fill=blue}}
\tikzset{gaugeg/.style={circle, draw,inner sep=2.5pt,fill=teal}}
\tikzset{gaugegoodteal/.style={circle, draw,inner sep=2.5pt,fill=goodteal}}
\tikzset{gaugem/.style={circle, draw,inner sep=2.5pt,fill=magenta}}
\tikzset{hasse/.style={circle, fill,inner sep=2pt}}
\tikzset{d2/.style={circle, fill,inner sep=1.3pt}}
\tikzset{shrinky/.style={circle, fill,inner sep=1pt}}
\tikzset{sized/.style={circle, draw, inner sep=1.5pt}}
\tikzset{seven/.style={circle, draw,inner sep=3pt}}

\tikzset{gaugebl/.style={circle,draw=black,fill=black,inner sep=1.5pt}}
\tikzset{gaugeblnormal/.style={circle,draw=black,fill=black,inner sep=2.5pt}}
\tikzset{hasse/.style={circle, fill,inner sep=2pt}}
\tikzstyle{dashed_brane}=[thick, dashed]
\tikzstyle{dotted_brane}=[thick, dotted]
\tikzstyle{O3plus}=[thick, color=teal]
\tikzstyle{O3minustilde}=[thick, color=blue]
\tikzstyle{O3plustilde}=[thick, color=red]
\tikzset{D5/.style={cross out, draw=black, minimum size=7, inner sep=0pt, outer sep=0pt}, cross/.default={1pt}}
\tikzset{flavor/.style={regular polygon,regular polygon sides=4,inner sep=2.5pt, label = {}, draw}}
\tikzset{redflavor/.style={regular polygon,regular polygon sides=4,inner sep=2.5pt, color=red, label = {}, draw}}
\tikzset{redgauge/.style={inner sep=1mm,color=red,draw=none,minimum size=2mm,circle, draw}}
\tikzset{blueflavor/.style={regular polygon,regular polygon sides=4,inner sep=2.5pt, color=blue, label = {}, draw}}
\tikzset{bluegauge/.style={inner sep=1mm,color=blue,draw=none,minimum size=2mm,circle, draw}}

\makeatletter
\DeclareRobustCommand{\rvdots}{%
  \vbox{
    \baselineskip4\p@\lineskiplimit\z@
    \kern-\p@
    \hbox{.}\hbox{.}\hbox{.}
  }}
\makeatother

\newcommand{\convexpath}[2]{
  [   
  create hullcoords/.code={
    \global\edef\namelist{#1}
    \foreach [count=\counter] \nodename in \namelist {
      \global\edef\numberofnodes{\counter}
      \coordinate (hullcoord\counter) at (\nodename);
    }
    \coordinate (hullcoord0) at (hullcoord\numberofnodes);
    \pgfmathtruncatemacro\lastnumber{\numberofnodes+1}
    \coordinate (hullcoord\lastnumber) at (hullcoord1);
  },
  create hullcoords
  ]
  ($(hullcoord1)!#2!-90:(hullcoord0)$)
  \foreach [
  evaluate=\currentnode as \previousnode using \currentnode-1,
  evaluate=\currentnode as \nextnode using \currentnode+1
  ] \currentnode in {1,...,\numberofnodes} {
    let \p1 = ($(hullcoord\currentnode) - (hullcoord\previousnode)$),
    \n1 = {atan2(\y1,\x1) + 90},
    \p2 = ($(hullcoord\nextnode) - (hullcoord\currentnode)$),
    \n2 = {atan2(\y2,\x2) + 90},
    \n{delta} = {Mod(\n2-\n1,360) - 360}
    in 
    {arc [start angle=\n1, delta angle=\n{delta}, radius=#2]}
    -- ($(hullcoord\nextnode)!#2!-90:(hullcoord\currentnode)$) 
  }
}

\usepackage[all]{xy}

\tikzset{gaugebl/.style={circle,draw=black,fill=black,inner sep=1.5pt}}
\tikzset{gaugeblnormal/.style={circle,draw=black,fill=black,inner sep=2.5pt}}
\tikzset{hasse/.style={circle, fill,inner sep=2pt}}
\tikzstyle{dashed_brane}=[thick, dashed]
\tikzstyle{dotted_brane}=[thick, dotted]
\tikzstyle{O3plus}=[thick, color=teal]
\tikzstyle{O3minustilde}=[thick, color=blue]
\tikzstyle{O3plustilde}=[thick, color=red]
\tikzset{D5/.style={cross out, draw=black, minimum size=7, inner sep=0pt, outer sep=0pt}, cross/.default={1pt}}
\tikzset{flavor/.style={regular polygon,regular polygon sides=4,inner sep=2.5pt, label = {}, draw}}
\tikzset{redflavor/.style={regular polygon,regular polygon sides=4,inner sep=2.5pt, color=red, label = {}, draw}}
\tikzset{redgauge/.style={inner sep=1mm,color=red,draw=none,minimum size=2mm,circle, draw}}
\tikzset{blueflavor/.style={regular polygon,regular polygon sides=4,inner sep=2.5pt, color=blue, label = {}, draw}}
\tikzset{bluegauge/.style={inner sep=1mm,color=blue,draw=none,minimum size=2mm,circle, draw}}
\allowdisplaybreaks

\tikzset{gaugebl/.style={circle,draw=black,fill=black,inner sep=1.5pt}}
\tikzset{gaugeblnormal/.style={circle,draw=black,fill=black,inner sep=2.5pt}}
\tikzset{hasse/.style={circle, fill,inner sep=2pt}}
\tikzstyle{dashed_brane}=[thick, dashed]
\tikzstyle{dotted_brane}=[thick, dotted]
\tikzstyle{O3plus}=[thick, color=green]
\tikzstyle{O3minustilde}=[thick, color=blue]
\tikzstyle{O3plustilde}=[thick, color=red]
\tikzset{D5/.style={cross out, draw=black, minimum size=7, inner sep=0pt, outer sep=0pt}, cross/.default={1pt}}
\tikzset{flavor/.style={regular polygon,regular polygon sides=4,inner sep=2.5pt, label = {}, draw}}
\tikzset{redflavor/.style={regular polygon,regular polygon sides=4,inner sep=2.5pt, color=red, label = {}, draw}}
\tikzset{redgauge/.style={inner sep=1mm,color=red,draw=none,minimum size=2mm,circle, draw}}
\tikzset{blueflavor/.style={regular polygon,regular polygon sides=4,inner sep=2.5pt, color=blue, label = {}, draw}}
\tikzset{bluegauge/.style={inner sep=1mm,color=blue,draw=none,minimum size=2mm,circle, draw}}
\allowdisplaybreaks

\newcommand{\TabBTwoAC}{
\begin{tabular}{|c|c|}
    \hline
    \multicolumn{2}{|c|}{$\mathfrak{g}=\mathfrak{b}_2 \cong \mathfrak{c}_2$} \\\hline
    \boldmath$\rho$\unboldmath & \boldmath$d_{SD}\left(\rho\right)$\unboldmath \\\hline
     $\left(1^{5}\right)$ & $\left(4\right)$ \\\hline
    \cellcolor{gray!25} $\left(2^{2},1\right)$ & \cellcolor{gray!25} $\left(2,1^{2}\right)$ \\\hline
    \cellcolor{gray!25} $\left(3,1^{2}\right)$ & \cellcolor{gray!25} $\left(2^{2}\right)$ \\\hline
    $\left(5\right)$ & $\left(1^{4}\right)$ \\\hline
\end{tabular}
}

\newcommand{\TabBThreeAC}{
\begin{tabular}{|c|c|}
    \hline
    \multicolumn{2}{|c|}{$\mathfrak{g}=\mathfrak{b}_3$} \\\hline
    \boldmath$\rho$\unboldmath & \boldmath$d_{SD}\left(\rho\right)$\unboldmath \\\hline
    $\left(1^{7}\right)$ & $\left(6\right)$ \\\hline
    \cellcolor{gray!25} $\left(2^{2},1^{3}\right)$ & \cellcolor{gray!25} $\left(4,1^{2}\right)$ \\\hline
    \cellcolor{gray!25} $\left(3,1^{4}\right)$ & \cellcolor{gray!25} $\left(4,2\right)$ \\\hline
    $\left(3,2^{2}\right)$ & $\left(3^{2}\right)$ \\\hline
    $\left(3^{2},1\right)$ & $\left(2^{3}\right)$ \\\hline
    $\left(5,1^{2}\right)$ & $\left(2^{2},1^{2}\right)$ \\\hline
    $\left(7\right)$ & $\left(1^{6}\right)$ \\
    \hline
    \multicolumn{2}{c}{} \\
    \hline
    \multicolumn{2}{|c|}{$\mathfrak{g}=\mathfrak{c}_3$} \\
    \hline
    \boldmath$\rho$\unboldmath & \boldmath$d_{SD}\left(\rho\right)$\unboldmath \\\hline
    $\left(1^{6}\right)$ & $\left(7\right)$ \\\hline
    \cellcolor{gray!15}$\left(2,1^{4}\right)$ & $\bullet$ \\\hline
    \cellcolor{gray!15}$\left(2^2,1^{2}\right)$ & $\left(5,1^2\right)$ \\\hline
    $\left(2^{3}\right)$ & $\left(3^{2},1\right)$ \\\hline
    $\left(3^{2}\right)$ & $\left(3,2^{2}\right)$ \\\hline
    \cellcolor{gray!25}$\left(4,1^{2}\right)$ & \cellcolor{gray!25}$\left(2^2,1^3\right)$ 
    \\ \hline
     \cellcolor{gray!25}$\left(4,2\right)$ & \cellcolor{gray!25}$\left(3,1^{4}\right)$ 
    \\ \hline
    $\left(6\right)$ & $\left(1^{7}\right)$ 
    \\  \hline
\end{tabular}
}

\newcommand{\TabBFourAC}{
\begin{tabular}{|c|c|}
    \hline
    \multicolumn{2}{|c|}{$\mathfrak{g}=\mathfrak{b}_4$} \\\hline
    \boldmath$\rho$\unboldmath & \boldmath$d_{SD}\left(\rho\right)$\unboldmath \\\hline
    $\left(1^{9}\right)$ & $\left(8\right)$ \\\hline
    \cellcolor{gray!25} $\left(2^{2},1^{5}\right)$ & \cellcolor{gray!25} $\left(6,1^{2}\right)$ \\\hline
    \cellcolor{gray!25} $\left(3,1^{6}\right)$ & \cellcolor{gray!25} $\left(6,2\right)$ \\\hline
    \cellcolor{gray!15}$\left(2^{4},1\right)$ & $\bullet$ \\\hline
   \cellcolor{gray!15} $\left(3,2^{2},1^{2}\right)$ & $\left(4^{2}\right)$ \\\hline
    $\left(3^{2},1^{3}\right)$ & $\left(4,2^{2}\right)$ \\\hline
    $\left(3^{3}\right)$ & $\left(3^{2},2\right)$ \\\hline
    $\left(5,1^{4}\right)$ & $\left(4,2,1^{2}\right)$ \\
    \hline
    $\left(5,2^{2}\right)$ & $\left(3^{2},1^{2}\right)$ \\
    \hline
    \cellcolor{gray!25}$\left(4^{2},1\right)$ & \cellcolor{gray!25}$\left(2^{3},1^{2}\right)$ \\
    \hline
    \cellcolor{gray!25}$\left(5,3,1\right)$ & \cellcolor{gray!25}$\left(2^{4}\right)$ \\\hline
    $\left(7,1^{2}\right)$ & $\left(2^{2},1^{4}\right)$ \\\hline
    $\left(9\right)$ & $\left(1^{8}\right)$ \\
    \hline
 \multicolumn{2}{c}{} \\
 \hline
    \multicolumn{2}{|c|}{$\mathfrak{g}=\mathfrak{c}_4$} \\\hline
    \boldmath$\rho$\unboldmath & \boldmath$d_{SD}\left(\rho\right)$\unboldmath \\\hline
    $\left(1^{8}\right)$ & $\left(9\right)$ \\\hline
     \cellcolor{gray!15}$\left(2,1^{6}\right)$ & $\bullet$ \\\hline
     \cellcolor{gray!15} $\left(2^2,1^{4}\right)$ & $\left(7,1^2\right)$ \\\hline
    \cellcolor{gray!25}$\left(2^{3},1^2\right)$ & \cellcolor{gray!25} $\left(4^2,1 \right)$ \\\hline
    \cellcolor{gray!25}$\left(2^{4}\right)$ &\cellcolor{gray!25} $\left(5,3,1 \right)$ \\\hline
    \cellcolor{gray!15}$\left(4,1^{4}\right)$ & $\bullet$ \\\hline
    $\left(3^{2},1^2\right)$ & $\left(5,2^{2}\right)$ \\\hline
    $\left(3^{2},2\right)$ & $\left(3^{3}\right)$ \\\hline
   \cellcolor{gray!15} $\left(4,2,1^{2}\right)$ & $\left(5,1^{4}\right)$ \\\hline
    $\left(4,2^{2}\right)$ & $\left(3^{2},1^{3}\right)$ \\\hline
    $\left(4^{2}\right)$ & $\left(3,2^{2},1^{2}\right)$ \\\hline
    \cellcolor{gray!25} $\left(6,1^2\right)$ & \cellcolor{gray!25} $\left(2^2,1^{5}\right)$ \\\hline
    \cellcolor{gray!25} $\left(6,2\right)$ & \cellcolor{gray!25} $\left(3,1^{6}\right)$ \\\hline
    $\left(8\right)$ & $\left(1^{9}\right)$ \\\hline
\end{tabular}
}

\newcommand{\TabGAC}{
\begin{tabular}{|c|c|}
    \hline
    \multicolumn{2}{|c|}{$\mathfrak{g}=\mathfrak{g}_2$} \\\hline
    \boldmath$\rho$\unboldmath & \boldmath$d_{SD}\left(\rho\right)$\unboldmath \\\hline
    $\varnothing$ & $G_2$ \\\hline
    \cellcolor{green!25} $A_1$ & \cellcolor{green!25} $A_1$ \\\hline
    \cellcolor{green!25} $\tilde{A}_1$ & \cellcolor{green!25} $\tilde{A}_1$ \\\hline
    \cellcolor{green!25} $G_2\left(a_1\right)$ & \cellcolor{green!25} $G_2\left(a_1\right)$ \\\hline
    $G_2$ & $\varnothing$ \\\hline
\end{tabular}
}

\newcommand{\TabFAC}{
\begin{tabular}{|c|c|}
    \hline
    \multicolumn{2}{|c|}{$\mathfrak{g} =\mathfrak{f}_4$} \\\hline
    \boldmath$\rho$\unboldmath & \boldmath$d_{SD}\left(\rho\right)$\unboldmath \\\hline
    $\varnothing$ & $F_4$ \\\hline
    \cellcolor{gray!15} $A_1$ & $\bullet$ \\\hline
    \cellcolor{gray!15} $\tilde{A}_1$ & $F_4\left(a_1\right)$ \\\hline
    $A_1+\tilde{A}_1$ & $F_4\left(a_2\right)$ \\\hline
    $A_2$ & $B_3$ \\\hline
    $\tilde{A}_2$ & $C_3$ \\\hline
    \cellcolor{blue!20} $A_2+\tilde{A}_1$ & \cellcolor{blue!20} $A_2+\tilde{A}_1$ \\\hline
    \cellcolor{blue!20} $B_2$ & \cellcolor{blue!20} $B_2$ \\\hline
    \cellcolor{blue!20} $\tilde{A}_2+A_1$ & \cellcolor{blue!20} $\tilde{A}_2+A_1$ \\\hline
    \cellcolor{blue!20} $C_3\left(a_1\right)$ & \cellcolor{blue!20} $C_3\left(a_1\right)$ \\\hline
    \cellcolor{blue!20} $F_4\left(a_3\right)$ & \cellcolor{blue!20} $F_4\left(a_3\right)$ \\\hline
    $C_3$ & $\tilde{A}_2$ \\\hline
    $B_3$ & $A_2$ \\\hline
    $F_4\left(a_2\right)$ & $A_1+\tilde{A}_1$ \\\hline
    $F_4\left(a_1\right)$ &  \cellcolor{gray!15} $\tilde{A}_1$ \\\hline
    $F_4$ & $\varnothing$ \\\hline
\end{tabular}}
    
\newcommand{\TabESixAC}{
\begin{tabular}{|c|c|}
    \hline
    \multicolumn{2}{|c|}{$\mathfrak{g} =E_6$} \\\hline
    \boldmath$\rho$\unboldmath & \boldmath$d_{SD}\left(\rho\right)$\unboldmath \\\hline
    $\varnothing$ & $E_6$ \\\hline
    $A_1$ & $E_6\left(a_1\right)$\\\hline
    $2A_1$ & $D_5$\\\hline
    \cellcolor{gray!25} $3A_1$ & \cellcolor{gray!25} $A_5$ \\\hline
    \cellcolor{gray!25} $A_2$ & \cellcolor{gray!25} $E_6\left(a_3\right)$ \\\hline
    $A_2+A_1$ & $D_5\left(a_1\right)$\\\hline
    $2A_2$ & $D_4$\\\hline
    $A_2+2A_1$ & $A_4+A_1$\\\hline
    $A_3$ & $A_4$\\\hline
    \cellcolor{green!25} $2A_2+A_1$ & \cellcolor{green!25} $2A_2+A_1$\\\hline
    \cellcolor{green!25} $A_3+A_1$ & \cellcolor{green!25} $A_3+A_1$ \\\hline
    \cellcolor{green!25} $D_4\left(a_1\right)$ & \cellcolor{green!25} $D_4\left(a_1\right)$ \\\hline
    $A_4$ & $A_3$ \\\hline
    $D_4$ & $2A_2$\\\hline
    $A_4+A_1$ & $A_2+2A_1$\\\hline
    $D_5\left(a_1\right)$ & $A_2+A_1$\\\hline
    \cellcolor{gray!25} $A_5$ & \cellcolor{gray!25} $3A_1$ \\\hline
    \cellcolor{gray!25} $E_6\left(a_3\right)$ & \cellcolor{gray!25} $A_2$ \\\hline
    $D_5$ & $2A_1$\\\hline
    $E_6\left(a_1\right)$ & $A_1$\\\hline
    $E_6$ & $\varnothing$\\\hline
    \end{tabular}}
    
\newcommand{\TabESevenAC}{
\begin{tabular}{|c|c|}
    \hline
    \multicolumn{2}{|c|}{$\mathfrak{g} =E_7$} \\\hline
    \boldmath$\rho$\unboldmath & \boldmath$d_{SD}\left(\rho\right)$\unboldmath \\\hline
    $\varnothing$ & $E_7$ \\\hline
    $A_1$ & $E_7\left(a_1\right)$\\\hline
    $2A_1$ & $E_7\left(a_2\right)$\\\hline
    $\left(3A_1\right)''$ & $E_6$\\\hline
    \cellcolor{gray!25} $\left(3A_1\right)'$ & \cellcolor{gray!25} $D_6$\\\hline
    \cellcolor{gray!25} $A_2$ & \cellcolor{gray!25} $E_7\left(a_3\right)$\\
    \hline
    \cellcolor{gray!15} $4A_1$ & $\bullet$\\\hline
    \cellcolor{gray!15} $A_2+A_1$ & $E_6\left(a_1\right)$\\
    \hline
    $A_2+2A_1$ & $E_7\left(a_4\right)$\\\hline
    $A_3$ & $D_6\left(a_1\right)$\\\hline
    $2A_2$ & $D_5+A_1$\\\hline
    $A_2+3A_1$ & $A_6$\\\hline
    $\left(A_3+A_1\right)''$ & $D_5$ \\\hline
    \cellcolor{green!25} $2A_2+A_1$ & \cellcolor{green!25} $A_5+A_1$\\\hline
    \cellcolor{green!25} $\left(A_3+A_1\right)'$ & \cellcolor{green!25} $D_6\left(a_2\right)$\\\hline
    \cellcolor{green!25} $D_4\left(a_1\right)$ & \cellcolor{green!25} $E_7\left(a_5\right)$ \\\hline
     \cellcolor{gray!25} $A_3+2A_1$ & \cellcolor{gray!25}$A_{5}'$\\\hline
      \cellcolor{gray!25} $D_4\left(a_1\right)+A_1$ & \cellcolor{gray!25} $E_6\left(a_3\right)$\\\hline
    $D_4$ & $\left(A_5\right)''$\\\hline
    $A_3+A_2$ & $D_5\left(a_1\right)+A_1$\\\hline
    $A_4$ &  \cellcolor{gray!15}  $D_5\left(a_1\right)$\\\hline
    $A_3+A_2+A_1$ & $A_4+A_2$\\\hline
    $A_{5}''$ & $D_4$\\\hline
    \cellcolor{gray!15} $D_4+A_1$ & $\bullet$\\\hline
    $A_4+A_1$ & $A_4+A_1$\\
    \hline
     \cellcolor{gray!15} $D_5\left(a_1\right)$ &  $A_4$\\\hline
    $A_4+A_2$ & $A_3+A_2+A_1$ \\\hline
    $D_5\left(a_1\right)+A_1$ & $A_3+A_2$ \\\hline
     \cellcolor{gray!25} $A_{5}'$ &  \cellcolor{gray!25} $A_3+2A_1$\\\hline
    \cellcolor{gray!25} $E_6\left(a_3\right)$ & \cellcolor{gray!25} $D_4\left(a_1\right)+A_1$\\\hline
     \cellcolor{green!25} $A_5+A_1$ & \cellcolor{green!25} $2A_2+A_1$\\\hline
     \cellcolor{green!25} $D_6\left(a_2\right)$ & \cellcolor{green!25} $\left(A_3+A_1\right)'$\\\hline
    \cellcolor{green!25} $E_7\left(a_5\right)$ & \cellcolor{green!25} $D_4\left(a_1\right)$\\\hline
    $D_5$ & $\left(A_3+A_1\right)''$\\\hline
    $A_6$ & $A_2+3A_1$\\\hline
    $D_5+A_1$ & $2A_2$\\\hline
    $D_6\left(a_1\right)$ & $A_3$\\\hline
    $E_7\left(a_4\right)$ & $A_2+2A_1$\\\hline
    $E_6\left(a_1\right)$ &  \cellcolor{gray!15}  $A_2+A_1$\\\hline
     \cellcolor{gray!25} $D_6$ & \cellcolor{gray!25} $\left(3A_1\right)'$\\\hline
    \cellcolor{gray!25} $E_7\left(a_3\right)$ & \cellcolor{gray!25} $A_2$\\\hline
    $E_6$ & $\left(3A_1\right)''$\\\hline
    $E_7\left(a_2\right)$ & $2A_1$\\\hline
    $E_7\left(a_1\right)$ & $A_1$\\\hline
    $E_7$ & $\varnothing$\\\hline
    \end{tabular}}
    
\newcommand{\TabEEightFirstBitAC}{
\begin{tabular}{|c|c|}
    \hline
    \multicolumn{2}{|c|}{$\mathfrak{g} =E_8$} \\\hline
    \boldmath$\rho$\unboldmath & \boldmath$d_{SD}\left(\rho\right)$\unboldmath \\\hline
    $\varnothing$ & $E_8$ \\\hline
    $A_1$ & $E_8\left(a_1\right)$\\\hline
    $2A_1$ & $E_8\left(a_2\right)$\\
    \hline
    \cellcolor{gray!25} $3A_1$ & \cellcolor{gray!25} $E_7$\\\hline
    \cellcolor{gray!25}$A_2$ & \cellcolor{gray!25} $E_8\left(a_3\right)$ \\
    \hline
    \cellcolor{gray!15}$4A_1$ & $\bullet$ \\ \hline
    \cellcolor{gray!15}$A_2+A_1$ & $E_8\left(a_4\right)$\\
    \hline
    $A_2+2A_1$ & $E_8\left(b_4\right)$\\\hline
    $A_3$ & $E_7\left(a_1\right)$\\\hline
     \cellcolor{gray!25}$A_2+3A_1$ & \cellcolor{gray!25} $D_7$\\
     \hline
     \cellcolor{gray!25}$2A_2$ &  \cellcolor{gray!25} $E_8\left(a_5\right)$\\\hline
    \cellcolor{green!25}$2A_2+A_1$ & \cellcolor{green!25}$E_6+A_1$\\\hline
    \cellcolor{green!25} $A_3+A_1$ & \cellcolor{green!25}$E_7\left(a_2\right)$\\\hline
    \cellcolor{green!25}$D_4\left(a_1\right)$ & \cellcolor{green!25}$E_8\left(b_5\right)$\\\hline
    $D_4$ & $E_6$\\\hline
    \cellcolor{green!15}$2A_2+2A_1$ & $\bullet$\\\hline
    \cellcolor{green!15}$A_3+2A_1$ & $\bullet$\\\hline
    \cellcolor{green!15}$D_4\left(a_1\right)+A_1$ & $E_8\left(a_6\right)$\\\hline
    $A_3+A_2$ & $D_7\left(a_1\right)$\\\hline
    $A_4$ & $E_7\left(a_3\right)$\\\hline
    \cellcolor{gray!25} $A_3+A_2+A_1$ & \cellcolor{gray!25} $A_7$\\\hline
    \cellcolor{gray!25}$D_4\left(a_1\right)+A_2$ & \cellcolor{gray!25}$E_8\left(b_6\right)$\\\hline
    \cellcolor{gray!15}$D_4+A_1$ & $\bullet$ \\\hline
    $A_4+A_1$ & $E_6\left(a_1\right)+A_1$\\\hline
    \cellcolor{gray!15}$2A_3$ &$\bullet$ \\\hline
\cellcolor{gray!15}$D_5\left(a_1\right)$ & $E_6\left(a_1\right)$\\\hline
    \cellcolor{gray!15}$A_4+2A_1$ & $D_7\left(a_2\right)$\\\hline
    $A_4+A_2$ & $D_5+A_2$\\\hline
    $D_5\left(a_1\right)+A_1$ & $E_7\left(a_4\right)$\\\hline
    $A_4+A_2+A_1$ & $A_6+A_1$\\\hline
     \cellcolor{gray!25} $A_5$ & \cellcolor{gray!25} $D_5+A_1$\\\hline
     \cellcolor{gray!25}  $E_6\left(a_3\right)$ &  \cellcolor{gray!25} $D_6\left(a_1\right)$\\
      \hline
    $D_4+A_2$ & $A_6$ \\\hline
    $D_5$ &  $D_5$\\\hline
    \cellcolor{red!25} $A_4+A_3$ & \cellcolor{red!25} $A_4+A_3$ \\\hline
    \cellcolor{red!25} $A_5+A_1$ & \cellcolor{red!25} $A_5+A_1$ \\\hline
    \cellcolor{red!25} $D_5\left(a_1\right)+A_2$ & \cellcolor{red!25} $D_5\left(a_1\right)+A_2$ \\\hline
    \cellcolor{red!25} $D_6\left(a_2\right)$ & \cellcolor{red!25} $D_6\left(a_2\right)$\\\hline
    \cellcolor{red!25} $E_6\left(a_3\right)+A_1$ & \cellcolor{red!25} $E_6\left(a_3\right)+A_1$\\\hline
     \cellcolor{red!25} $E_7\left(a_5\right)$ & \cellcolor{red!25} $E_7\left(a_5\right)$\\\hline
      \cellcolor{red!25} $E_8\left(a_7\right)$ & \cellcolor{red!25} $E_8\left(a_7\right)$\\\hline
    \end{tabular}}
    
\newcommand{\TabEEightSecondBitAC}{
\begin{tabular}{|c|c|}
    \hline
    \multicolumn{2}{|c|}{$\mathfrak{g} =E_8$} \\\hline
    \boldmath$\rho$\unboldmath & \boldmath$d_{SD}\left(\rho\right)$\unboldmath \\\hline
    \cellcolor{gray!25} $D_5+A_1$ & \cellcolor{gray!25} $A_5$\\\hline \cellcolor{gray!25}$D_6\left(a_1\right)$ & \cellcolor{gray!25} $E_6\left(a_3\right)$\\\hline
    $A_6$ & $D_4+A_2$\\\hline
    $A_6+A_1$ & $A_4+A_2+A_1$\\\hline
    $E_7\left(a_4\right)$ & $D_5\left(a_1\right)+A_1$\\\hline
    $E_6\left(a_1\right)$ & \cellcolor{gray!15}$D_5\left(a_1\right)$\\\hline
    $D_5+A_2$ & $A_4+A_2$\\\hline
     $E_6$ & $D_4$\\\hline
   \cellcolor{gray!15} $D_6$ & $\bullet$ \\\hline
     $D_7\left(a_2\right)$ & \cellcolor{gray!15}$A_4+2A_1$\\\hline
     $E_6\left(a_1\right)+A_1$ & $A_4+A_1$ \\\hline
    \cellcolor{gray!25}$A_7$ &\cellcolor{gray!25}$A_3+A_2+A_1$ \\\hline
\cellcolor{gray!25}$E_8\left(b_6\right)$ & \cellcolor{gray!25}$D_4\left(a_1\right)+A_2$\\\hline
   \cellcolor{gray!15} $E_7\left(a_3\right)$ & $A_4$\\\hline
    $D_7\left(a_1\right)$ & $A_3+A_2$\\\hline
    \cellcolor{green!25} $E_6+A_1$ & \cellcolor{green!25} $2A_2+A_1$\\\hline
    \cellcolor{green!25}$E_7\left(a_2\right)$ & \cellcolor{green!25} $A_3+A_1$\\\hline
    $E_8\left(a_6\right)$ & \cellcolor{green!15}$D_4\left(a_1\right)+A_1$\\\hline
\cellcolor{green!25}$E_8\left(b_5\right)$ & \cellcolor{green!25} $D_4\left(a_1\right)$\\\hline
    \cellcolor{gray!25}$D_7$ & \cellcolor{gray!25} $A_2+3A_1$ \\\hline
\cellcolor{gray!25}$E_8\left(a_5\right)$ &\cellcolor{gray!25} $2A_2$\\\hline
    $E_7\left(a_1\right)$ & $A_3$\\\hline
    $E_8\left(b_4\right)$ & $A_2+2A_1$\\\hline
    $E_8\left(a_4\right)$ &  \cellcolor{gray!15}$A_2+A_1$\\\hline
    \cellcolor{gray!25}$E_7$ & \cellcolor{gray!25}$3A_1$\\\hline
    \cellcolor{gray!25} $E_8\left(a_3\right)$ & \cellcolor{gray!25} $A_2$\\\hline
    $E_8\left(a_2\right)$ & $2A_1$\\\hline
    $E_8\left(a_1\right)$ & $A_1$\\\hline
    $E_8$ & $\varnothing$\\\hline
    \end{tabular}}
    
\usepackage{todonotes}

\title{Quiver Maps, Nilpotent Orbits and Special Pieces of Nilcones}

\author{Sam Bennett,}
\author{Amihay Hanany,}
\author{Rudolph Kalveks}

\affiliation{Abdus Salam Centre for Theoretical Physics, Imperial College London,\\ Prince Consort Road
London, SW7 2AZ, UK}
\emailAdd{samuel.bennett18@imperial.ac.uk}
\emailAdd{a.hanany@imperial.ac.uk}
\emailAdd{rudolph.kalveks09@imperial.ac.uk}

\preprint{Imperial/TP/26/AH/03}
\abstract{This paper explores 3d $\mathcal{N}=4$ quiver gauge theories whose moduli spaces represent nilpotent orbits, S\l odowy slices or, more generally, S\l odowy intersections, which span the Special Pieces of nilcones of Classical or Exceptional algebras. We introduce a map between magnetic and electric quivers containing symmetric group actions, such as wreathings (or loops), bouquets, and/or non-simply laced foldings, which can be related to symmetric subgroups of Lusztig's canonical quotient groups for Special Pieces. The map on quivers induces a map on nilpotent orbits that partially resolves the obstruction to quiver dualities presented by the non-involutive nature of the Lusztig Spaltenstein and Barbasch Vogan maps. We use Coulomb and Higgs branch quiver methods complemented by localisation formulae. Some new quivers for intersections within Exceptional nilcones are presented.}

\begin{document}
\maketitle
\listoftables
\section{Introduction}
\label{sec:intro}
This paper continues the exploration of the group symmetries, structures and dualities that involve $3d$ $\mathcal{N}=4$ SUSY quiver gauge theories whose Higgs and/or Coulomb branches evaluate as (a) the closures of orbits within the nilpotent cones (or {\it nilcones}) of Lie algebras, (b) their transverse spaces, known as S\l odowy {\it slices}, or more generally, (c) the S\l odowy {\it intersections} defined by pairs of nilpotent {\it orbits} \cite{Hanany:2016gbz, Hanany:2017ooe, Cabrera:2018ldc, Hanany:2019tji}.\footnote{These moduli spaces are often referred to as, as {\it orbits}, {\it slices}, or {\it intersections}.} 

These moduli spaces arise in a variety of contexts, including three-dimensional $T_{\sigma}^{\rho}(G)$ theories \cite{Cremonesi:2014uva, Cabrera:2017ucb}, four-dimensional Class-$S$ theories \cite{CHACALTANA_2013, chacaltana2013tinkertoys}, and five- and six-dimensional theories constructed using branes \cite{Hanany:2022itc,Cabrera:2019_6d2}. The Hilbert series for these moduli spaces also have well-defined group theoretic constructions using localisation formulae \cite{Hanany:2016gbz, Hanany:2017ooe, Cabrera:2018ldc, Hanany:2019tji}. The identification of the symmetries and dualities of the corresponding quiver gauge theories is not only relevant for their evaluation, but also provides insights into subtle relationships between gauge theories.

The closures of the nilpotent orbits of any Lie algebra are ordered according to their inclusion relations within the nilcone. This information is conventionally captured using a Hasse diagram, on which a variety of order reversing maps, such as those of Lusztig/Spaltenstein ($d_{LS}$) and Barbasch/Vogan ($d_{BV}$) \cite{barbasch_vogan, 2020arXiv201016089B}, have been defined. However, with the exception of the $\mathfrak{a}$ algebras, these maps are not involutive; they serve to partition orbits between those termed {\it special}, which return to themselves under $d^{2}$, and those that are not. It is a general result that any {\it non-special} orbit is related to some parent {\it special} orbit by certain finite symmetric group factors and forms part of this parent orbit's {\it special piece} \cite{collingwood1993nilpotent}.

The focus herein is on the study of quiver theories that involve orbits from such {\it special pieces} of nilcones, whose symmetric group factors encode remarkable dualities.

The analysis of such quiver theories involving exceptional algebras is particularly interesting. While group-theoretic localization formulae are known for their orbits, slices and intersections \cite{Hanany:2017ooe, Hanany:2019tji}, there is no general prescription for identifying the electric or magnetic quivers whose respective Higgs or Coulomb branches evaluate to yield such moduli spaces. This is in contrast to the situation for classical algebras, where many prescriptions and quiver constructions exist in the literature \cite{Cremonesi:2014uva,Gaiotto:2008ak,Benini:2010uu}. For $A$-type algebras, a comprehensive set of constructions is available, based on either the Higgs or Coulomb branches of unitary quivers. For $BCD$-type algebras a set of Higgs branch constructions is available for quivers with orthosymplectic gauge and flavour nodes \cite{kraft1982geometry, Hanany:2019tji}. A subset of $BCD$-type intersections admit Coulomb branch constructions using unitary or orthosymplectic quivers \cite{Hanany:2016gbz, Cabrera:2018ldc, Hanany:2019tji,Braverman:2016pwk,Braverman:2016wma,braverman2018coulombbranches3dmathcal}; generically, these theories may be related by quiver subtractions \cite{Cabrera:2018ann}, foldings and symmetrisations \cite{Bourget:2020bxh}, or dualities such as 3d mirror symmetry.

Unsurprisingly, $A$-type algebras offer the simplest realisation of much of the characteristic phenomena -- their nilpotent orbits are in 1:1 correspondence with partitions of $n$ (equal of course to the number of irreducible representations of $S_n$). As a corollary, the simple action of transposing partitions $\lambda \rightarrow \lambda^{\top}$ provides a 1:1 order reversing involution on the poset of nilpotent orbits \cite{collingwood1993nilpotent}. This action is involutive on $A$-type nilpotent cones, and as a consequence all $A$-type orbits are {\it special}. \footnote{In this paper, as in much of the nilpotent orbit literature, an orbit $\sigma$ is \emph{special} with respect to a map $d$ iff $d^{2}(\sigma) = \sigma$.}

Moreover, any pair of $A$-type orbits is dual generically to some other pair of orbits, such that the set of $A$-type intersections is partitioned into pairs that are dual to each other. The result is that nilpotent orbit closures become dual to S\l odowy slices. \footnote{In terms of the Hasse diagram, this is simply the notion that the symplectic singularity constructed by the slice from the trivial orbit (or origin of the nilcone) to an orbit $\sigma$ is dual to the slice constructed from a related orbit, $d_{\text{}}(\sigma)$, to the regular (or maximal) orbit. 
} Remarkably, the unitary quiver whose Coulomb branch evaluates to an $A$-type algebra intersection has a Higgs branch which evaluates as the dual intersection. These moduli spaces are of course exchanged under $3d$ mirror symmetry \cite{Intriligator:1996ex, deBoer:1996mp, Borokhov:2002cg, Benini:2010uu}.

In contrast to the $A$-type case, $BCD$-type and Exceptional algebras do not possess such an order reversing involution on the entire poset of orbits. For these algebras, the $d_{LS}$ and $d_{BV}$ maps partition orbits into {\it special} orbits, for which $d^2(\cal O)={\cal O}$, and {\it non-special} orbits, for which $d^2(\cal O) \neq {\cal O}$. Each {\it non-special} orbit ${\cal O}_{n.s.}$ is related to a parent {\it special} orbit ${\cal O}_{spec}$ by the relation $d^2({\cal O}_{n.s.}) = {\cal O}_{spec}$ and by certain $S_n$ symmetric group actions. The latter are described in terms of Lusztig's Canonical Quotient Group $\bar A({\cal O})$ and its sub-groups, the component group $A({\cal O})$ and the Sommers-Achar group $C({\cal O})$ \cite{collingwood1993nilpotent, CHACALTANA_2013}. These finite groups are, in turn, sub-groups of the Weyl reflection group of the algebra $\mathfrak{g}$, which is incorporated into each orbit in a distinct manner, according to its characteristic. Remarkably, these same $S_n$ finite group symmetries manifest themselves in the magnetic and electric quivers for orbits, slices and intersections involving {\it special pieces}.

This paper aims to explicate the manner in which these finite group symmetries are encoded in the magnetic and electric quivers for orbits, slices and intersections involving {\it special pieces}, and the related dualities. We focus on cases where one of the pair of orbits defining an intersection lies inside a special piece while the other lies outside. This exposes a rich interplay between the symmetric group factors defined by the special piece and other group symmetries encoded in the nilcone.

Numerous examples are studied, drawing upon both unitary and orthosymplectic, magnetic and/or electric quivers. The pattern of finite group symmetries that surrounds {\it special pieces} facilitates the generation of families of quivers once some members are known. The examples herein build upon quivers identified, inter alia, in \cite{Bourget:2020bxh, Hanany:2022itc, Bennett:2024loi}. The paper is organised as follows:

Sections \ref{subsec:SI} and \ref{subsec:Local} recapitulate the definition of S\l odowy intersections between orbits of Classical and Exceptional algebras, their group theoretic localisation formulae, and the {\it special duality} \cite{Hanany:2019tji} that arises between certain intersections when constructed as the Higgs and Coulomb branches of particular quivers.

Section \ref{subsec:Specials} concerns the {\it special pieces} of nilcones of Classical and Exceptional algebras and on the role therein of finite symmetry groups, Lusztig's canonical quotient group $\bar A({\cal O})$, and its key sub-groups, the component (or fundamental) group $A({\cal O})$ and the Sommers-Achar group $C({\cal O})$.

Section \ref{subsec:SD} draws on Lusztig's canonical quotient group $\bar A({\cal O})$ (and its sub-groups) to modify the $d_{LS}$ and $d_{BV}$ maps, which are only 1:1 for {\it special} orbits, to incorporate the topology of orbits inside {\it special pieces}. This leads to a map $d_{SD}$ that yields a 1:1 involution for any {\it non-special} orbit whose dual special piece shares the same Lusztig's canonical quotient group.

Section \ref{subsec:LL} explores how the $A({\cal O})$ and $C({\cal O})$ subgroups associated to orbits inside a special piece are related to features of the magnetic or electric quivers for orbits, slices or intersections which involve the special piece. The differences in $A({\cal O})$ and $C({\cal O}))$ for magnetic and electric quivers are used to define a map between them, termed the {\it Loop Lace} map. An outline is given for how the {\it special duality} of orbits, slices and intersections under $d_{SD}$ are related to this {\it Loop Lace} map. Similar treatments can be applied to both unitary and $OSp$ quivers.

Section \ref{sec:bcspecial} provides some simple unitary quiver examples taken from the $\mathfrak{bc}$ non-simply laced algebras. These involve $S_2$ special pieces only, but illustrate the interplay between the $d_{SD}$ and {\it Loop Lace} maps, referred to as {\it special duality}.

Sections \ref{sec:g2special} through \ref{sec:e8special} examine numerous unitary and orthosymplectic examples involving orbits taken from the {\it special pieces} of the Exceptional algebras. These involve $S_2, S_3, S_4,\text{ and } S_5$ special pieces, and include some quiver constructions not previously encountered in the Literature.

We end with a summary discussion of the findings and identify avenues for future work.\\

Appendix \ref{app:OrbitData} contains data on nilpotent orbits of the Exceptional and some Classical algebras that is relevant to our study. The tables provide details on Bala-Carter labels, partitions, dimensions, Characteristics, LS/BV duals, transverse symmetries and fugacity maps, along with other relevant information. For further background on nilpotent orbits the reader is referred to the Bibliography.\\

\paragraph{Notation and Terminology}
The group $G$ has the Lie algebra $\mathfrak{g}$. The symplectic group of rank-$k$ is labelled as $C_k$ inside a quiver and $\sprm(k)$ otherwise; its algebra is referred to as $\mathfrak{c}_k$. We refer to nilpotent orbits ${\mathcal{O}}^{\mathfrak{g}}_{\rho}$ by their algebras and their vector/fundamental partitions in the case of Classical algebras, or by the accepted Bala-Carter labels in the case of Exceptional algebras.\footnote{We do not utilise the group theoretic information encoded in Bala-Carter labels.} We may refer to canonical nilpotent orbits in various ways; $\mathcal{N}^{\mathfrak{g}}$ labels the nilcone of the semi-simple Lie algebra $\mathfrak{g}$, equivalently the closure of the regular orbit $ \bar{\mathcal{O}}^{\mathfrak{g}}_{\mathcal N}$ or $\overline{reg.\mathfrak{g}}$ and equivalently the S{\l}odowy slice from the trivial orbit $\mathcal{S}^{\mathfrak{g}}_{\mathcal{N},\;\varnothing}$. We may abbreviate the S{\l}odowy intersection $\mathcal{S}^{\mathfrak{g}}_{\mathcal{N},\;\rho}$ as the slice $\mathcal{S}^{\mathfrak{g}}_{\rho}$.  The subset of {\it special} nilpotent orbits is labelled $\left\{{\mathcal{O}}^{\mathfrak{g}}_{\mathrm{spec}}\right\}$. We refer to special pieces by their parent orbit ${\mathcal{O}}^{\mathfrak{g}}_{\mathrm{spec}}$. For clarity, we may also list the parent along with other orbits within the special piece.


\section{Nilcones and their Special Pieces}
\label{sec:Theory}

\subsection{S\l odowy Intersections and Quiver Dualities}
\label{subsec:SI}

S\l odowy intersections, which build upon the theory of nilpotent orbits, arise in various contexts, perhaps most familiarly as the moduli spaces of $T^{\sigma}_{\rho}(G)$ theories \cite{Benini:2010uu,Cremonesi:2014vla,Cremonesi:2014uva}. For any algebra, its S\l odowy intersections are defined by pairs of nilpotent orbits whose closures have a well-defined ordering in the Hasse diagram of its nilcone, such that the closure of one orbit contains the other. The dimension of the intersection is simply the difference between the dimensions of the two orbits in the pair. Its global symmetry is generally that of the S\l odowy slice to the orbit of smaller dimension.

Canonical cases arise (a) when one of the pair is taken as the trivial orbit, in which case the intersection is simply the closure of the higher dimensioned nilpotent orbit in the pair, or (b) when one of the pair is taken as the regular orbit, in which case the intersection is a space (``S\l odowy slice")\footnote{For brevity, we generally use the term S\l odowy slice to refer to its restriction to the nilcone.} transverse to the lower dimensioned orbit in the pair, or (c) when the pair is defined by the trivial and regular orbits, in which case the S\l odowy intersection yields the closure of the maximal nilpotent orbit or nilcone of the algebra.

\begin{figure}[htp]
\centering
\begin{displaymath}
    \xymatrix{
  &  {\cal Q} \left( {T}_{\rho}^{\sigma^{T}} \right) \ar[dl]|{Coulomb} \ar[dr]|{Higgs} 
  & 
  \\ 
 {{\cal S}^{A}_{\sigma, \rho}}  
 & \text{\scriptsize \it 3d~Mirror Symmetry} \ar[d] \ar[u] 
 &  {{\cal S}^{A}_{\rho^T, \sigma^T}}   
 \\
   & {\cal Q} \left( {T}_{\sigma^{T}}^{\rho} \right) \ar[ul]|{Higgs} \ar[ur]|{Coulomb} 
   &  }
\end{displaymath}

\caption[$A$ Series 3d Mirror Symmetry]{$A$ series $3d$ mirror symmetry. ${\cal Q} \left({T}_{\sigma}^{\rho} \right)$ is a linear unitary quiver from a ${T}_{\sigma}^{\rho}$ theory. Under $A$ series {\it special duality}, the nilpotent orbit partitions $\rho$ and $\sigma$ are dualised under the Lusztig-Spaltenstein map, to $\rho^T$ and $\sigma^T$, and then interchanged to yield the dual intersection $\left( {{{\cal S}^{A}}_{\sigma ,\rho }}\right)^{\vee} \equiv {{{\cal S}^{A}}_{\rho^T, \sigma^T}}$. All the quiver constructions yield refined Hilbert series.}
\label{fig:Ams}
\end{figure}
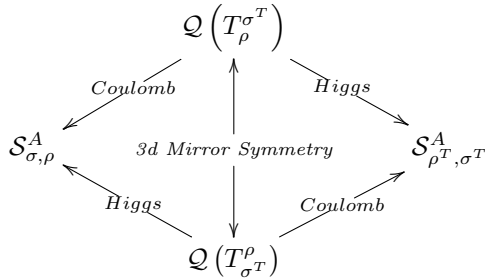

In \cite{Cabrera:2018ldc, Hanany:2019tji} it was shown how, for classical algebras, a variety of quiver constructions can be given for their S\l odowy intersections. For $A$-type algebras, this can be done on either the Coulomb or the Higgs branches of a family of unitary quivers that have linear gauge nodes with attached flavours. Taken together with the 1:1 $d_{LS}$ map between orbits and their duals, these constructions reveal a remarkable duality between quivers and their 3d mirror pairs, between their Coulomb and Higgs branches, and between their S\l odowy intersections and duals. This is neatly encapsulated in the map in figure \ref{fig:Ams}, adapted from \cite{Hanany:2019tji}.

For BCD and Exceptional algebras, these duality relationships are complicated by the absence of a 1:1 map between orbits and their duals, as will be elaborated, and a comparable complete map cannot be given.

\begin{figure}[htp]
\centering
\begin{displaymath}
    \xymatrix{
  &  {\cal Q}_{EM} \ar[dl]|{Coulomb} \ar[dr]|{Higgs} 
  & 
  \\ 
 {{\cal S}^{\mathfrak{g}}_{\sigma, \rho}}  
 &  
 \text{\scriptsize \it Special Duality} \ar[l] \ar[r] 
 &  {{\cal S}^{{\mathfrak{g}}^{\vee}}_{d(\rho), d(\sigma)}}   
 }
\end{displaymath}
\caption[Electric Magnetic Special Duality]{Electric Magnetic Special Duality. ${{\cal Q}_{EM}}$ is a quiver. Under Electric Magnetic Special Duality, the nilpotent orbits $\rho$ and $\sigma$ are dualised under the Lusztig-Spaltenstein map (if ${\mathfrak{g}}$ is $ADE$ type) or the Barbasch-Vogan map (if ${\mathfrak{g}}$ is $BC$ type), to $d(\rho)$ and $d(\sigma)$, and then interchanged to yield the dual intersection $\left({{{\cal S}^{\mathfrak{g}}}_{\sigma ,\rho }}\right)^{\vee} \equiv {{{\cal S}^{{\mathfrak{g}}^{\vee}}}_{d(\rho), d(\sigma)}}$, which lies in in the GNO dual algebra ${\mathfrak{g}}^{\vee}$.}
\label{fig:GSD}
\end{figure}
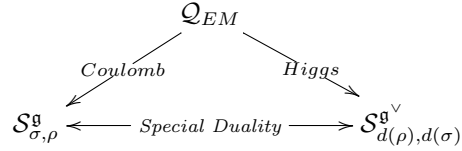

Nonetheless, for BCD algebras there exists a family of orthosymplectic quivers that provides a Higgs branch construction for any desired S\l odowy intersection \cite{Hanany:2019tji}. Also, a number of unitary or orthosymplectic quivers are known, whose Coulomb branches yield a subset of S\l odowy intersections of BCD and Exceptional algebras.\footnote{One limitation of the Coulomb branch constructions is that the monopole formula always yields well-formed palindromic Hilbert series and therefore can only provide the normalisation (including some non-nilpotent elements) of a non-normal nilpotent orbit.} For certain such quivers, denoted ${{\cal Q}_{EM}}$, the Higgs branches evaluate as dual intersections of their Coulomb branches, such that the top half of the map in figure \ref{fig:Ams} is satisfied, as illustrated in figure \ref{fig:GSD}.

Instances of this duality are found amongst magnetic quivers based on unitary Dynkin diagrams for closures of nilpotent orbits that have a characteristic height of 2 \cite{Hanany:2017ooe}. Well-known examples include the magnetic quivers for minimal and next to minimal orbits of ADE type algebras; when treated instead as electric quivers, these evaluate, respectively, as the sub-regular, and sub-sub regular S\l odowy slices of their algebras.

This relationship does not, however, readily generalise to all intersections of $BCD$ or Exceptional algebras, even if their magnetic quivers are known. Obstructions include (a) the absence of a 1:1 map between orbits and their duals, (b) the absence of a well-defined Higgs branch for non-simply laced quivers\footnote{The Higgs branches of non-simply laced quivers, which are known to be non-Lagrangian, have no known interpretation.}, and (c) complications surrounding non-normal nilpotent orbits.

Nonetheless, in later sections we aim to navigate these obstructions to show how some families of magnetic and electric quivers can be constructed around dual pairs of S\l odowy intersections that each involve an orbit lying within dual {\it special pieces} of a Hasse diagram.

\subsection{Localisation}
\label{subsec:Local}

In all cases, it is helpful to be able to validate the quiver constructions of S\l odowy intersections against direct mathematical calculations of their Hilbert series and this is facilitated by various localisation formulae that appear in different guises across the Literature \cite{Hanany:2017ooe, Gadde:2011uv, Cremonesi:2014kwa, Cremonesi:2014uva, Hanany:2019tji}. These are relations of the Weyl Character formula \cite{fuchs2003symmetries} and of the modified Hall Littlewood polynomials \cite{macdonald1998symmetric}. It is helpful to state a self-consistent set that can in principle be used to construct Hilbert series from the S\l odowy intersections of any algebra. In practice, the localisation computations can be burdensome, depending on the order of the Weyl group of the algebra. (Naturally, this motivates quiver based methods that may replicate mathematical results in an efficient manner.)

The localisation calculation for a S\l odowy intersection ${{\cal S}^{\mathfrak{g}}_{\sigma,\rho}}$ can be carried out in three steps: (a) calculation of the (normalisation of) the closure of the superior nilpotent orbit ${{\bar{\cal O}}_{\sigma} ^{{\mathfrak{g}}~norm}}$ in the pair, (b) calculation of the S\l odowy slice ${{\cal S}^{\mathfrak{g}}_{{\cal N},\rho }}$ to the lower dimensioned orbit in the pair, and (c) their combination to yield the intersection ${{\cal S}^{\mathfrak{g}}_{\sigma,\rho}}$.

The formula for the generating function that yields the Hilbert series  of the closure of a normal nilpotent orbit ${{\bar{\cal O}}_{\sigma, \bf{[n]}}^{{\mathfrak{g}}~norm}}$ (or the normalisation of the closure of a non-normal orbit) can be stated as \cite{Hanany:2017ooe}:
\begin{equation}
\label{eq:NOL}
g_{HS}^{{\bar{\cal O}}_{\sigma,{\bf[n]}}^{{\mathfrak{g}}~norm}}\left( {{\bf x},t} \right) 
\equiv  \sum\limits_{w \in W_{G}} {w \cdot \left( {\bf x}^{\bf [n]} 
\prod\limits_{\scriptsize \begin{array}{c} {\alpha  \in \Phi _{\mathfrak{g}}^{+} (\sigma)}  \end{array}}
 {\frac{1}{{1 - {{\bf z(x)}^{\bf \alpha} }t^2}}} \prod\limits_{\beta  \in \Phi_{\mathfrak{g}}^{+} }{\frac{1}{{1 - {{\bf z(x)}^{ - \bf \beta }}}}}\right)}.
\end{equation}

Here, the weight space of $G$ is parameterised by Cartan Subalgebra (``CSA") fugacities $\bf x$ and Dynkin labels $\bf [n]$ (which can be viewed as representing background charges); the fugacities for the simple roots of ${\mathfrak{g}}$ are monomials $\bf z(x)$; $w$ ranges over the Weyl reflection group $W_G$, acting upon the CSA fugacities as $w:{\bf x} \to w \cdot {\bf x} $; $\alpha$ and $\beta$ range over subsets of the positive root space $ \Phi _{{\mathfrak{g}}}^{+}$; and $\sigma$ identifies the nilpotent orbit. The subset $ \Phi _{{\mathfrak{g}}}^ +(\sigma)$ contains those roots that have a Characteristic height ${[\alpha] \ge 2} $, where $[\alpha] \equiv \alpha \cdot \bf q(\sigma)$, with $\bf q(\sigma)$ being the Characteristic \cite{dynkin1960semisimple} of the orbit.\\

The key ingredient in distinguishing between orbits is the Characteristic, which determines those roots $\alpha$ of its algebra that seed the nilpotent orbit. In the case that $\sigma$ is taken as the maximal orbit (or nilcone ${\cal N}$), and $[\bf{n}]$ is taken as the singlet irrep, then \ref{eq:NOL} evaluates as the familiar form:

\begin{equation}
\label{eq:Nilcone}
\begin{aligned}
g_{HS}^{{{\cal N}^{\mathfrak{g}}}}({\bf x},t) \equiv g_{HS}^{{\bar{\cal O}_{\cal N}^{\mathfrak{g}}}}({\bf x},t) & = PE\left[  \chi _{adjoint}^{G}({\bf x}){~}{t^{2}} - \sum\limits_{i = 1}^r {{t^{2{d_i}}}}  \right],
\end{aligned}
\end{equation}
where $ \chi_{adjoint}^{G}$ is the adjoint representation and $d_i$ ranges over the degrees of the Casimirs of $G$.\\

The S\l odowy slice formula modifies \ref{eq:Nilcone} to  extract the space transverse to a given orbit. The symmetry group $F$ of this transverse space is the commutant of the $SU(2)$ nilpotent orbit embedding in $G$. The key ingredient in the calculation is the fugacity map that describes the branching $\psi$ from characters of $G$ to the $SU(2)$ of the nilpotent orbit and the $F$ of the transverse space:

\begin{equation}
\label{eq:CSAmap}
\begin{aligned}
\psi(\rho):\chi _{adjoint}^G (\bf x) &\to \mathop  \bigoplus \limits_{[k],{\bf [m]}} {a_{[k],{\bf [m]}}}\left( {\chi _{[k]}^{SU(2)}(t) \otimes \chi _{\bf [m]}^F (\bf y)} \right),
\end{aligned}
\end{equation}
where the ${a_{[k],{\bf [m]}}}$ are the resulting branching coefficients for the adjoint representation. The Hilbert series for the S\l odowy slice is obtained by replacing the $SU(2)$ characters by their highest weights in $t$ and substituting into \ref{eq:Nilcone}:

\begin{equation}
\label{eq:Slodslice}
\begin{aligned}
g_{HS}^{{\cal S}^{\mathfrak{g}}_{{\cal N},\rho }}({\bf y},t) &= PE\left[ {\mathop  \bigoplus \limits_{[k], \bf [m]}  a_{[k],\bf [m]} {~} \chi _{\bf [m]}^F({\bf y}){t^{k + 2}} - \sum\limits_{i = 1}^r {{t^{2{d_i}}}} } \right].\\
\end{aligned}
\end{equation}

The S\l odowy intersection formula is obtained from \ref{eq:NOL}, \ref{eq:Nilcone} and \ref{eq:Slodslice} by restricting the slice according to the superior orbit in the intersection pair. This involves a composition of the slice with the orbit following a quotient by the nilcone:
\begin{equation}
\label{eq:SV14}
\begin{aligned}
g_{HS}^{{\cal S}^{\mathfrak{g}}_{\sigma,\rho,{\bf [n]}}}({\bf y},t)  = g_{HS}^{{\cal S}^{\mathfrak{g}}_{{\cal N}, \rho}}({\bf y}, t) {\left. {\frac{g_{HS}^{ {\bar{\cal O}}^{\mathfrak{g}}_{\sigma, {\bf [n]}} }({\bf x},t)}{g_{HS}^{{\cal N}^{\mathfrak{g}}} ({\bf x}, t)}} \right|_{\psi(\rho)(\bf x \to y)}}.
\end{aligned}
\end{equation}
In the examples that follow, intersections are generally calculated at the origin of the weight lattice of $G$ and carry singlet Dynkin labels $[{\bf n}]=[0, \dots]$ (generally dropped for brevity).

One important limitation is that in the case of an intersection ending on a non-normal orbit, the localisation formulae \ref{eq:NOL} and \ref{eq:SV14}  yield a normalisation. Such a Hilbert series describes a symplectic singularity, but typically contains elements outside the nilcone. This normalisation is generally the same as the result obtained by Coulomb branch methods when the magnetic quivers are known.\footnote{If the Hilbert series for a non-normal intersection is desired, methods beyond the scope of this paper are required.} For further detail on S\l odowy intersections the reader is referred to \cite{Hanany:2019tji}.\\
%

\subsection{Special Pieces of Nilcones}
\label{subsec:Specials}
Much of the structure associated with the nilpotent orbits of a semi-simple Lie algebra $\mathfrak{g}$ arises from the Springer correspondence, which identifies their irreducible, equivariant local systems with irreducible representations of the Weyl group $\mathcal{W}_{\mathfrak{g}}$ \cite{achar2007specialpiecesspringercorrespondence}. Famously, this reduces in $A$-type algebras to the observation that both the irreducible representations of the permutation group $S_{n}\simeq \mathcal{W}_{A_{n-1}}$ and nilpotent orbits are labelled by partitions of $n$. For $BCD$-type algebras, which require far greater apparatus, Lusztig used the correspondence to develop the theory of special orbits from an associated set of special Weyl group representations \cite{LUSZTIG1979323,lusbook}.

The combinatorics of these special orbits led Spaltenstein to identify an involution $d_{\text{LS}}:\left\{ \bar{\mathcal{O}}^{\mathfrak{g}}_{\mathrm{sp}}\right\}\rightarrow\left\{ \bar{\mathcal{O}}^{\mathfrak{g}}_{\mathrm{sp}}\right\}$ that preserves their partial order under the natural inclusion relation. Barbasch and Vogan later specified a variant of Spaltenstein's map, relevant for $B$- and $C$-type algebras, termed $d_{BV}$, that draws on their duality under the Langlands correspondence. The $d_{BV}$ map made contact with gauge theories in \cite{CHACALTANA_2013}, in which it was conjectured that $d_{\text{BV}}$ interpolates between the Higgs and Coulomb branches of related $T^{\rho}_{\sigma}$ theories. The reader is referred to \cite{Cabrera:2017njm} for background material on these various maps.

Every {\it non-special} orbit is associated with a parent {\it special} orbit by virtue of sharing a common $d_{LS}$ dual (or $d_{BV}$ dual in the case of $BC$-type algebras); each such subset of orbits under a common parent constitutes a {\it special piece}. The tables in Appendix \ref{app:OrbitData} list the orbits of Exceptional and selected $BCD$ algebras along with their key data.

Importantly, the orbits in each {\it special piece} are related by subgroups of Lusztig's canonical quotient group. This group, denoted $\bar{A}( {\mathcal{O}}^{\mathfrak{g}}_{\mathrm{sp}})$, is determined by the parent orbit in the {\it special piece} and is an $S_n$ symmetric group (or products thereof). The non-trivial symmetric groups that appear amongst the canonical quotients of orbits of Exceptional algebras are $S_2$, $S_3$, $S_4$ and $S_5$. In the nilcones of $BCD$-type algebras only Lusztig canonical quotients of the form $S_2^{n}$ (for any $n$) appear.\footnote{The cyclic group $\mathbb{Z}_3$ also appears amongst the fundamental groups of $E_6$ orbits.} For each non-special orbit inside the parent special orbit, its fundamental group $A({\cal O})$, and the Sommers-Achar group, $C({\cal O})$, are parabolic subgroups of the parent special orbit's Lusztig canonical quotient \cite{collingwood1993nilpotent, chacaltana2015tinkertoys,achar_lcq}.

Physically, the Lusztig canonical quotient and component group conjugacy class appear in the quiver itself either as outer automorphisms, wreathings or foldings. This paper illustrates this interaction and seeks to provide a systematic account.

To clarify the relationships between these subgroups of the canonical quotient group, it should be noted that the orbits within each $S_n$ {\it special piece} are in bijection with partitions of $n$, such that the elements of the partition are the degrees of the $C(\cal{O})$ subgroups and the multiplicities of the elements are the degrees of the $A(\cal{O})$ subgroups.

\begin{sidewaystable}[htp]
\begin{center}
\begin{tabular}{ c c c c }
{  
 \xymatrixcolsep{-4pc} \xymatrix{  { (1^2) \cong (S_2, 1)}  \ar@{-}[d]|{}     \\ 
   { (2) \cong (1, S_2)}     \\  }}
 &
 \xymatrixcolsep{-4pc}   \xymatrix{
  { (1^3) \cong (S_3, 1)}  \ar@{-}[d]|{}     \\ 
   { (2,1) \cong (1, S_2)}  \ar@{-}[d]|{}    \\ 
     { (3) \cong (1, S_3)}     \\  }    
 & 
 \xymatrixcolsep{-3.5pc}  \xymatrix{
 &  {(1^4) \cong (S_4, 1)}  \ar@{-}[d]|{}  &   \\ 
  & { (2,1^2) \cong (S_2, S_2)} \ar@{-}[dl]|{} \ar@{-}[dr]|{}   &   \\ 
 {  (2^2) \cong (S_2, S_2 \times S_2)}   \ar@{-}[dr]|{}  &  & { (3,1) \cong (1, S_3)}  \ar@{-}[dl]|{}\\
     &{ (4) \cong (1, S_4)}   &  \\ }
  &
 \xymatrixcolsep{-3.5pc} \xymatrix{
 &  {(1^5) \cong (S_5, 1)}  \ar@{-}[d]|{}  &   \\ 
  & { (2,1^3) \cong (S_3, S_2)} \ar@{-}[dl]|{} \ar@{-}[dr]|{}   &   \\ 
{ (2^2,1)\cong (S_2, S_2 \times S_2)} \ar@{-}[d]|{} \ar@{-}[drr]|{}   &  & { (3,1^2) \cong (S_2, S_3)}  \ar@{-}[d]|{} \ar@{-}[dll]|{} \\
 {  (3,2) \cong (1,S_3 \times S_2)}   \ar@{-}[dr]|{}  &  & { (4,1) \cong (1, S_4)}  \ar@{-}[dl]|{}\\
     &{ (5) \cong (1, S_5)}   &  \\ }
\end{tabular}
\end{center}
\caption[Topology of $S_n$ Special Pieces]{Topology of $S_n$ Special Pieces. For each orbit in the {\it special pieces} for $S_2$, $S_3$, $S_4$ and $S_5$, the corresponding partition is shown, together with the $(A({\cal O}), C({\cal O}))$ subgroups. The trivial group $S_1$ is denoted by 1.}
\label{tab:SnSP}
\end{sidewaystable}
As a result, both the topology of orbits in a {\it special piece} and their $(A(\cal O)$, $C( \cal O))$ subgroups are determined by the inclusion relations amongst the parabolic subgroups of $S_n$. This topology is shown in table \ref{tab:SnSP} for $S_2$ through $S_5$. The topologies for $S_2$, $S_3$ and $S_4$ are sub-diagrams of $S_5$ obtained by the simple expedient of truncating the partitions. These topologies can be related to the quiver diagrams for S\l odowy intersections involving these orbits, as will be illustrated by the examples in later sections.

If the parent of the {\it special piece} is $d_{LS}/d_{BV}$ dual to itself, the entire {\it special piece} can be considered as self-dual. If the parent of the {\it special piece} is $d_{LS}/d_{BV}$ dual to the parent of a different {\it special piece} then, providing both share the same $S_n$ canonical quotient group, the two {\it special pieces} are dual to each other. Orbits in the remaining {\it special pieces} of nilcones are $d_{LS}/d_{BV}$ dual to isolated {\it special} orbits whose {\it special pieces} are trivially $S_1$.

It should be noted that only the top and bottom orbits in a {\it special piece} exhibit the full $S_n$ symmetry of the {\it special piece} and these are (usually) normal. The orbits internal to {\it special pieces}, which arise for $S_3$ and above, are non-normal, so that their localisation formulae (and magnetic quivers) give their normalisations, which contain non-nilpotent elements.

\FloatBarrier
\subsection{Special Duality}
\label{subsec:SD}

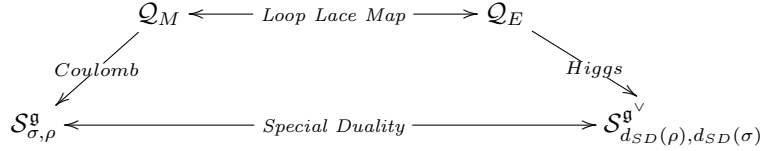
\begin{figure}[htp]
\centering
\begin{displaymath}
    \xymatrix{
  & {\cal Q}_{M} \ar[dl]|{Coulomb}  &  \text{\scriptsize \it Loop Lace Map} \ar[l] \ar[r] &  {\cal Q}_{E} \ar[dr]|{Higgs} &   \\ 
 {{\cal S}^{\mathfrak g}_{\sigma, \rho}}   & &  \text{\scriptsize \it Special Duality} \ar[ll] \ar[rr]  & & {{\cal S}^{{\mathfrak g}^{\vee}}_{d_{SD}(\rho), d_{SD}(\sigma)}}   
 }
\end{displaymath}
\caption[Electric Magnetic Special Duality in Special Pieces]{Electric Magnetic Special Duality in Special Pieces. ${{\cal Q}_{M}}$ and ${{\cal Q}_{E}}$ are quivers that are related by a Loop Lace map. The nilpotent orbit $\sigma$ lies in a $S_n$ {\it special piece} that is either self-dual or dual to another $S_n$ {\it special piece}. The nilpotent orbits $\rho$ and $\sigma$ are dualised under the {\it special duality} map (see text), to $d_{SD}(\rho)$ and $d_{SD}(\sigma)$, and then interchanged to yield the dual intersection $\left({{{\cal S}^{\mathfrak g}}_{\sigma ,\rho }}\right)^{\vee} \equiv {{{\cal S}^{{\mathfrak g^{\vee}}}}_{d_{SD}(\rho), d_{SD}(\sigma)}}$, which lies in in the GNO dual algebra ${\mathfrak g}^{\vee}$.}
\label{fig:GSDSP}
\end{figure}

In order to make the connection from the $S_n$ symmetries of {\it special pieces} to quiver theories, we propose two new maps. The first is an extension to the $d_{LS}$ and $d_{BV}$ maps which yields a full involution for orbits within {\it special pieces} that satisfy certain conditions, and which we refer to as the $d_{SD}$ map. The second is a map between certain quivers that exhibit $S_n$ symmetries, which we refer to as the {\it Loop Lace} map. Using these two maps we obtain an extension of the {\it special duality} relation described earlier (see figure \ref{fig:GSD}), which now embraces {\it non-special} orbits, as shown in figure \ref{fig:GSDSP}.

From a gauge-theory perspective, $d_{SD}$ relates the Coulomb and Higgs branches of \emph{distinct} quiver theories ${{\cal Q}_{M}}$ and ${{\cal Q}_{E}}$; in this sense, $d_{SD}$ differs fundamentally from the action of $d_{LS}$ or $d_{BV}$ between $T_{\sigma}^{\rho}$ theories. In this extension of {\it special duality}, electric and magnetic quivers are treated as different objects and only one moduli space of each participates.

We define the $d_{SD}$ map as in figure \ref{fig:SDO}. The map acts on orbits ${\cal O}_{\sigma}^{\mathfrak g}$, but it is convenient to notate the map in terms of its action on the partition data or other labels $\sigma$ that uniquely identify orbits.

\begin{figure}[htp]
\centering
\begin{displaymath}
 \xymatrix{
\sigma \ar[rrr] | {d_{SD}} \ar[d] {}
 & & &
 {d_{SD}}(\sigma) \ar[d]
 \\ 
 {\left(\sigma_{sp}, \lambda \right)} \ar[u]  \ar[rrr] | {d_{SD}} 
 & & &
  {\left(d_{LS/BV}(\sigma_{sp}), \lambda  \right)} \ar[u] {}
 }
\end{displaymath}
\caption[Special Duality Map on Orbits]{Special Duality Map on Orbits. An orbit $\sigma$ is identified by the couple consisting of (a) the parent $\sigma_{sp}$ of its special piece together with (b) the partition data $\lambda$ from its canonical quotient group $S_n$. The $d_{SD}$ map applies the $d_{LS}$ (or $d_{BV}$ map for $\mathfrak{bc}$-type algebras) to $\sigma_{sp}$, while the partition data $\lambda$ is kept invariant. The new couple identifies the orbit ${d_{SD}}(\sigma)$.
 }
\label{fig:SDO}
\end{figure}
The $d_{SD}(\sigma)$ map yields an involution providing $d_{LS}(\sigma_{sp})$ (or $d_{BV}(\sigma_{sp})$ for $\mathfrak{b}$- and $\mathfrak{c}$-type algebras) has the same $S_n$ canonical quotient group as $\sigma_{sp}$. In the case that this canonical quotient group is trivially $S_1$ the $d_{SD}$ map collapses to $d_{LS}$ (or $d_{BV}$ respectively). 

\begin{table}[h!]
    \centering
    \subfloat[][]{\TabBTwoAC}
    \quad \quad
    \subfloat[][]{\TabBThreeAC}
    \quad \quad
    \subfloat[][]{\TabBFourAC}
\caption[The $d_{SD}$ map on nilpotent orbits of $\mathfrak{b}_{2,3,4}$ and  $\mathfrak{c}_{2,3,4}$]{The $d_{SD}$ map on nilpotent orbits of $\mathfrak{b}_{2,3,4}$ and  $\mathfrak{c}_{2,3,4}$. For {\it special} orbits with trivial {\it special pieces}, the map reduces to $d_{BV}$. Non-trivial {\it special pieces} with $S_{2}$ symmetries are distinguished by shading, according to whether or not they have $S_{2}$ duals. Orbits for which the image of the $d_{SD}$ map is outside the nilcone are indicated by $\bullet$.}
\label{fig:AC_Map_Data_BC}
\end{table}

\begin{table}[h!]
    \centering
    \subfloat[][]{\TabGAC}
    \quad \quad
    \subfloat[][]{\TabFAC}
    \quad \quad
    \subfloat[][]{\TabESixAC}
\caption[The $d_{SD}$ map on nilpotent orbits of $\mathfrak{g}_2$, $\mathfrak{f}_4$ and $\mathfrak{e}_6$]{The $d_{SD}$ map on the nilpotent orbits of $\mathfrak{g}_2$, $\mathfrak{f}_4$ and $\mathfrak{e}_6$. For {\it special} orbits with trivial {\it special pieces}, the map reduces to $d_{LS}$. Non-trivial {\it special pieces} with $S_{k > 1}$ symmetries are distinguished by shading according to degree $k$. Orbits for which the $d_{SD}$ map is outside the nilcone are indicated by $\bullet$.}
\label{fig:AC_Map_Data_GFE6}
\end{table}

\begin{table}[h!]
\centering
\resizebox{\columnwidth}{!}{
    \subfloat[][]{\TabESevenAC}
    \quad \quad
    \subfloat[][]{\TabEEightFirstBitAC}
    \quad \quad
    \subfloat[][]{\TabEEightSecondBitAC}}
\caption[The $d_{SD}$ map on the nilpotent orbits of $\mathfrak{e}_7$ and $\mathfrak{e}_8$.]{The $d_{SD}$ map on the nilpotent orbits of $\mathfrak{e}_7$ and $\mathfrak{e}_8$. For {\it special} orbits with trivial {\it special pieces}, the map reduces to $d_{LS}$. Orbits within non-trivial {\it special pieces} with $S_{k>1}$ symmetries are distinguished by shading according to degree $k$. Orbits for which the $d_{SD}$ map is not an involution are indicated by $\bullet$. Not all special pieces are adjacent in this ordering.}
\label{fig:AC_Map_Data_E7E8}
\end{table}

The action of $d_{SD}$ on the orbits of $\mathfrak{b}_k$ and $\mathfrak{c}_k$ algebras for $k=2,3,4$ is shown in Table \ref{fig:AC_Map_Data_BC}. The orientation preserving action of $d_{SD}$ on orbits in {\it special pieces} is manifest: pairs of {\it special pieces} with $d_{BV}$ dual parents and the same $S_2$ symmetries are mapped to each other.

For $\mathfrak{b}_2 \cong \mathfrak{c}_2$, the $S_2$ special piece is dual to itself under $d_{SD}$ and the group isomorphism. The algebras $\mathfrak{b}_3$ and $\mathfrak{c}_3$ each contain $S_2$ special pieces that are mapped to each other under $d_{SD}$. The algebras $\mathfrak{b}_4$ and $\mathfrak{c}_4$ each contain two $S_2$ special pieces that are mapped to each other under $d_{SD}$. In addition $\mathfrak{c}_3$, $\mathfrak{b}_4$ and $\mathfrak{c}_4$ contain $S_2$ special pieces that are $d_{BV}$ dual to special orbits with trivial special pieces, and for which $d_{SD}$ fails to generate complete images.

The action of $d_{SD}$ on the orbits of Exceptional algebras is shown in Tables \ref{fig:AC_Map_Data_GFE6}
and \ref{fig:AC_Map_Data_E7E8}. The orientation preserving action of $d_{SD}$ on orbits in {\it special pieces} is manifest: pairs of {\it special pieces} with $d_{LS}$ dual parents and the same $S_n$ symmetries are exchanged, while self-dual {\it special pieces} remain unchanged.

For both $\mathfrak{g}_2$ and $E_6$ the $d_{SD}$ map is an involution on the whole nilcone. In the case of $\mathfrak{f}_4$, the $d_{SD}$ map identifies a self-dual $S_4$ {\it special piece}, but the map is not defined on all orbits: the non-trivial $S_2$ special piece associated with the $\tilde{A}_1$ parent orbit is mapped by $d_{LS}$ to the trivial $S_1$ special piece $F_4\left(a_1\right)$, and this leaves the $A_1$ orbit without an image inside the nilcone under $d_{SD}$.

In the case of $E_7$, the nilcone contains one pair of $S_3$ and two pairs of $S_2$ {\it special pieces} over which the $d_{SD}$ map yields an involution. However, two $S_2$ {\it special pieces} are dual to trivial $S_1$ {\it special pieces} under $d_{LS}$, and so two orbits are left without images inside the nilcone under $d_{SD}$.

In the case of $E_8$, the nilcone contains one self-dual $S_5$, one pair of $S_3$ and four pairs of $S_2$ {\it special pieces} over which the $d_{SD}$ map yields an involution. However, one $S_3$ and four $S_2$ {\it special pieces} are dual to trivial $S_1$ {\it special pieces} under $d_{LS}$, and so six orbits are left without images inside the nilcone under $d_{SD}$.

\FloatBarrier

\subsection{Loop Lace Map}
\label{subsec:LL}

Permutation symmetries of the form $S_n$ arise in various ways such as (a) a bouquet of $n$ identical nodes (or legs) centered around some node, (b) an $S_n$ wreathing of $n$ identical nodes (or legs), which may be realised by a loop in certain magnetic quivers, or (c) a folding of $n$ identical nodes (or legs), realised by a non-simply laced link in a magnetic quiver. Naturally, combinations of the above are possible \cite{Bourget:2020bxh} and these can be enumerated in terms of partitions of $n$. Although both magnetic and electric quivers can contain loops and multiple links, no Higgs branch interpretation is known for non-simply laced edges.

The Loop Lace map, as defined in the following, can be applied to quivers that exhibit symmetries and acts to exchange different manifestations of $S_n$ symmetry. In essence, loops and non-simple lacings in a magnetic quiver are exchanged with bouquets and loops, respectively, to define a related electric quiver. The inverse Loop Lace map from an electric to a magnetic quiver simply reverses the above procedure.

The canonical cases involve symmetries of gauge nodes around some central node in a unitary or orthosymplectic quiver, as shown in Table \ref{tab:AC_Quiver_Map}. The magnetic quivers are related by foldings and by symmetrisations (or wreathing) discussed in \cite{Bourget:2020bxh,Bourget:2022tmw}. The electric quivers are related by a variant of the Higgs branch subtraction algorithm introduced in \cite{Bennett:2024loi}. Starting with the the electric quiver in the bottom row, the Higgs subtraction algorithms give the other electric quivers in the table. The magnetic quivers are related by Coulomb branch quiver subtractions as in \cite{Bourget_2022instanton}, with the important proviso that their Coulomb branches generate normalisations of the non-normal orbits internal to special pieces. The Loop Lace map integrates these related electric and magnetic quiver transitions into a single framework.

\begin{table}[h!]
    \centering
    \begin{tabular}{|c|c|c|c|c|}
    \hline
 \multirow{2}{*}{$S_n{~}\mathrm{ Partition}$} & \multicolumn{2}{c|}{Unitary} & \multicolumn{2}{c|}{Orthosymplectic} \\
     \cline{2-5}
    & $Q_{\text{M}}$ & $Q_{\text{E}}$ & $Q_{\text{M}}$ & $Q_{\text{E}}$  \\
   \hline
   $(1^n)$
   &
    \raisebox{-0.5\height}
    {\begin{tikzpicture}
    \node[gauge] (Q) at (0,0){$Q$};
    \node[gauge, label=below:$n$] (1) at (1,0){};
    \draw (1) to[out=-45,in=45,looseness=8](1);
    \draw[-] (Q)--(1);
    \end{tikzpicture}} 
    &
    \raisebox{-0.5\height}
    {\begin{tikzpicture}
    \node[gauge] (Q) at (0,0){$Q$};
    \node[gauge, label=right:$1$] (1) at (1,0.75){};
    \node[gauge, label=right:$1$] (2) at (1,-0.75){};
    \node[] (vdots) at (1,0.1) {$\vdots$};
    \draw[-] (Q)--(1) (Q)--(2);
    \draw[red] (1) circle (0.2cm);
    \draw[red] (2) circle (0.2cm);
    \draw [decorate, decoration = {brace, raise=5pt, amplitude=5pt}] (1.5,1) --  (1.5,-1) node[pos=0.5,right=10pt,black]{$n$};
    \end{tikzpicture}}
    &
    \raisebox{-0.5\height}
    {\begin{tikzpicture}
    \node[gauge] (Q) at (0,0){$Q$};
    \node[gaugeb, label=below:$C_n$] (1) at (1,0){};
    \draw (1) to[out=-45,in=45,looseness=8](1);
    \draw[-] (Q)--(1);
    \end{tikzpicture}} 
    &
     \raisebox{-0.5\height}{\begin{tikzpicture}
    \node[gauge] (Q) at (0,0){$Q$};
    \node[gaugeb, label=right:$C_1$] (1) at (1,0.75){};
    \node[gaugeb, label=right:$C_1$] (2) at (1,-0.75){};
    \node[] (vdots) at (1,0.1) {$\vdots$};
    \draw[-] (Q)--(1) (Q)--(2);
    \draw[green] (1) circle (0.2cm);
    \draw[green] (2) circle (0.2cm);
    \draw [decorate, decoration = {brace, raise=5pt, amplitude=5pt}] (1.5,1) --  (1.5,-1) node[pos=0.5,right=10pt,black]{$n$};
    \end{tikzpicture}}
    \\
    \hline
    $\begin{aligned}
(r_{1}^{\lambda_{1}},\cdots,r_{k}^{\lambda_{k}})\\
    \end{aligned}$ & 
    \raisebox{-0.5\height}
    {\begin{tikzpicture}
    \node[gauge] (Q) at (0,0){$Q$};
    \node[gauge, label=above:$\lambda_{1}$] (1) at (1,0.75){};
    \draw[-] (0.375,0.45)--(0.6,0.45)--(0.6,0.225);
    \node[] at (0.25,0.6) {\small $r_1$};
    \node[] at (0.25,-0.6) {\small $r_k$};
    \draw[-] (0.375,-0.45)--(0.6,-0.45)--(0.6,-0.225);
    \node[gauge, label=below:$\lambda_{k}$] (2) at (1,-0.75){};
    \node[] (vdots) at (1,0.1) {$\vdots$};
    \draw[-] (Q)--(1) (Q)--(2);
    \draw (1) to[out=45,in=-45,looseness=8](1);
    \draw (2) to[out=45,in=-45,looseness=8](2);
    \draw[red] (1) circle (0.2cm);
    \draw[red] (2) circle (0.2cm);
    \draw [decorate, decoration = {brace, raise=5pt, amplitude=5pt}] (1.5,1) --  (1.5,-1) node[pos=0.5,right=10pt,black]{$k$};
    \end{tikzpicture}} 
    & \raisebox{-0.5\height}
    {\begin{tikzpicture}
    \node[gauge] (Q) at (0,0){$Q$};
    \node[gauge, label=left:$r_{1}$] (1) at (0.3,0.9){};
    \node[gauge, label=right:$r_{1}$] (2) at (1.3,0.9){};
    \node[gauge, label=right:$r_{k}$] (3) at (1.3,-0.9){};
    \node[gauge, label=left:$r_{k}$] (4) at (0.3,-0.9){};
    \node[] (vdots) at (1,0.1) {$\vdots$};
    \node[] (vdots) at (0.825,0.9) {$\dots$};
    \node[] (vdots) at (0.825,-0.9) {$\dots$};
    \draw[-] (Q)--(1) (Q)--(2) (Q)--(3) (Q)--(4);
    \draw[red] (1) circle (0.2cm);
    \draw[red] (2) circle (0.2cm);
    \draw[red] (3) circle (0.2cm);
    \draw[red] (4) circle (0.2cm);
    \draw (1) to[out=45,in=135,looseness=8](1);
    \draw (2) to[out=45,in=135,looseness=8](2);
    \draw (3) to[out=-45,in=-135,looseness=8](3);
    \draw (4) to[out=-45,in=-135,looseness=8](4);
    \draw [decorate, decoration = {brace, raise=5pt, amplitude=5pt}] (0.1,1.2) --  (1.5,1.2) node[pos=0.5,above=10pt,black]{$\lambda_{1}$};
    \draw [decorate, decoration = {brace, raise=5pt, mirror,amplitude=5pt}] (0.1,-1.2) --  (1.5,-1.2) node[pos=0.5,below=10pt,black]{$\lambda_{k}$};
    \end{tikzpicture}}
    & 
    \raisebox{-0.5\height}
    {\begin{tikzpicture}
    \node[gauge] (Q) at (0,0){$Q$};
    \node[gaugeb, label=above:$C_{\lambda_{1}}$] (1) at (1,0.75){};
    \draw[-] (0.375,0.45)--(0.6,0.45)--(0.6,0.225);
    \node[] at (0.25,0.6) {\small $r_1$};
    \node[] at (0.25,-0.6) {\small $r_k$};
    \draw[-] (0.375,-0.45)--(0.6,-0.45)--(0.6,-0.225);
    \node[gaugeb, label=below:$C_{\lambda_{k}}$] (2) at (1,-0.75){};
    \node[] (vdots) at (1,0.1) {$\vdots$};
    \draw[-] (Q)--(1) (Q)--(2);
    \draw (1) to[out=45,in=-45,looseness=8](1);
    \draw (2) to[out=45,in=-45,looseness=8](2);
    \draw[green] (1) circle (0.2cm);
    \draw[green] (2) circle (0.2cm);
    \draw [decorate, decoration = {brace, raise=5pt, amplitude=5pt}] (1.5,1) --  (1.5,-1) node[pos=0.5,right=10pt,black]{$k$};
    \end{tikzpicture}} & \raisebox{-0.5\height}
    {\begin{tikzpicture}
    \node[gauge] (Q) at (0,0){$Q$};
    \node[gaugeb, label=left:$C_{r_{1}}$] (1) at (0.3,0.9){};
    \node[gaugeb, label=right:$C_{r_{1}}$] (2) at (1.3,0.9){};
    \node[gaugeb, label=right:$C_{r_{k}}$] (3) at (1.3,-0.9){};
    \node[gaugeb, label=left:$C_{r_{k}}$] (4) at (0.3,-0.9){};
    \node[] (vdots) at (1,0.1) {$\vdots$};
    \node[] (vdots) at (0.825,0.9) {$\dots$};
    \node[] (vdots) at (0.825,-0.9) {$\dots$};
    \draw[-] (Q)--(1) (Q)--(2) (Q)--(3) (Q)--(4);
    \draw[green] (1) circle (0.2cm);
    \draw[green] (2) circle (0.2cm);
    \draw[green] (3) circle (0.2cm);
    \draw[green] (4) circle (0.2cm);
    \draw (1) to[out=45,in=135,looseness=8](1);
    \draw (2) to[out=45,in=135,looseness=8](2);
    \draw (3) to[out=-45,in=-135,looseness=8](3);
    \draw (4) to[out=-45,in=-135,looseness=8](4);
    \draw [decorate, decoration = {brace, raise=5pt, amplitude=5pt}] (0.1,1.2) --  (1.5,1.2) node[pos=0.5,above=10pt,black]{$\lambda_{1}$};
    \draw [decorate, decoration = {brace, raise=5pt, mirror,amplitude=5pt}] (0.1,-1.2) --  (1.5,-1.2) node[pos=0.5,below=10pt,black]{$\lambda_{k}$};
    \end{tikzpicture}}
    \\
    \hline 
 $(n)$ &   \raisebox{-0.5\height}
    {\begin{tikzpicture}
    \node[gauge] (Q) at (0,0){$Q$};
    \node[gauge, label=below:$1$] (1) at (1.6,0){};
    \draw[transform canvas={yshift=1.3pt}] (Q)--(1);
    \draw[transform canvas={yshift=-1.3pt}] (Q)--(1);
    \node at (0.8,0.4) {$n$};
    \draw[-] (0.7,-0.2)--(0.9,0)--(0.7,0.2);
    \end{tikzpicture}} 
    &
    \raisebox{-0.5\height}{\begin{tikzpicture}
    \node[gauge] (Q) at (0,0){$Q$};
    \node[gauge, label=below:$n$] (1) at (1,0){};
    \draw (1) to[out=-45,in=45,looseness=8](1);
    \draw[-] (Q)--(1);
    \end{tikzpicture}}
    & 
     \raisebox{-0.5\height}{\begin{tikzpicture}
    \node[gauge] (Q) at (0,0){$Q$};
    \node[gaugeb, label=below:$C_1$] (1) at (1.6,0){};
    \draw[transform canvas={yshift=1.3pt}] (Q)--(1);
    \draw[transform canvas={yshift=-1.3pt}] (Q)--(1);
    \node at (0.8,0.4) {$n$};
    \draw[-] (0.7,-0.2)--(0.9,0)--(0.7,0.2);
    \end{tikzpicture}}
      & 
     \raisebox{-0.5\height}{\begin{tikzpicture}
    \node[gauge] (Q) at (0,0){$Q$};
    \node[gaugeb, label=below:$C_n$] (1) at (1,0){};
    \draw (1) to[out=-45,in=45,looseness=8](1);
    \draw[-] (Q)--(1);
    \end{tikzpicture}} 
    \\
    \hline
    \end{tabular}
    \caption[Loop Lace Map on Quivers]{Loop Lace Map. The Loop Lace map identifies pairs of magnetic and electric quivers with embedded $S_n$ symmetries. The partitions are subject to $ \sum_{i=1}^{k} \lambda_{i}r_{i} = n$. $\mathcal{Q}$ refers to the rest of the quiver, which acts as a spectator to the $S_n$ action. The decorations should be interpreted in the Higgs branch sense introduced in \cite{Bennett:2024loi}. The unitary quivers are presented unframed, with implicit $U(1)$ framing on any one of the long nodes. }
    \label{tab:AC_Quiver_Map}
\end{table}

In the unitary case, a magnetic quiver containing a $\urm(n)$ node with $S_n$ symmetrisation, realised by an adjoint hypermultiplet (or loop), maps to a bouquet of $n$ decorated nodes of $\urm(1)$ in the electric quiver. A $\urm(1)$ node linked by a non-simply laced edge with $n$ directed links in the magnetic quiver maps to a node of $\urm(n)$ with an $\surm(n)$ adjoint hypermultiplet (or loop) in the electric quiver.

In the orthosymplectic case, a magnetic quiver containing a node of $\sprm(n)$ with a second-rank antisymmetric hypermultiplet (or loop) maps to $n$ decorated $\sprm(1)$ nodes in the electric quiver. An $\sprm(1)$ non-simply laced node with $n$ directed links in the magnetic quiver maps to an $\sprm(n)$ node with a hypermultiplet (or loop) transforming in the $[0,1,0,…]$ antisymmetric irrep of $\sprm(n)$ in the electric quiver.

The cases shown for partitions $(1^n)$ and $(n)$ correspond to the canonical quotient subgroup partition data at the top and bottom of {\it special pieces}, which describes $(A(\cal O),C( \cal O))$ subgroups $(S_n,1)$ and $(1,S_n)$ respectively (see Table \ref{tab:SnSP}). In between, the $C(\cal O))$ subgroups correspond to the lacing numbers of non-simply laced edges (in a magnetic quiver) or the ranks of nodes with hypermultiplets (in an electric quiver); and the $A(\cal O))$ subgroups correspond to the ranks of nodes carrying hypermultiplets (in a magnetic quiver) or the multiplicities of bouquets (in an electric quiver).

Considering dimensions, unit step changes in partitions of $n$ entail that the Coulomb branch transitions between adjacent magnetic quivers $Q_M$ in the heirarchy of partitions (as in Table \ref{tab:AC_Quiver_Map}) have complex dimension 2, corresponding to Kraft Procesi transitions of Kleinian $ADE$, $a_1$ or $m$ type. Dimensional calculations show that this is also true for transitions between adjacent electric quivers $Q_E$. Thus, for a family of such quivers, defined by partitions of $n$, the  difference in complex dimension between the top and bottom quivers is $2(n-1)$. In the case of the magnetic quivers the dimension of the Coulomb branch of the top exceeds that of the bottom by $2(n-1)$, and in the case of the electric quivers, the dimension of the Higgs branch of the bottom exceeds that of the top by $2(n-1)$. Thus the sum of the Coulomb branch plus Higgs branch dimensions is invariant across the heirarchy of partitions.

Clearly, the quivers $Q_E$ can also be evaluated on their Coulomb branches ${\cal C}(Q_E)$. These are given by $S_n$ symmetrisations or wreathings of the bouquet in $Q_E(1^n)$, and, as noted in \cite{Bourget:2020bxh}, these generate covering spaces of ${\cal C}(Q_E(n))$. These have the same dimensions but the volumes of their Hilbert series are in inverse proportion to the cardinality of the symmetrisations. This volumetric covering group ratio $|G_{{\cal C}}|$ corresponds to the order of the finite group quotient $G_{{\cal C}(Q_E)}$, where:

\begin{equation}
G_{{\cal C}(Q_E(\bf{ r^\lambda}))} \equiv {S_n}/\left({\prod_{i=1}^{k} (S_{r_i})^{\lambda_i}}\right),
{~~~~~~~}
|G_{{\cal C}(Q_E(\bf{ r^\lambda}))}| = {n!}/\left({\prod_{i=1}^{k} ({r_i}!)^{\lambda_i}}\right).
\end{equation}

It may be recalled from \cite{Bourget:2020bxh} that symmetrisations via wreathing or folding are not limited to those based on quiver legs with a single gauge node of rank one, but can be applied to quiver legs more generally. While such general symmetrisations can be used to define families of magnetic quivers in a similar manner to that outlined above, their relationship with the electric quivers that are dual to such wreathings is not straightforward. Indeed, for a general magnetic quiver leg being folded/symmetrised, only the top dual electric quiver is known. The resulting partial Loop Lace map takes the form in Table \ref{tab:PartialAC_Quiver_Map}.

\begin{table}[h!]
    \centering
    \begin{tabular}{|c|c|c|}
    \hline
   {$S_n{~}\mathrm{ Partition}$} & $Q_{\text{M}}$ & $Q_{\text{E}}$\\
   \hline
   $(1^n)$
   &
    \raisebox{-0.5\height}
    {\begin{tikzpicture}
    \node[gauge] (Q) at (0,0){$Q$};
    \node[gauge] (1) at (2,0){$q$};
    \node[] at (1.5,0){$[$};
    \node[] at (2.5,0){$]$};
    \node[] at (3.0,0){$\; \wr \; S_n$};
    \draw[-] (Q)--(1);
    \end{tikzpicture}} 
    &
    \raisebox{-0.5\height}
    {\begin{tikzpicture}
    \node[gauge] (Q) at (0,0){$Q$};
    \node[gauge] (1) at (1,0.75){$q$};
    \node[gauge] (2) at (1,-0.75){$q$};
    \node[] (vdots) at (1,0.1) {$\vdots$};
    \draw[-] (Q)--(1) (Q)--(2);
    \draw[red] (1) circle (0.3cm);
    \draw[red] (2) circle (0.3cm);
    \draw [decorate, decoration = {brace, raise=5pt, amplitude=5pt}] (1.5,1) --  (1.5,-1) node[pos=0.5,right=10pt,black]{$n$};
    \end{tikzpicture}} \\
    \hline
    $\begin{aligned}  (r_{1}^{\lambda_{1}},\cdots,r_{k}^{\lambda_{k}})\\
    \end{aligned}$ 
    & 
    \raisebox{-0.5\height}
    {\begin{tikzpicture}
    \node[gauge] (Q) at (0,0){$Q$};
    \node[gauge] (1) at (1,0.75){$q$};
    \draw[-] (0.375,0.45)--(0.6,0.45)--(0.6,0.225);
    \node[] at (0.25,0.6) {\small $r_1$};
    \node[] at (0.25,-0.6) {\small $r_k$};
    \draw[-] (0.375,-0.45)--(0.6,-0.45)--(0.6,-0.225);
    \node[] at (0.5,0.75){$[$};
    \node[] at (1.5,0.75){$]$};
    \node[] at (2.0,0.75){$\; \wr \; S_{\lambda_1}$};
    \node[gauge] (2) at (1,-0.75){$q$};
     \node[] at (0.5,-0.75){$[$};
    \node[] at (1.5,-0.75){$]$};
    \node[] at (2.0,-0.75){$\; \wr \; S_{\lambda_k}$};
    \node[] (vdots) at (2,0.1) {$\vdots$};
    \draw[-] (Q)--(1) (Q)--(2);
    \draw[red] (1) circle (0.3cm);
    \draw[red] (2) circle (0.3cm);
    \draw [decorate, decoration = {brace, raise=5pt, amplitude=5pt}] (2.5,1) --  (2.5,-1) node[pos=0.5,right=10pt,black]{$k$};
    \end{tikzpicture}} 
    & \\
    \hline 
 $(n)$ &   
 \raisebox{-0.5\height}
    {\begin{tikzpicture}
    \node[gauge] (Q) at (0,0){$Q$};
    \node[gauge] (1) at (2,0){$q$};
    \draw[transform canvas={yshift=1.3pt}] (Q)--(1);
    \draw[transform canvas={yshift=-1.3pt}] (Q)--(1);
    \node at (1,0.4) {$n$};
    \draw[-] (0.7,-0.2)--(0.9,0)--(0.7,0.2);
    \end{tikzpicture}} 
    & \\
    \hline
    \end{tabular}
    \caption[Partial Loop Lace Map on Quivers]{Partial Loop Lace Map. The partial Loop Lace map identifies pairs of magnetic and electric quivers with embedded $S_n$ symmetries. $q$ is some general ``good" magnetic quiver leg that is being symmetrised. $\mathcal{Q}$ refers to the rest of the quiver, which acts as a spectator to the $S_n$ action. The partitions are subject to $ \sum_{i=1}^{k} \lambda_{i}r_{i} = n$. The decorations should be interpreted in the Higgs branch sense introduced in \cite{Bennett:2024loi}.}
    \label{tab:PartialAC_Quiver_Map}
\end{table}

Remarkably, the intersection/transition between the orbits in a nilcone $\mathcal{N}^{\mathfrak{g}}$ represented by the $S_{2,3,4,5}$ wreathed and the $S_{2,3,4,5}$ folded variants of a magnetic quiver $q$, where $\rm{dim}_{\mathbb H}(q) = m$, is given by
\begin{equation}
    \frac{\mathrm{Sym}^{2,3,4,5}\left(\mathbb{H}^{m}\right)}{\mathbb{H}^{m}}
\label{quiv:2345_wreath}
\end{equation}
Examples familiar to the quiver literature include $\urm(2)$, $\urm(3)$, $\urm(4)$ and $\urm(5)$ each with an adjoint hypermultiplet, shown below.
\begin{equation}
\raisebox{-0.5\height}{\begin{tikzpicture}
\node (a0) at (-3.5,0) {\begin{tikzpicture}
    \node[gauge] (Q) at (0,0){$Q$};
    \node[gauge, label=below:$2$] (1) at (1,0){};
    \draw (1) to[out=-45,in=45,looseness=8](1);
    \draw[-] (Q)--(1);
\end{tikzpicture}};
\node (b0) at (-3.5,-3) {\begin{tikzpicture}
    \node[gauge] (Q) at (0,0){$Q$};
    \node[gauge, label=below:$1$] (1) at (1,0){};
    \draw[transform canvas={yshift=1pt}] (Q)--(1);
    \draw[transform canvas={yshift=-1pt}] (Q)--(1);
    \draw[-] (0.5,-0.2)--(0.7,0)--(0.5,0.2);
\end{tikzpicture}};
\node (a1) at (0,0) {\begin{tikzpicture}
    \node[gauge] (Q) at (0,0){$Q$};
    \node[gauge, label=below:$3$] (1) at (1,0){};
    \draw (1) to[out=-45,in=45,looseness=8](1);
    \draw[-] (Q)--(1);
\end{tikzpicture}};
\node (b1) at (0,-3) {\begin{tikzpicture}
    \node[gauge] (Q) at (0,0){$Q$};
    \node[gauge, label=below:$1$] (1) at (1,0){};
    \draw[transform canvas={yshift=1.3pt}] (Q)--(1);
    \draw[transform canvas={yshift=-1.3pt}] (Q)--(1);
    \draw[-] (Q)--(1);
    \draw[-] (0.5,-0.2)--(0.7,0)--(0.5,0.2);
\end{tikzpicture}};
\node (a2) at (3.5,0) {\begin{tikzpicture}
    \node[gauge] (Q) at (0,0){$Q$};
    \node[gauge, label=below:$4$] (1) at (1,0){};
    \draw (1) to[out=-45,in=45,looseness=8](1);
    \draw[-] (Q)--(1);
\end{tikzpicture}};
\node (b2) at (3.5,-3) {\begin{tikzpicture}
    \node[gauge] (Q) at (0,0){$Q$};
    \node[gauge, label=below:$1$] (1) at (1,0){};
    \draw[transform canvas={yshift=1.5pt}] (Q)--(1);
    \draw[transform canvas={yshift=-1.5pt}] (Q)--(1);
    \draw[transform canvas={yshift=0.5pt}] (Q)--(1);
    \draw[transform canvas={yshift=-0.5pt}] (Q)--(1);
    \draw[-] (0.5,-0.2)--(0.7,0)--(0.5,0.2);
\end{tikzpicture}};
\node (a3) at (7,0) {\begin{tikzpicture}
    \node[gauge] (Q) at (0,0){$Q$};
    \node[gauge, label=below:$5$] (1) at (1,0){};
    \draw (1) to[out=-45,in=45,looseness=8](1);
    \draw[-] (Q)--(1);
\end{tikzpicture}};
\node (b3) at (7,-3) {\begin{tikzpicture}
    \node[gauge] (Q) at (0,0){$Q$};
    \node[gauge, label=below:$1$] (1) at (1,0){};
    \draw (1) to[out=-45,in=45,looseness=8](1);
     \draw[transform canvas={yshift=2pt}] (Q)--(1);
    \draw[transform canvas={yshift=-2pt}] (Q)--(1);
    \draw[transform canvas={yshift=1pt}] (Q)--(1);
    \draw[transform canvas={yshift=-1pt}] (Q)--(1);
    \draw[-] (Q)--(1);
    \draw[-] (0.5,-0.2)--(0.7,0)--(0.5,0.2);
\end{tikzpicture}};
\draw[-] (a0)--(b0) node[midway, right]{$\frac{\mathrm{Sym^{2}\left(\mathbb{H}\right)}}{\mathbb{H}}$};
\draw[-] (a1)--(b1) node[midway, right]{$\frac{\mathrm{Sym^{3}\left(\mathbb{H}\right)}}{\mathbb{H}}$};
\draw[-] (a2)--(b2) node[midway, right]{$\frac{\mathrm{Sym^{4}\left(\mathbb{H}\right)}}{\mathbb{H}}$};
\draw[-] (a3)--(b3) node[midway, right]{$\frac{\mathrm{Sym^{5}\left(\mathbb{H}\right)}}{\mathbb{H}}$};
\end{tikzpicture}}
\end{equation}
Interestingly, it is shown in Sections \ref{sec:bcspecial} to \ref{sec:e8special} that there is in fact a larger set of theories for which the transverse space between a wreathing and folding recovers a symmetric product space. Although such transverse spaces between wreathed and folded quivers are in general not computable using current techniques, the nilcone structure leads in certain circumstances to an identification via the localisation formula. Table \ref{tab:SpecialTrans} shows the various transitions across special pieces in exceptional nilcones, alongside choice examples drawn from classical nilcones. The results are consistent with those in \cite{fu2024localgeometryspecialpieces}.

\begin{sidewaystable}[h!]
 \centering
     \begin{tabular}{|c|c|c|c|c|c|}
     \hline
     Group  & Special Piece & $\rm{dim}_{\mathbb{C}}$ & $\bar A(\cal O)$  & Transition &  Hilbert Series \\
 \hline
     $B_3 $ & $\{(3,1^4),(2^2,1^3)\}$ & $2$ & $S_2$ & $\frac{\text{Sym}^2(\mathbb{H})}{\mathbb{H}}\equiv c_1$ & $\frac{1+t^2}{\left(1-t^2\right)^2}$ \\
\hline

     $B_4 $ & $\{(3,2^2,1^2),(2^4,1)\}$ &
     \multirow{4}{*}{ $4$ } &
     \multirow{4}{*}{ $S_2$} &
     \multirow{4}{*}{$\frac{\text{Sym}^2\left(\mathbb{H}^2\right)}{\mathbb{H}^2}\equiv c_2$} &
     \multirow{4}{*}{ $\frac{1+6 t^2+t^4}{\left(1-t^2\right)^4} $} \\
 \cline {1-2}

     $C_4 $ & $\{(4,2,1^2),(4,1^4)\}$ & & & & \\
 \cline {1-2}

    $E_8 $ & $\{A_4+2A_1,2A_3\}$ & & & & \\
  \cline {1-2}
   $E_8 $ & $\{E_7(a_3),D_6\}$ & & & & \\

  \hline
       $G_2 $ & $\{G_2(a_1), A_1\}$ & \multirow{2}{*} {$4$} & \multirow{2}{*}{ $S_3$ }& \multirow{2}{*}{$\frac{\text{Sym}^3(\mathbb{H})}{\mathbb{H}}$} & \multirow{2}{*}{$\frac{1+t^2+2 t^3+t^4+t^6}{\left(1-t^2\right)^2 \left(1-t^3\right)^2}$ }\\
 \cline{1-2}

 \hline

   $F_4 $ & $\{\tilde A_1,A_1\}$ & \multirow{2}{*}{$6$} & \multirow{2}{*}{$S_2$} & \multirow{2}{*}{$\frac{\text{Sym}^2\left(\mathbb{H}^3\right)}{\mathbb{H}^3}\equiv c_3$} &\multirow{2}{*} {$\frac{\left(1+t^2\right) \left(1+14 t^2+t^4\right)}{\left(1-t^2\right)^6} $} \\
  \cline{1-2}

   $E_8 $ & $\{D_5(a_1),D_4+A_1\}$ & & & & \\
  \hline
  
   $F_4 $ & $\{F_4(a_3),A_2+\tilde A_1\}$ & $6$ & $S_4$ & $\frac{\text{Sym}^4(\mathbb{H})}{\mathbb{H}}  $   &$\frac{\left(\scriptsize{\begin{array}{c}1+t^2+2 t^3+4t^4+2 t^5+4 t^6+\\2 t^7+4 t^8+ 2 t^9+ t^{10}+   t^{12}\end{array}}\right)}{\left(1-t^2\right)^2  \left(1-t^3\right)^2  \left(1-t^4\right)^2}$ \\
  \hline

  $E_8 $ & $\{ A_2+A_1,4A_1 \}$ & $8$ & $S_2$ &$\frac{\text{Sym}^2\left(\mathbb{H}^4\right)}{\mathbb{H}^4}\equiv c_4$  & $\frac{1+28 t^2+70 t^4+28 t^6+  t^8}{\left(1-t^2\right)^8}$ \\
  \hline
  
    $E_8 $ & $\{ D_4(a_1)+A_1,2A_2+2A_1 \}$ & $8$ & $S_3$ & $\frac{\text{Sym}^3\left(\mathbb{H}^2\right)}{\mathbb{H}^2}$ & $\frac{\left(\scriptsize{\begin{array}{c}1+6 t^2+16 t^3+21 t^4+36 t^5+56 t^6+\\ 36 t^7+21 t^8+16 t^9+6 t^{10}+    t^{12} \end{array}}\right)} {\left(1-t^2\right)^4 \left(1-t^3\right)^4}$ \\
  \hline
  
      $E_8 $ & $\{ E_8(a_7),A_4+A_3 \}$ & $8$ & $S_5$ & $\frac{\text{Sym}^5\left(\mathbb{H}\right)}{\mathbb{H}}$ & $\frac{\left(\scriptsize{\begin{array}{c} 1+t^2+2 t^3+4 t^4+6 t^5+7 t^6+8 t^7+\\12 t^8  +12 t^9  +14 t^{10} +12 t^{11} +12 t^{12}+8 t^{13}\\ +7 t^{14}+6 t^{15}+4 t^{16} +2 t^{17}+t^{18}+     t^{20} \end{array}}\right)} {\left(1-t^2\right)^2 \left(1-t^3\right)^2 \left(1-t^4\right)^2 \left(1-t^5\right)^2}$ \\
  \hline
     \end{tabular}
 \caption[Special Piece Transitions]{Special Piece Transitions. For each special piece, its canonical quotient group and the intersection between the top and bottom orbits is given, along with its dimension and Hilbert Series. The special piece transitions generically looks like the $k^{\text{th}}$ symmetric product of a vector space $\mathbb{H}^{n}$.}
 \label{tab:SpecialTrans}
 \end{sidewaystable}
 \FloatBarrier
 \FloatBarrier
The resulting combination of the transitions between the top and bottom of a special piece into a single symplectic singularity is consistent with our use of $d_{SD}$ to decouple the internal structure of special pieces from the action of the Lusztig Spaltenstein map.

We do not present Loop Lace maps for transitions that are entirely internal to Special Pieces. Other than singularities of the type $c_k$, many of the transitions involved are non-normal and their quivers are not known.

The examples in Sections \ref{sec:bcspecial} to \ref{sec:e8special} examine the action of the Loop Lace map on a selection of electric and magnetic quivers for orbits, S\l odowy slices and S\l odowy intersections. The evaluation of their Higgs and Coulomb branches identifies relationships that correspond to the $d_{SD}$ map. In effect, the examples illustrate how the Loop Lace map on quivers induces the $d_{SD}$ map on nilpotent orbits.

\paragraph{Loop Lace Example Conventions}
\begin{enumerate}
\item The examples are organised around families of quivers. Each family is defined by a parent magnetic quiver $Q_M(\rho_{sp},\sigma)$ for a S\l odowy intersection $S_{\rho_{sp},\sigma}^{\mathfrak g}$, where $\rho_{sp}$ is the parent of a $S_{\ge 2}$ {\it special piece}. The orbit ${\sigma}$ generally lies below this {\it special piece} in ${\cal N}^{\mathfrak{g}}$.  
    
\item The Loop Lace map, together with the partition data for the {\it special piece}, determines the other magnetic quivers $Q_M$ and electric quivers $Q_E$ in the family.
    
\item Each orbit in the special piece is identified by its label $\rho$, along with its component group $A\left(\mathcal{O}\right)$ and Somers-Achar group $C\left(\mathcal{O}\right)$.

\item The Coulomb and Higgs branches of $Q_M$ and $Q_E$, respectively, are S\l odowy intersections, with complex dimensions as shown. The dimensions of the intersections $S_{\rho,\sigma}^{\mathfrak g}$ and their Special Duals $S_{d_{SD}(\sigma),d_{SD}(\rho)}^{\mathfrak g}$ are given.

\item $G_{\mathcal{C}}(Q_E)$ identifies the cover by ${\cal C}(Q_{\text{E}})$ of ${\cal C}(Q_{\text{M}}(\rho_{sp},\sigma))$.

\item In the absence of explicit framing, unitary quivers are presented unframed, with implicit $U(1)$ framing on any one of the long nodes.

\item In addition to the usual flavour node symmetries, the electric quivers carry Higgs branch symmetries $A_1$ on each hypermultiplet.

\item In the case of a partial Loop Lace map, only one electric quiver is shown due to its trivial {\it special piece}.

\item Subtraction rules for Coulomb branches are as given in \cite{Bourget_2022instanton}. 

\end{enumerate}

\FloatBarrier
\section{Special Pieces in Classical Nilcones}
\label{sec:bcspecial}

Non-trivial special pieces occur in classical nilcones of types $\mathfrak{b}_{k}$, $\mathfrak{c}_{k}$ and $\mathfrak{d}_{k}$. These nilcones, which are of complex dimension $2k^{2}$, $2k^{2}$ and $2k(k-1)$ respectively, represent an infinite family of spaces whose associated orbits admit an explicit combinatorics. Classical nilpotent orbits are labelled by a partition $\lambda$ of the dimension $n$ of the vector representation ($n= 2k+1$, $n= 2k$ or $n=2k$, respectively). These partitions satisfy further conditions according to algebra type \cite{collingwood1993nilpotent}.

Both the Lusztig-Spaltenstein $d_{LS}$ and Barbasch-Vogan $d_{BV}$ maps provide order reversing involutions on the {\it special} orbits within Classical nilcones. They are both algorithmic, based on the transposition of partitions combined with the addition/removal/shifting of partition elements \cite{Cabrera:2017njm}. The action of these maps on non-special orbits is not involutive, but serves to identify the parent orbits of the special pieces to which they belong. Only $(S_2)^{n}$ special pieces arise in Classical algebras. The $d_{BV}$ map differs from the $d_{LS}$ map in the important respect that $d_{BV}$ maps between partitions for $\mathfrak{b}_{k}$ and $\mathfrak{c}_{k}$ orbits. This implementation of GNO duality is necessary to obtain the {\it special duality} relations (see figure \ref{fig:GSD}) between intersections of $\mathfrak{b}_{k}$ and $\mathfrak{c}_{k}$ algebras.

As is well-known, the minimal orbits of $\mathfrak{b}_{k}$ and $\mathfrak{c}_{k}$ are not special. Indeed, they invariably form part of $S_2$ special pieces, of which the next to minimal orbits are the parents. In the case of $\mathfrak{b}_{k}$, these $S_2$ special pieces are dual to $S_2$ special pieces in $\mathfrak{c}_{k}$ and full Loop Lace maps can be constructed.

Thus, the well-known quiver in \eqref{quiv:Bk_example} realises $\overline{n.min.\mathfrak{b}_{k}}$ as its Coulomb branch. The corresponding partition is $(3,1^{2k-2})$. The presence of a single adjoint hypermultiplet on the $\urm(2)$ gauge node is in keeping with the prescription of Section \ref{sec:Theory} -- the $\overline{n.min.\mathfrak{b}_{k}}$ orbit is an $S_2$ quotient of $\overline{min.\mathfrak{d}_{k+1}}$, which has trivial component group.
\begin{equation}
\raisebox{-0.5\height}{\begin{tikzpicture}
    \node[gauge, label=below:$2$] (1) at (0,0){};
    \draw (1) to[out=-135,in=135,looseness=8](1);
    \node[gauge, label=below:$2$] (2) at (1,0){};
    \node[] (cdots) at (2,0){$\cdots$};
    \node[gauge, label=below:$2$] (3) at (3,0){};
    \node[gauge, label=below:$1$] (4) at (4,0){};
    \node[gauge, label=right:$1$] (5) at (3,1){};
    \draw[-] (1)--(2)--(cdots)--(3)--(4) (3)--(5);
\end{tikzpicture}}
\label{quiv:Bk_example}
\end{equation}

A Loop Lace map constructed around \eqref{quiv:Bk_example} generates the quivers in Table \ref{tab:loop-2-...-2-1-1}. The Coulomb and Higgs branches of these quivers generate the closures of orbits and S\l odowy slices which are related by the action of $d_{SD}$, as shown.
The special piece $\{\left(3,1^{2k-2}\right),\;\left(2^{2},1^{2k-3}\right)\}$ in the $\mathfrak{b}_{k}$ nilcone is mapped by $d_{SD}$ to $\{\left(2k-2,2\right),\;\left(2k-2,1^{2}\right)\}$ in the $\mathfrak{c}_{k}$ nilcone. The mappings under $d_{BV}$ and $d_{SD}$ are compared below.
\begin{align}d_{SD}\left[\left(3,1^{2k-2}\right)\right]&=d_{BV}\left[\left(3,1^{2k-2}\right)\right]= \left(2k-2,2\right)\\
    d_{SD}\left[\left(2^{2},1^{2k-3}\right)\right]&= \left(2k-2,1^{2}\right)\\
    d_{BV}\left[\left(2^{2},1^{2k-3}\right)\right]&= \left(2k-2,2\right)
\end{align}

Whereas under $d_{BV}$, the minimal and next to minimal orbits of $\mathfrak{b}_{k}$ are both mapped to the subregular orbit of $\mathfrak{c}_{k}$, the $d_{SD}$ map distinguishes these images to identify the subregular and sub-subregular orbits of $\mathfrak{c}_{k}$, consistent with the quivers in the Loop Lace map. The Higgs branch of the top electric quiver $Q_E$ yields the Kleinian $D_{k+1}$ singularity for all ranks $k$. The transition between the minimal and next to minimal orbits of $\mathfrak{b}_{k}$ is $c_1$ and the transition between the sub-subregular and subregular orbits of $\mathfrak{c}_{k}$ is Kleinian $A_1$.

\begin{table}[h!]
\centering
\scalebox{0.8}{\begin{tabular}{|c|c|c|c|c|c|}
\hline
    $\rho$ & $\left(\left|{\cal C}(Q_M)\right|, \left|{\cal H}(Q_E)\right|\right)$ & $(A(\cal O), C(\cal O))$ & $Q_{\text{M}}\left[\bar{\mathcal{O}}^{\mathfrak{b}_k}_{\rho}\right]$ & $Q_{\text{E}}\left[\mathcal{S}^{\mathfrak{c}_k}_{d_{SD}\left(\rho\right)}\right]$ & $G_{\mathcal{C}}$ \\
    \hline
    $\left(3,1^{2k-2}\right)$ & 
    $\left(4k-2,2\right)$ & $(S_2,1)$ &  
    \raisebox{-0.5\height}
    {\begin{tikzpicture}
    \node[gauge, label=below:$2$] (1) at (0,0){};
    \draw (1) to[out=45,in=-45,looseness=8](1);
    \node[gauge, label=below:$2$] (2) at (-1,0){};
    \node[] (cdots) at (-2,0){$\cdots$};
    \node[gauge, label=below:$2$] (3) at (-3,0){};
    \node[gauge, label=below:$1$] (4) at (-4,0){};
    \node[gauge, label=right:$1$] (5) at (-3,1){};
    \draw[-] (1)--(2)--(cdots)--(3)--(4) (3)--(5);
    \draw [decorate, decoration = {brace, raise=5pt, mirror,amplitude=5pt}](-3.25,-0.5)-- (-0.75,-0.5)   node[pos=0.5,below=10pt,black]{$k-2$};
    \end{tikzpicture}}
    & \raisebox{-0.5\height}
    {\begin{tikzpicture}
    \node[gauge, label=below:$1$] (1) at (0,0){};
    \node[gauge, label=left:$1$] (6) at (1,1){};
    \node[gauge, label=below:$2$] (2) at (1,0){};
    \node[] (cdots) at (2,0){$\cdots$};
    \node[gauge, label=below:$2$] (3) at (3,0){};
    \node[gauge, label=below:$1$] (4) at (4,0){};
    \node[gauge, label=right:$1$] (5) at (3,1){};
    \draw[-] (1)--(2)--(cdots)--(3)--(4) (3)--(5) (2)--(6);
    \draw[green] (4) circle (0.2cm);
    \draw[green] (5) circle (0.2cm);
    \draw [decorate, decoration = {brace, raise=5pt, mirror,amplitude=5pt}] (0.75,-0.5) --  (3.25,-0.5) node[pos=0.5,below=10pt,black]{$k-2$};
    \end{tikzpicture}} & $S_2$\\ 
    \hline
    $\left(2^{2},1^{2k-3}\right)$ &
    $\left(4k-4,4\right)$ & $(1,S_2)$ & \raisebox{-0.5\height}
    {\begin{tikzpicture}
    \node[gauge, label=below:$1$] (1) at (0,0){};
    \node[gauge, label=below:$2$] (2) at (-1,0){};
    \node[] (cdots) at (-2,0){$\cdots$};
    \node[gauge, label=below:$2$] (3) at (-3,0){};
    \node[gauge, label=below:$1$] (4) at (-4,0){};
    \node[gauge, label=right:$1$] (5) at (-3,1){};
    \draw[-] (2)--(cdots)--(3)--(4) (3)--(5);
    \draw[transform canvas={yshift=1.3pt}] (1)--(2);
    \draw[transform canvas={yshift=-1.3pt}] (1)--(2);
    \draw[-] (-0.6,-0.2)--(-0.4,0)--(-0.6,0.2);
    \draw [decorate, decoration = {brace, raise=5pt, mirror,amplitude=5pt}] (-3.25,-0.5) -- (-0.75,-0.5)  node[pos=0.5,below=10pt,black]{$k-2$};
    \end{tikzpicture}} 
    & \raisebox{-0.5\height}{\begin{tikzpicture}
    \node[gauge, label=below:$2$] (1) at (0,0){};
    \draw (1) to[out=45,in=-45,looseness=8](1);
    \node[gauge, label=below:$2$] (2) at (-1,0){};
    \node[] (cdots) at (-2,0){$\cdots$};
    \node[gauge, label=below:$2$] (3) at (-3,0){};
    \node[gauge, label=below:$1$] (4) at (-4,0){};
    \node[gauge, label=right:$1$] (5) at (-3,1){};
    \draw[-] (1)--(2)--(cdots)--(3)--(4) (3)--(5);
    \draw [decorate, decoration = {brace, raise=5pt, mirror,amplitude=5pt}](-3.25,-0.5)-- (-0.75,-0.5)   node[pos=0.5,below=10pt,black]{$k-2$};
    \end{tikzpicture}}
    & $1$\\ \hline
\end{tabular}}
\caption[Electric and Magnetic Quivers for Orbits and Slices to $S_2$ Special Pieces in ${\mathfrak{b}_k}$ and ${\mathfrak{c}_k}$ nilcones]{Electric and Magnetic Quivers for Orbits and Slices to $S_2$ Special Pieces in ${\mathfrak{b}_k}$ and ${\mathfrak{c}_k}$ nilcones. The ${\mathfrak{b}_k}$ special piece $\{(3, 1^{2k-2}), (2^2, 1^{2k-3})\}$, whose parent is the next to minimal orbit of ${\mathfrak{b}_k}$, is dual under $d_{SD}$ to the special piece $\{(2k-2, 2), (2k-2, 1^{2})\}$ of ${\mathfrak{c}_k}$.}
\label{tab:loop-2-...-2-1-1}
\end{table}

The situation for the quivers surrounding the $S_2$ special piece whose parent is the next to minimal orbit of $\mathfrak{c}_{k}$ differs somewhat, since the $d_{BV}$ dual is a special orbit that only has a trivial special piece. Consequently, only a partial Loop Lace map can be constructed, as in Table \ref{tab:ckn2m}. The transition between the minimal and next to minimal orbits of $\mathfrak{c}_{k}$ is $c_{k-1}$. The Higgs branch of the electric quiver yields the Kleinian $A_{2k-1}$ singularity for all ranks $k$.

\begin{table}[h!]
\centering
\scalebox{0.75}{
\begin{tabular}{|c|c|c|c|c|c|}
\hline
    $\rho$ & 
     $\left(\left|{\cal C}(Q_M)\right|, \left|{\cal H}(Q_E)\right|\right)$ 
     & $(A(\cal O), C(\cal O))$
    & $Q_{\text{M}}\left[\bar{\mathcal{O}}^{\mathfrak{c}_k}_{\rho}\right]$ & $Q_{\text{E}}\left[\mathcal{S}^{\mathfrak{b}_k}_{d_{SD}\left(\rho\right)}\right]$ 
    & $G_{\mathcal{C}}(Q_E)$ \\
    \hline
    $\left(2^2,1^{2k-4}\right)$ 
    & $\left(4k+2,2\right)$ 
    & $(S_2,1)$ &
    \raisebox{-0.5\height}
    {\begin{tikzpicture}
    \node[gauge, label=below:$1$] (1) at (0,0){};
    \node[gauge, label=below:$1$] (2) at (-1,0){};
    \node[] (cdots) at (-2,0){$\cdots$};
    \node[gauge, label=below:$1$] (3) at (-3,0){};
    \node[flavour, label=below:$1$] (4) at (-4,0){};
    \draw[-](1)--(2)-- (cdots) --(3)--(4);
    \node[] at (-0.5,0){$]$};
    \node[] at (-4.5,0){$[$};
    \node[] at (-5.0,0){$S_2 \; \wr \;$};
    \draw [decorate, decoration = {brace, raise=5pt, mirror,amplitude=5pt}] (-3.25,-0.5)--(-0.75,-0.5) node[pos=0.5,below=10pt,black]{$k-1$};
    \end{tikzpicture}}  
    &
    \raisebox{-0.5\height}{\begin{tikzpicture}
    \node[gauge, label=below:$1$] (1) at (0,0){};
    \node[gauge, label=below:$1$] (2) at (1,0){};
    \node[] (cdots) at (2,0){$\cdots$};
    \node[gauge, label=below:$1$] (4) at (3,0){};
    \node[gauge, label=below:$1$] (5) at (4,0){};
    \node[flavour, label=right:$1$] (6) at (0,1){};
    \node[flavour, label=right:$1$] (7) at (4,1){};
    \draw[-] (6)--(1)--(2)--(cdots)--(4)--(5)--(7);
    \draw[green] \convexpath{6,2,1} {0.2cm};
    \draw[green] \convexpath{4,7,5} {0.2cm};
    \draw [decorate, decoration = {brace, raise=5pt, mirror, amplitude=5pt}](0.75,-0.5)--(3.25,-0.5) node[pos=0.5,below=10pt,black]{$2k-3$};
    \end{tikzpicture}} 
    & $S_2$\\ 
    \hline
    $\left(2,1^{2k-2} \right)$ & 
    $\left(2k, \textrm{N/A}\right)$ 
    \raisebox{-0.5\height}
    & $(1,S_2)$ &
    \raisebox{-0.5\height}{\begin{tikzpicture}
    \node[gauge, label=below:$1$] (1) at (0,0){};
    \node[gauge, label=below:$1$] (2) at (-1,0){};
    \node[gauge, label=below:$1$] (3) at (-3,0){};
    \node[flavour, label=below:$1$] (4) at (-4,0){};
    \node[] (cdots) at (-2,0){$\cdots$};
    \draw[-] (2)--(cdots)--(3)--(4);
    \draw[transform canvas={yshift=1.3pt}] (1)--(2);
    \draw[transform canvas={yshift=-1.3pt}] (1)--(2);
    \draw[-] (-0.4,0.2)--(-0.6,0)--(-0.4,-0.2);
    \draw [decorate, decoration = {brace, raise=5pt, mirror,amplitude=5pt}] (-3.25,-0.5)--(-0.75,-0.5) node[pos=0.5,below=10pt,black]{$k-1$};
    \end{tikzpicture}} 
    & \raisebox{-0.5\height}{$\textrm{N/A}$} & \textrm{N/A} \\ \hline
\end{tabular}}
\caption[Electric and Magnetic Quivers for Orbits and Slices to $S_2$ Special Piece in ${\mathfrak{c}_k}$ nilcone]{Electric and Magnetic Quivers for Orbits and Slices to $S_2$ Special Piece in ${\mathfrak{c}_k}$ nilcone. The special piece $\{\left(2^2,1^{2k-4}\right), \left(2,1^{2k-2} \right)  \}$, whose parent is the next to minimal orbit of ${\mathfrak{c}_k}$, is dual under $d_{BV}$ to the subregular orbit $(2k-1,1^2)$ of ${\mathfrak{b}_k}$, which has a trivial special piece.}
\label{tab:ckn2m}
\end{table}

In the case of rank 4, the  ${\mathfrak{b}_4}$ special piece $\{(5,3,1), (4^2,1)\}$ maps to the ${\mathfrak{c}_4}$ special piece $\{(2^4), (2^3,1^2)\}$ under $d_{SD}$, and their related intersections can be arranged in a Loop Lace map, as in Table \ref{tab:2-2-loopbc4}. The Coulomb branch of the parent magnetic quiver is $\rm{Sym}^2({\mathbb C_2}/{\mathbb Z_2})$, while its Higgs branch evaluates as the reduced $2\;\surm(2)$ instanton moduli space on $\mathbb{C}^{2}$.

\begin{table}[h!]
\centering
\begin{tabular}{|c|c|c|c|c|c|}
\hline
$\rho$ & $\left(\left|{\cal C}(Q_M)\right|, \left|{\cal H}(Q_E)\right|\right)$ & $(A(\cal O), C(\cal O))$ & $Q_{\text{M}}\left[\mathcal{S}^{\mathfrak{b}_4}_{\rho,\;(3^3)}\right]$ & $Q_{\text{E}}\left[\mathcal{S}^{\mathfrak{c}_4}_{(3^2,2),\;d_{SD}(\rho)}\right]$ & $G_{\mathcal{C}}(Q_E)$\\
\hline
$\left(5,3,1 \right)$ & $\left(4,4\right)$ & $(S_2,1)$ & \raisebox{-0.5\height}{\begin{tikzpicture}
    \node[flavour, label=below:$2$] (1) at (0,0){};
    \node[gauge, label=below:$2$] (2) at (1,0){};
    \draw (2) to[out=-45,in=45,looseness=8](2);
    \draw[-] (1)--(2);
    \end{tikzpicture}} & \raisebox{-0.5\height}{\begin{tikzpicture}
    \node[flavour, label=below:$2$] (1) at (0,0){};
    \node[gauge, label=right:$1$] (2) at (1,0.5){};
    \node[gauge, label=right:$1$] (3) at (1,-0.5){};
    \draw[-] (1)--(2) (1)--(3);
    \draw[green] (2) circle (0.2cm);
    \draw[green] (3) circle (0.2cm);
    \end{tikzpicture}} & $S_2$ \\
    \hline
    $\left(4^2,1 \right)$ & $\left(2,6\right)$ & $(1,S_2)$ & \raisebox{-0.5\height}{\begin{tikzpicture}
    \node[flavour, label=below:$2$] (1) at (0,0){};
    \node[gauge, label=below:$1$] (2) at (1,0){};
    \draw[transform canvas={yshift=1.3pt}] (1)--(2);
    \draw[transform canvas={yshift=-1.3pt}] (1)--(2);
    \draw[-] (0.4,0.2)--(0.6,0)--(0.4,-0.2);
    \end{tikzpicture}} & \raisebox{-0.5\height}{\begin{tikzpicture}
    \node[flavour, label=below:$2$] (1) at (0,0){};
    \node[gauge, label=below:$2$] (2) at (1,0){};
    \draw (2) to[out=-45,in=45,looseness=8](2);
    \draw[-] (1)--(2);
    \end{tikzpicture}} & $1$\\ \hline
\end{tabular}
\caption[Electric and Magnetic Quivers for Intersections involving $S_2$ Special Pieces in ${\mathfrak{b}_4}$ and ${\mathfrak{c}_4}$ nilcones]{Electric and Magnetic Quivers for Intersections involving $S_2$ Special Pieces in ${\mathfrak{b}_4}$ and ${\mathfrak{c}_4}$ nilcones. The $\{(5,3,1),\; (4^2,1) \}$ special piece of ${\mathfrak{b}_4}$ is $d_{SD}$ dual to the $\{(2^4),\; (2^3,1^2) \}$ special piece of ${\mathfrak{c}_4}$. The intersections are taken with the orbits $(3^{3})$ and $(3^2,2)$ of the respective algebras. The Coulomb branch global symmetry of the magnetic quivers is $A_1$. The Higgs branch global symmetry of the electric quivers is $A_1 \times A_1$.}
\label{tab:2-2-loopbc4}
\end{table}

The $S_2$ special piece in ${\mathfrak{b}_4}$ associated with the parent orbit ${\cal O}^{\mathfrak{b}_4}_{(3,2^2,1^2)}$, along with some of the examples above are part of larger families for higher rank that involve near to minimal orbits of $\mathfrak{b}_{2n}$. These form a two parameter family, as shown in tables \ref{tab:bcfamilyI} and \ref{tab:bcfamilyII}.

\begin{table}[h!]
\centering
\scalebox{0.8}{\begin{tabular}{|c|c|c|c|c|c|}
\hline
    $\rho$ & $\left(\left|{\cal C}(Q_M)\right|, \left|{\cal H}(Q_E)\right|\right)$ & $(A(\cal O), C(\cal O))$ & $Q_{\text{M}}\left[\bar{\mathcal{O}}^{\mathfrak{b}_{2n}}_{\rho}\right]$ & $Q_{\text{E}}\left[\mathcal{S}^{\mathfrak{c}_{2n}}_{d_{SD}\left(\rho\right)}\right]$ & $G_{\mathcal{C}}$ \\
    \hline
    $\left(3,2^{2n-2},1^{2}\right)$ & 
    $\left(2n(2n+1),\;2n\right)$ & $(S_2,1)$ &  
    \raisebox{-0.5\height}
    {\begin{tikzpicture}
    \node[gauge, label=below:$1$] (1) at (0,0){};
    \node[] (cdots) at (1,0){$\cdots$};
    \node[gauge, label=below:$2n-1$] (2) at (2,0){};
    \node[] at (2.5,0){$[$};
    \node[gauge, label=below:$n$] (3) at (3,0){};
    \node[flavour, label=below:$1$] (4) at (4,0){};
    \node[] at (4.5,0){$]$};
    \node[] (S2) at (5,0){$\wr \; S_2 \;$};
    \draw[-] (1)--(cdots)--(2)--(3)--(4);
    \end{tikzpicture}}
    & \raisebox{-0.5\height}
    {\begin{tikzpicture}
    \node[gauge, label=below:$1$] (1) at (0,0){};
    \node[] (cdots) at (1,0){$\cdots$};
    \node[gauge, label=right:$2n-1$] (2) at (2,0){};
    \node[gauge, label=left:$n$] (3a) at (2,1){};
    \node[flavour, label=below:$1$] (4a) at (3,1){};
    \node[gauge, label=left:$n$] (3b) at (2,-1){};
    \node[flavour, label=below:$1$] (4b) at (3,-1){};
    \draw[green] \convexpath{3a,4a} {0.2cm};
    \draw[green] \convexpath{3b,4b} {0.2cm};
    \draw[-] (1)--(cdots)--(2)--(3a)--(4a) (2)--(3b)--(4b);
    \end{tikzpicture}} & $S_2$\\ 
    \hline
    $\left(2^{2n},1\right)$ &
    $\left(4n^{2},\;\textrm{N/A}\right)$ & $(1,S_2)$ & \raisebox{-0.5\height}
    {\begin{tikzpicture}
    \node[gauge, label=below:$1$] (1) at (0,0){};
    \node[] (cdots) at (1,0){$\cdots$};
    \node[gauge, label=below:$2n-1$] (2) at (2,0){};
    \node[gauge, label=below:$n$] (3) at (3,0){};
    \node[flavour, label=below:$1$] (4) at (4,0){};
    \draw[-] (1)--(cdots)--(2) (3)--(4);
    \draw[transform canvas={yshift=1.3pt}] (2)--(3);
    \draw[transform canvas={yshift=-1.3pt}] (2)--(3);
    \draw[-] (2.4,0.2)--(2.6,0)--(2.4,-0.2);
    \end{tikzpicture}} 
    & \raisebox{-0.5\height}{\textrm{N/A}}
    & \raisebox{-0.5\height} {\textrm{N/A}} \\ \hline
\end{tabular}}
\caption[Electric and Magnetic Quivers for Orbits and Slices to $S_2$ Special Pieces in ${\mathfrak{b}_{2n}}$ nilcone]{Electric and Magnetic Quivers for Orbits and Slices to $S_2$ Special Pieces in ${\mathfrak{b}_{2n}}$ nilcone. The $\{(3,2^{2n-2},1^{2}),\; (2^{2n},1) \}$ special piece of ${\mathfrak{b}_{2n}}$ is $d_{BV}$ dual to the $((2n)^2)$ orbit of ${\mathfrak{c}_{2n}}$. The Coulomb branch global symmetry of the magnetic quivers is $B_{2n}$. The Higgs branch global symmetry of the electric quivers is $U(1)$. When treated as a magnetic quiver, $Q_E$ evaluates to the $(2^{2n},1^2)$ orbit of $d_{2n+1}$.}
\label{tab:bcfamilyI}
\end{table}

\begin{table}[h!]
\centering
\scalebox{0.72}{\begin{tabular}{|c|c|c|c|c|c|}
\hline
    $\rho$ & $\left(\left|{\cal C}(Q_M)\right|, \left|{\cal H}(Q_E)\right|\right)$ & $(A(\cal O), C(\cal O))$ & $Q_{\text{M}}\left[\bar{\mathcal{O}}^{\mathfrak{b}_k}_{\rho}\right]$ & $Q_{\text{E}}\left[\mathcal{S}^{\mathfrak{c}_k}_{d_{SD}\left(\rho\right)}\right]$ & $G_{\mathcal{C}}$ \\
    \hline
    $\left(3,2^{2n-2},1^{2k+2}\right)$ & 
    $\left(2n(2n+2k+1),\;2n\right)$ & $(S_2,1)$ &  
    \raisebox{-0.5\height}
    {\begin{tikzpicture}
    \node[gauge, label=below:$1$] (1) at (0.5,0){};
    \node[] (cdotsa) at (1.25,0){$\cdots$};
    \node[gauge, label=below:$2n$] (2) at (2,0){};
    \node[] (cdotsb) at (2.75,0){$\cdots$};
    \node[gauge, label=below:$2n$] (3) at (3.5,0){};
    \node[] at (4,0){$[$};
    \node[gauge, label=below:$n$] (4) at (4.5,0){};
    \node[] at (5,0){$]$};
    \node[] (S2) at (5.5,0){$\wr \; S_2 \;$};
    \node[flavour, label=above:$1$] (5) at (2,1){};
    \draw[-] (1)--(cdotsa)--(2)--(cdotsb)--(3)--(4) (2)--(5);
    \draw [decorate, decoration = {brace, raise=5pt, mirror,amplitude=5pt}] (1.75,-0.5)--(3.75,-0.5) node[pos=0.5,below=10pt,black]{$k$};
    \end{tikzpicture}}
    & \raisebox{-0.5\height}
    {\begin{tikzpicture}
    \node[gauge, label=below:$1$] (1) at (0.5,0){};
    \node[] (cdotsa) at (1.25,0){$\cdots$};
    \node[gauge, label=below:$2n$] (2) at (2,0){};
    \node[] (cdotsb) at (2.75,0){$\cdots$};
    \node[gauge, label=below:$2n$] (3) at (3.5,0){};
    \node[gauge, label=right:$n$] (4a) at (4.25,0.5){};
    \node[gauge, label=right:$n$] (4b) at (4.25,-0.5){};
    \node[flavour, label=above:$1$] (5) at (2,1){};
    \draw[-] (1)--(cdotsa)--(2)--(cdotsb)--(3)--(4a) (3)--(4b) (2)--(5);
    \draw [decorate, decoration = {brace, raise=5pt, mirror,amplitude=5pt}] (1.75,-0.5)--(3.75,-0.5) node[pos=0.5,below=10pt,black]{$k$};
    \draw[green] (4a) circle (0.2cm);
    \draw[green] (4b) circle (0.2cm);
    \end{tikzpicture}} & $S_2$\\ 
    \hline
    $\left(2^{2n},1^{2k+1}\right)$ &
$\left(4n(n+k),\;\textrm{N/A}\right)$ & $(1,S_2)$ & \raisebox{-0.5\height}
    {\begin{tikzpicture}
    \node[gauge, label=below:$1$] (1) at (0.5,0){};
    \node[] (cdotsa) at (1.25,0){$\cdots$};
    \node[gauge, label=below:$2n$] (2) at (2,0){};
    \node[] (cdotsb) at (2.75,0){$\cdots$};
    \node[gauge, label=below:$2n$] (3) at (3.5,0){};
    \node[gauge, label=below:$n$] (4) at (4.5,0){};
    \node[flavour, label=above:$1$] (5) at (2,1){};
    \draw[-] (1)--(cdotsa)--(2)--(cdotsb)--(3) (2)--(5);
    \draw [decorate, decoration = {brace, raise=5pt, mirror,amplitude=5pt}] (1.75,-0.5)--(3.75,-0.5) node[pos=0.5,below=10pt,black]{$k$};
    \draw[transform canvas={yshift=1.3pt}] (4)--(3);
    \draw[transform canvas={yshift=-1.3pt}] (4)--(3);
    \draw[-] (3.9,0.2)--(4.1,0)--(3.9,-0.2);
    \end{tikzpicture}} 
    & \raisebox{-0.5\height}{\textrm{N/A}}&  \raisebox{-0.5\height}{\textrm{N/A}}\\ \hline
\end{tabular}}
\caption[Electric and Magnetic Quivers for Orbits and Slices to $S_2$ Special Pieces in ${\mathfrak{b}_{2n+k}}$ nilcone]{Electric and Magnetic Quivers for Orbits and Slices to $S_2$ Special Pieces in ${\mathfrak{b}_{2n+k}}$ nilcone. The $\{(3,2^{2n-2},1^{2k+2}),\; (2^{2n},1^{2k+1}) \}$ special piece of ${\mathfrak{b}_{2n+k}}$ is $d_{BV}$ dual to the orbit $(2n+2k,2n)$ of ${\mathfrak{c}_{2n+k}}$, where $k \ge 1$. The Coulomb branch global symmetry of the magnetic quivers is $B_{2n+k}$. When treated as a magnetic quiver, $Q_E$ evaluates to the orbit $(2^{2n},1^{2k+2})$ of $d_{2n+k+1}$.}
\label{tab:bcfamilyII}
\end{table}

The remaining $S_2$ special piece in ${\mathfrak{c}_4}$ is associated with the parent orbit ${\cal O}^{\mathfrak{c}_4}_{(4,2,1^2)}$. We have not been able to identify a partial Loop Lace map example. Nonetheless, this $S_2$ special piece fits into the general schema in Table \ref{tab:SpecialTrans}, in which the transition across a $S_2$ special piece of $\rm{dim}_{\mathbb H} = 2$ is always the minimal orbit $c_2$.

The overall number of $S_2$ special pieces increases with rank and, from rank $4$ upwards, there are special pieces of  ${\mathfrak{b}_k}$ and ${\mathfrak{c}_k}$ that are not covered by the examples herein. Some of these are $d_{SD}$ dual between ${\mathfrak{b}_k}$ and ${\mathfrak{c}_k}$, so that Loop Lace maps can be given; while in other cases, they map under $d_{BV}$ to special orbits with trivial special pieces.

\FloatBarrier

\section{Special Piece in the $\mathfrak{g}_2$ Nilcone}
\label{sec:g2special}
The first example of an exceptional special piece arises in the $\mathfrak{g}_2$ nilcone (See Appendix \ref{app:OrbitData} for details). Of the five orbits in the nilcone, shown in \Figref{fig:AC_Map_Data_GFE6} and Table \ref{tab:orbitsg2}, only the 10 dimensional orbit $G_2(a_1)$ has non-trivial component group $A(\cal O)$. This orbit, which is both special and normal, is self-dual and is the image under $d_{LS}$ of the entire $S_3$ {\it special piece} $\{G_2(a_1),\;\tilde{A}_1,\;A_1\}$, of which it is the parent. By contrast, the $d_{SD}$ map acts as the identity, $d_{SD}(\rho) = \rho$, on these orbits.

The unitary magnetic quivers whose Coulomb branches construct the $\mathfrak{g}_2$ minimal orbit, and the normalisation of the 8 dimensional orbit $\tilde A_1$, are based on affine Dynkin diagrams, and have been known for some time \cite{Cremonesi:2014xha, Hanany:2017ooe}. The unitary magnetic quiver \eqref{quiv:G2_Example} for the orbit ${G_2(a_1)}$ was identified in \cite{Cremonesi:2014vla}. In \cite{Bennett:2024loi} it was shown that when treated as an electric quiver, \eqref{quiv:G2_Example} has as its Higgs branch the S\l odowy slice to the $\mathfrak{g}_2$ minimal orbit $A_1$, and that Higgs branch quiver subtractions from \eqref{quiv:G2_Example} generate electric quivers for the S\l odowy slices from the orbits $\tilde A_1$ and ${G_2(a_1)}$. 

Taken together with the above observations, these quivers for the $S_3$ special piece of $\mathfrak{g}_2$ can be arranged, consistent with $d_{SD}$ and the  Loop Lace map, as shown in Table \ref{Tab:G2_special_piece}.

\begin{equation}
    \raisebox{-0.5\height}{\begin{tikzpicture}
    \node[gauge, label=below:$1$] (0) at (0,0){};
    \node[gauge, label=below:$2$] (1) at (1,0){};
    \node[gauge, label=below:$3$] (2) at (2,0){};
    \draw[-] (0)--(1)--(2);
    \draw (2) to[out=-45,in=45,looseness=8](2);
    \end{tikzpicture}}
\label{quiv:G2_Example}
\end{equation}

\begin{table}[h!]
\centering
\begin{tabular}{|c|c|c|c|c|c|}
\hline
     $\rho$ &
$\left(\left|{\cal C}(Q_M)\right|, \left|{\cal H}(Q_E)\right|\right)$ 
     & $(A(\cal O), C(\cal O))$ &  $Q_{\text{M}}\left[\bar{\mathcal{O}}^{\mathfrak{g}_2}_{\rho}\right]$ & $Q_{\text{E}}\left[\mathcal{S}^{\mathfrak{g}_2}_{\rho}\right]$ & $G_{\mathcal{C}(Q_E)}$  \\
     \hline
     $G_2\left(a_1\right)$ & $\left(10,2\right)$ & $(S_3,1)$ & \raisebox{-0.5\height}{\begin{tikzpicture}
        \node[gauge, label=below:$1$] (0) at (0,0){};
        \node[gauge, label=below:$2$] (1) at (1,0){};
        \node[gauge, label=below:$3$] (2) at (2,0){};
        \draw[-] (0)--(1)--(2);
        \draw (2) to[out=-45,in=45,looseness=8](2);
        \end{tikzpicture}} & \raisebox{-0.5\height}{\begin{tikzpicture}
        \node[gauge, label=below:$1$] (0) at (0,0){};
        \node[gauge, label=below:$2$] (1) at (1,0){};
        \node[gauge, label=right:$1$] (2) at (2,0.5){};
        \node[gauge, label=right:$1$] (3) at (2,0){};
        \node[gauge, label=right:$1$] (4) at (2,-0.5){};
        \draw[-] (0)--(1)--(2) (3)--(1) (4)--(1);
        \draw[green] (2) circle (0.2cm);
        \draw[green] (3) circle (0.2cm);
        \draw[green] (4) circle (0.2cm);
        \end{tikzpicture}} & $S_3$ \\ \hline
     $\tilde{A_1}$ & $\left(8,4\right)$ & $(1,S_2)$ & \raisebox{-0.5\height}{\begin{tikzpicture}
        \node[gauge, label=below:$1$] (0) at (0,0){};
        \node[gauge, label=below:$2$] (1) at (1,0){};
        \node[gauge, label=below:$1$] (2) at (2,0){};
        \node[gauge, label=above:$1$] (3) at (1,1){};
        \draw[transform canvas={yshift=1.3pt}] (2)--(1);
        \draw[transform canvas={yshift=-1.3pt}] (2)--(1);
        \draw[-] (1.4,-0.2)--(1.6,0)--(1.4,0.2);
        \draw[-] (1)--(0) (1)--(3);
        \draw[red] (2) circle (0.2cm);
        \draw[red] (3) circle (0.2cm);
        \end{tikzpicture}} & \raisebox{-0.5\height}{\begin{tikzpicture}
        \node[gauge, label=below:$1$] (0) at (0,0){};
        \node[gauge, label=below:$2$] (1) at (1,0){};
        \node[gauge, label={[xshift=0.55cm, yshift=-0.35cm]$2$}] (2) at (2,0.25){};
        \node[gauge, label=right:$1$] (3) at (2,-0.25){};
        \draw[-] (0)--(1)--(2) (3)--(1);
        \draw[green] (2) circle (0.2cm);
        \draw[green] (3) circle (0.2cm);
        \draw (2) to[out=45,in=-45,looseness=8](2);
        \end{tikzpicture}} & $\frac{S_3}{S_2}$\\ \hline
     $A_1$ & $\left(6,6\right)$ & $(1,S_3)$ & \raisebox{-0.5\height}{\begin{tikzpicture}
        \node[gauge, label=below:$1$] (0) at (0,0){};
        \node[gauge, label=below:$2$] (1) at (1,0){};
        \node[gauge, label=below:$1$] (2) at (2,0){};
        \draw[transform canvas={yshift=1.3pt}] (2)--(1);
        \draw[transform canvas={yshift=-1.3pt}] (2)--(1);
        \draw[-] (1.4,-0.2)--(1.6,0)--(1.4,0.2);
        \draw[-] (0)--(1)--(2);
    \end{tikzpicture}} & \raisebox{-0.5\height}{\begin{tikzpicture}
        \node[gauge, label=below:$1$] (0) at (0,0){};
        \node[gauge, label=below:$2$] (1) at (1,0){};
        \node[gauge, label=below:$3$] (2) at (2,0){};
        \draw[-] (0)--(1)--(2);
        \draw (2) to[out=-45,in=45,looseness=8](2);
        \end{tikzpicture}} & $1$ \\ \hline
\end{tabular}
\caption[Electric and Magnetic Quivers for Orbits and Slices of $S_3$ Special Piece in ${\mathfrak{g}_2}$ nilcone]{Electric and Magnetic Quivers for Orbits and Slices to $S_3$ Special Piece in ${\mathfrak{g}_2}$ nilcone. The special piece is self-dual under $d_{SD}$. The orbit $\tilde A_1$ is non-normal and its Coulomb branch construction gives a normalisation.}
\label{Tab:G2_special_piece}
\end{table}

For each of the orbits labelled by $\rho$, the Loop Lace map identifies the electric and magnetic quivers that realise the closure of the orbit and its S\l odowy slice. The $A(\cal O)$ and $ C(\cal O)$ subgroups of the $S_3$ canonical quotient group are easily read from the quivers according to the prescriptions of Tables \ref{tab:SnSP} and \ref{tab:AC_Quiver_Map}.

Under the Loop Lace map, the magnetic quiver \eqref{quiv:G2_Example}, is mapped to $Q_{\text{E}}[\mathcal{S}^{\mathfrak{g}_2}_{G_2(a_1)}]$, whose Higgs branch is the $G_2$ singularity, also known as `$D_4$ with an $S_3$ action'. The $D_4$ Dynkin diagram is manifest in this electric quiver, as is the $S_3$ Higgs branch decoration. The Coulomb branch of this electric quiver is an $S_3$ (or 6 times) cover of $G_2(a_1)$.\footnote{The decoration has no effect on the evaluation of a Higgs or Coulomb branch, but is relevant to the transitions between an orbit/slice and its neighbours.}

By way of explanatory comment, the adjoint hypermultiplet on the $\urm(3)$ gauge node in the magnetic quiver \eqref{quiv:G2_Example} can be seen as an orbifolding of the magnetic quiver for $d_4$ (the affine $D_4$ Dynkin diagram) by $S_3$.\footnote{This theory has a straightforward physical realisation as the magnetic quiver associated to $\surm(2)$ with four flavours coupled to two tensors in six dimensions at infinite coupling. For a more careful description, the reader is directed to Section $3.3.1$ of \cite{Cabrera:2019_6d1}.} The fundamental group of the resulting $G_2(a_1)$ orbit is this $S_3$. This leaf in the $\mathfrak{g}_2$ nilcone is consequently associated with the $\left(1^{3}\right)$ partition, whose multiplicity is the degree of $A\left(\mathcal{O}\right) = S_3$.

The magnetic quiver $Q_{\text{M}}[\bar{\mathcal{O}}^{\mathfrak{g}_2}_{\tilde{A}_{1}}]$ for the orbit ${\tilde{A}_{1}}$ has non-trivial $C({\cal O})=S_2$, corresponding to the double linked non-simply laced edge in this quiver. The orbit ${\tilde{A}_{1}}$ is non-normal and the Coulomb branch calculation yields its normalisation.

Applying the Loop Lace map, the electric quiver $Q_{\text{E}}[\mathcal{S}^{\mathfrak{g}_2}_{\tilde{A}_{1}}]$ contains a corresponding node of rank 2, which carries a loop denoting an $\surm(2)$ adjoint hypermultiplet. The Higgs branch generates the slice $\mathcal{S}^{\mathfrak{g}_2}_{\tilde{A}_{1}}$, which retains a residual $A_1$ symmetry from the hypermultiplet. The Coulomb branch of $Q_{\text{E}}[\mathcal{S}^{\mathfrak{g}_2}_{\tilde{A}_{1}}]$ is an ${S_3}/{S_2}$ (or 3 times) cover of $G_2(a_1)$.

The magnetic quiver $Q_{\text{M}}[\bar{\mathcal{O}}^{\mathfrak{g}_2}_{A_{1}}]$, has no non-trivial $A(\cal O)$ component group, but instead a Somers-Achar subgroup $C({\cal O}) = S_3 $. These are associated, respectively, with the absence of any adjoint hypermultiplets, and with the non-simple lacing of the rightmost $\urm(1)$ gauge node with lacing number $3$. Applying the Loop Lace map, in the electric quiver $Q_{\text{E}}\left[\mathcal{S}^{\mathfrak{g}_2}_{A_{1}}\right]$, the trivial component group is associated with the lack of any identically decorated gauge nodes, while the $C(\cal O)$ subgroup gives rise to a rank 3 node carrying an $\surm(3)$ adjoint hypermultiplet.

Note that the $\mathfrak{g}_2$ nilcone appears amongst the S\l odowy intersections of various other nilcones. This gives rise to related instances of the Loop Lace map and {\it special duality} amongst the quivers for S\l odowy intersections in $E_6$, $E_7$ and $E_8$.

\FloatBarrier
\section{Special Pieces in the $\mathfrak{f}_4$ Nilcone}
\label{sec:f4special}
The $\mathfrak{f}_4$ nilcone contains 16 orbits and has two non-trivial special pieces, as shown in \Figref{fig:AC_Map_Data_GFE6} and Table \ref{tab:orbitsf4}. There is an $S_4$ special piece associated with the $F_4\left(a_3\right)$ orbit, which contains five orbits, four of which are non-special. There is also an $S_2$ special piece, comprised of one special and one non-special orbit, associated with the $\tilde{A}_1$ orbit. The $S_4$ special piece is self-dual under $d_{SD}$.

\subsection{$S_4$ Special Piece}
\label{subsec:f4specialS4}

Consider the orthosymplectic quiver \eqref{quiv:D1--C4_Loop}, which was discussed in \cite{Hanany:2022itc} as a magnetic quiver for the 40-$\rm{dim}_{\mathbb{C}}$ orbit $F_4\left(a_3\right)$.\footnote{Interestingly, this magnetic quiver arose while studying the $6d$ theory describing the Higgs branch at the SCFT point of four M5-branes on $\mathbb{C}^{2}/D_{4}$ with boundary conditions $\rho_{L}=(7,1)$, $\rho_{R}=(1^{8})$. Indeed, this Higgs branch is known to be $\bar{\mathcal{O}}^{\mathfrak{f}_4}_{F_4\left(a_3\right)}$.} The $A({\cal O})=S_4$ component group of this orbit follows from the fact that the hypermultiplet in the 
$[0100]$ of $\sprm(4)$, along with the node itself, can be replaced by four $\sprm(1)$ gauge nodes linked by bifundamental $\sorm(8)\times\sprm(1)$ matter, together with an $S_4$ symmetrisation, without changing the Coulomb branch.

\begin{equation}
    \raisebox{-0.5\height}{\begin{tikzpicture}
    \node[gauger, label=below:$D_1$] (0) at (0,0){};
    \node[gaugeb, label=below:$C_1$] (1) at (1,0){};
    \node[gauger, label=below:$D_2$] (2) at (2,0){};
    \node[gaugeb, label=below:$C_2$] (3) at (3,0){};
    \node[gauger, label=below:$D_3$] (4) at (4,0){};
    \node[gaugeb, label=below:$C_3$] (5) at (5,0){};
    \node[gauger, label=below:$D_4$] (6) at (6,0){};
    \node[gaugeb, label=below:$C_4$] (7) at (7,0){};
    \draw[-] (0)--(1)--(2)--(3)--(4)--(5)--(6)--(7);
    \draw (7) to[out=-45,in=45,looseness=8](7);
    \end{tikzpicture}}
\label{quiv:D1--C4_Loop}
\end{equation}

Similar to the case of the $S_3$ discrete quotient of \eqref{quiv:G2_Example}, ungauging the $S_4$ permutation symmetry in \eqref{quiv:D1--C4_Loop} results in an $S_4$ cover of the original Coulomb branch. Coulomb branch quiver subtraction on \eqref{quiv:D1--C4_Loop} goes some way to describing its full moduli space within the $\mathfrak{f}_4$ nilcone. In particular, the structure of the $\mathcal{O}^{\mathfrak{f}_4}_{F_{4}(a_3)}$ special piece can be captured using the Loop Lace map for an orthosymplectic quiver developed in section \ref{subsec:LL}, and is shown in Table \ref{tab:D1--C4-Loop_Tab}.

For each orbit, the component group $A(\cal O)$ and $C(\cal O)$ group are identifiable from the associated electric and magnetic quivers. As previously, the various hypermultiplets and short nodes in the magnetic quivers are related to these objects. A distinct feature is the presence of $C(\mathcal{O})=S_2\times S_2$, which generates the adjoint hypermultiplet on the short $\sprm(2)$ node of the magnetic quiver $Q_{\text{M}}[\bar{\mathcal{O}}^{\mathfrak{f}_4}_{B_2}]$. In the electric quiver $Q_{\text{E}}[\mathcal{S}^{\mathfrak{f}_4}_{B_2}]$, this is associated with the bouquet of two decorated $\sprm(2)$ gauge nodes with adjoint hypermultiplets.

Note that the orbits $\{ {A_2+\tilde{A}_1},\; B_2,\;  C_3(a_1)\}$ are internal to the $S_4$ {\it special piece} and are non-normal. The Coulomb branch constructions of these orbits yield their normalisations.

The electric quivers in Table \ref{tab:D1--C4-Loop_Tab} were evaluated on their Coulomb branches in \cite{Hanany:2022itc}; this recovered the various covers of the Coulomb branch of \eqref{quiv:D1--C4_Loop} in accordance with $G_{\mathcal{C}}(Q_E)$.

A {\it special piece} obviously has the potential to involve intersections with orbits other than the trivial or maximal. While there is no general prescription to identify the magnetic quivers, analogous to \ref{quiv:G2_Example} or \ref{quiv:D1--C4_Loop}, for the possible parent intersections, several examples can be identified by drawing upon known constructions and by applying the principles of quiver subtraction and the Loop Lace map.
\begin{subequations}
\begin{align}
\raisebox{-0.5\height}{\begin{tikzpicture}
    \node[gauge, label=below:$2$] (1) at (0,0){};
    \node[gauge, label=below:$4$] (2) at (1,0){};
    \draw[-] (1)--(2);
    \draw (2) to[out=-45,in=45,looseness=8](2);
\end{tikzpicture}}\\
\raisebox{-0.5\height}{\begin{tikzpicture}
    \node[gauge, label=below:$1$] (0) at (0,0){};
    \node[gauge, label=below:$2$] (1) at (1,0){};
    \node[gauge, label=below:$4$] (2) at (2,0){};
    \draw[-] (0)--(1)--(2);
    \draw (2) to[out=-45,in=45,looseness=8](2);
\end{tikzpicture}}
\end{align}
\label{quiv:2-4-Loop_1-2-4-Loop}
\end{subequations}
In \cite{Bennett:2024loi}, the investigation of Higgs branch quiver subtractions identified two unitary quivers relevant to the $S_4$ special piece of $\mathfrak{f}_4$, and these are shown in \eqref{quiv:2-4-Loop_1-2-4-Loop}. When treated as magnetic quivers, their Coulomb branches generate the 10-$\rm{dim}_{\mathbb{C}}$ and 12-$\rm{dim}_{\mathbb{C}}$ intersections, $S_{F_4(a_3),\;A_2)}^{\mathfrak{f}_4}$ and $S_{F_4(a_3),\;A_1+\tilde A_1)}^{\mathfrak{f}_4}$, respectively. The Loop Lace map guides the identification of the families of magnetic and electric quivers associated with each of these intersections. These are arranged in Tables \ref{tab:2-4-Loop} and \ref{tab:1-2-4-Loop}.

Importantly, these theories share the $S_4$ symmetry and therefore behave identically under $d_{SD}$ and the Loop Lace map to \eqref{quiv:D1--C4_Loop}; in each case, the interpretation of the component group and [conjugacy class] from adjoint hypermultiplets and short nodes in the quiver remains the same.

One feature worth noting is that the sum of the Coulomb plus Higgs branch dimensions remains a constant across each Loop Lace pair of quivers in a given family, and is determined by the lower dimensioned orbit defining the intersection, and its dual. For example, in the cases of Table \ref{tab:2-4-Loop} and Table \ref{tab:1-2-4-Loop} respectively, we have the relations:
\begin{equation}
\left|{\cal C}(Q_M)\right|+ \left|{\cal H}(Q_E)\right| = \rm{dim}_{\mathbb{C}}\left({\cal O}_{B_3}^{\mathfrak{f}_4}\right)-\rm{dim}_{\mathbb{C}}\left({\cal O}_{A_2}^{\mathfrak{f}_4}\right) = 12,
\end{equation}
\begin{equation}
\left|{\cal C}(Q_M)\right|+ \left|{\cal H}(Q_E)\right| = \rm{dim}_{\mathbb{C}}\left({\cal O}_{F_4(a_2)}^{\mathfrak{f}_4}\right)-\rm{dim}_{\mathbb{C}}\left({\cal O}_{A_1+\tilde A_1}^{\mathfrak{f}_4}\right) = 16,
\end{equation}
and this generalises to all {\it special pieces} where the $d_{SD}$ map has a defined image.

\begin{landscape}
\begin{table}[p]
\centering
\begin{tabular}{|c|c|c|c|c|c|}
\hline
    $\rho$ &
    $\left(\left|{\cal C}(Q_M)\right|, \left|{\cal H}(Q_E)\right|\right)$ 
     & $(A(\cal O), C(\cal O))$ 
    & $Q_{\text{M}}\left[\bar{\mathcal{O}}^{\mathfrak{f}_4}_{\rho}\right]$ & $Q_{\text{E}}\left[\mathcal{S}^{\mathfrak{f}_4}_{\rho}\right]$ & $G_{\mathcal{C}}(Q_E)$ \\
    \hline
    $F_4\left(a_3\right)$ & $\left(40,8\right)$ & $(S_4,1)$ & \raisebox{-0.5\height}{\begin{tikzpicture}
        \node[gauger, label=below:$D_1$] (0) at (0,0){};
        \node[] (cdots) at (1,0){$\cdots$};
        \node[gaugeb, label=below:$C_3$] (1) at (2,0){};
        \node[gauger, label=below:$D_4$] (2) at (3,0){};
        \node[gaugeb, label=below:$C_4$] (3) at (4,0){};
        \draw[-] (0)--(cdots)--(1)--(2)--(3);
        \draw (3) to[out=-45,in=45,looseness=8](3);
        \end{tikzpicture}} & \raisebox{-0.5\height}{\begin{tikzpicture}
        \node[gauger, label=below:$D_1$] (0) at (0,0){};
        \node[] (cdots) at (1,0){$\cdots$};
        \node[gaugeb, label=below:$C_3$] (1) at (2,0){};
        \node[gauger, label=below:$D_4$] (2) at (3,0){};
        \node[gaugeb, label=right:$C_1$] (3) at (4,0.75){};
        \node[gaugeb, label=right:$C_1$] (4) at (4,0.25){};
        \node[gaugeb, label=right:$C_1$] (5) at (4,-0.25){};
        \node[gaugeb, label=right:$C_1$] (6) at (4,-0.75){};
        \draw[-] (0)--(cdots)--(1)--(2)--(3) (2)--(4) (2)--(5) (2)--(6);
        \draw[green] (3) circle (0.2cm);
        \draw[green] (4) circle (0.2cm);
        \draw[green] (5) circle (0.2cm);
        \draw[green] (6) circle (0.2cm);
        \end{tikzpicture}} & $S_4$ \\ \hline
    $C_3(a_1)$ & $\left(38,10\right)$ & $(S_2,S_2)$ & \raisebox{-0.5\height}{\begin{tikzpicture}
        \node[gauger, label=below:$D_1$] (0) at (0,0){};
        \node[] (cdots) at (1,0){$\cdots$};
        \node[gaugeb, label=below:$C_3$] (1) at (2,0){};
        \node[gauger, label=below:$D_4$] (2) at (3,0){};
        \node[gaugeb, label=below:$C_1$] (3) at (4,0){};
        \node[gaugeb, label=right:$C_2$] (4) at (3,1){};
        \draw[-] (0)--(cdots)--(1)--(2)--(4);
        \draw (4) to[out=135,in=45,looseness=8](4);
        \draw[transform canvas={yshift=1.3pt}] (2)--(3);
        \draw[transform canvas={yshift=-1.3pt}] (2)--(3);
        \draw[-] (3.4,0.2)--(3.6,0)--(3.4,-0.2);
        \draw[red] (4) circle (0.2cm);
        \draw[red] (3) circle (0.2cm);
        \end{tikzpicture}} & \raisebox{-0.5\height}{\begin{tikzpicture}
        \node[gauger, label=below:$D_1$] (0) at (0,0){};
        \node[] (cdots) at (1,0){$\cdots$};
        \node[gaugeb, label=below:$C_3$] (1) at (2,0){};
        \node[gauger, label=below:$D_4$] (2) at (3,0){};
        \node[gaugeb, label=right:$C_1$] (3) at (4,0){};
        \node[gaugeb, label=right:$C_1$] (4) at (4,-0.5){};
        \node[gaugeb, label={[xshift=0.75cm, yshift=-0.4cm]$C_2$}] (5) at (4,0.5){};
        \draw[-] (0)--(cdots)--(1)--(2)--(3) (2)--(4) (2)--(5);
        \draw[green] (3) circle (0.2cm);
        \draw[green] (4) circle (0.2cm);
        \draw[green] (5) circle (0.2cm);
        \draw (5) to[out=45,in=-45,looseness=8](5);
        \end{tikzpicture}} & $\frac{S_4}{S_2}$\\ \hline
    $B_2$ & $\left(36,12\right)$ & $(S_2,S_2\times S_2)$ & \raisebox{-0.5\height}{\begin{tikzpicture}
        \node[gauger, label=below:$D_1$] (0) at (0,0){};
        \node[] (cdots) at (1,0){$\cdots$};
        \node[gaugeb, label=below:$C_3$] (1) at (2,0){};
        \node[gauger, label=below:$D_4$] (2) at (3,0){};
        \node[gaugeb, label=below:$C_2$] (3) at (4,0){};
        \draw[-] (0)--(cdots)--(1)--(2);
        \draw (3) to[out=-45,in=45,looseness=8](3);
        \draw[transform canvas={yshift=1.3pt}] (2)--(3);
        \draw[transform canvas={yshift=-1.3pt}] (2)--(3);
        \draw[-] (3.4,0.2)--(3.6,0)--(3.4,-0.2);
        \end{tikzpicture}} & \raisebox{-0.5\height}{\begin{tikzpicture}
        \node[gauger, label=below:$D_1$] (0) at (0,0){};
        \node[] (cdots) at (1,0){$\cdots$};
        \node[gaugeb, label=below:$C_3$] (1) at (2,0){};
        \node[gauger, label=below:$D_4$] (2) at (3,0){};
        \node[gaugeb, label=right:$C_2$] (3) at (4,0.25){};
        \node[gaugeb, label=right:$C_2$] (4) at (4,-0.25){};
        \draw[-] (0)--(cdots)--(1)--(2)--(3) (2)--(4);
        \draw[green] (3) circle (0.2cm);
        \draw[green] (4) circle (0.2cm);
        \draw (3) to[out=135,in=45,looseness=8](3);
        \draw (4) to[out=-135,in=-45,looseness=8](4);
        \end{tikzpicture}} & $\frac{S_4}{S_2\times S_2}$\\ \hline
   $\tilde{A}_2+A_1$ & $\left(36,12\right)$ & $(1,S_3)$ & \raisebox{-0.5\height}{\begin{tikzpicture}
        \node[gauger, label=below:$D_1$] (0) at (0,0){};
        \node[] (cdots) at (1,0){$\cdots$};
        \node[gaugeb, label=below:$C_3$] (1) at (2,0){};
        \node[gauger, label=below:$D_4$] (2) at (3,0){};
        \node[gaugeb, label=below:$C_1$] (3) at (4,0){};
        \node[gaugeb, label=right:$C_1$] (4) at (3,1){};
        \draw[-] (0)--(cdots)--(1)--(2)--(4) (2)--(3);
        \draw[transform canvas={yshift=1.3pt}] (2)--(3);
        \draw[transform canvas={yshift=-1.3pt}] (2)--(3);
        \draw[-] (3.4,0.2)--(3.6,0)--(3.4,-0.2);
        \draw[red] (3) circle (0.2cm);
        \draw[red] (4) circle (0.2cm);
        \end{tikzpicture}} & \raisebox{-0.5\height}{\begin{tikzpicture}
        \node[gauger, label=below:$D_1$] (0) at (0,0){};
        \node[] (cdots) at (1,0){$\cdots$};
        \node[gaugeb, label=below:$C_3$] (1) at (2,0){};
        \node[gauger, label=below:$D_4$] (2) at (3,0){};
        \node[gaugeb, label=right:$C_3$] (3) at (4,0.25){};
        \node[gaugeb, label=right:$C_1$] (4) at (4,-0.25){};
        \draw[-] (0)--(cdots)--(1)--(2)--(3) (2)--(4);
        \draw[green] (3) circle (0.2cm);
        \draw[green] (4) circle (0.2cm);
        \draw (3) to[out=135,in=45,looseness=8](3);
        \end{tikzpicture}} & $\frac{S_4}{S_3}$
        \\ \hline
    $A_2+\tilde{A}_1$ & $\left(34,14\right)$ & $(1,S_4)$ & \raisebox{-0.5\height}{\begin{tikzpicture}
        \node[gauger, label=below:$D_1$] (0) at (0,0){};
        \node[] (cdots) at (1,0){$\cdots$};
        \node[gaugeb, label=below:$C_3$] (1) at (2,0){};
        \node[gauger, label=below:$D_4$] (2) at (3,0){};
        \node[gaugeb, label=below:$C_1$] (3) at (4,0){};
        \draw[-] (0)--(cdots)--(1)--(2);
        \draw[transform canvas={yshift=2pt}] (2)--(3);
        \draw[transform canvas={yshift=-2pt}] (2)--(3);
        \draw[transform canvas={yshift=0.75pt}] (2)--(3);
        \draw[transform canvas={yshift=-0.75pt}] (2)--(3);
        \draw[-] (3.4,0.2)--(3.6,0)--(3.4,-0.2);
        \end{tikzpicture}} & \raisebox{-0.5\height}{\begin{tikzpicture}
        \node[gauger, label=below:$D_1$] (0) at (0,0){};
        \node[] (cdots) at (1,0){$\cdots$};
        \node[gaugeb, label=below:$C_3$] (1) at (2,0){};
        \node[gauger, label=below:$D_4$] (2) at (3,0){};
        \node[gaugeb, label=below:$C_4$] (3) at (4,0){};
        \draw[-] (0)--(cdots)--(1)--(2)--(3);
        \draw (3) to[out=-45,in=45,looseness=8](3);
        \end{tikzpicture}} & $1$ \\ \hline
\end{tabular}
\caption[Electric and Magnetic Quivers for Orbits and Slices to $S_4$ Special Piece in ${\mathfrak{f}_4}$ nilcone]{Electric and Magnetic Quivers for Orbits and Slices to $S_4$ Special Piece in ${\mathfrak{f}_4}$ nilcone. The special piece is self-dual under $d_{SD}$. The orbits $\{C_3(a_1), \;B_2,\; \tilde A_2+ A_1 \}$ are non-normal and their Coulomb branch constructions give normalisations.}
\label{tab:D1--C4-Loop_Tab}
\end{table}
\end{landscape}

\begin{landscape}
\begin{table}[h!]
\centering
\begin{tabular}{|c|c|c|c|c|c|}
\hline
    $\rho$ & 
      $\left(\left|{\cal C}(Q_M)\right|, \left|{\cal H}(Q_E)\right|\right)$ 
     & $(A(\cal O), C(\cal O))$ 
     & $Q_{\text{M}}\left[\mathcal{S}^{\mathfrak{f}_4}_{\rho,\;A_2}\right]$ & $Q_{\text{E}}\left[\mathcal{S}^{\mathfrak{f}_4}_{B_3,\;\rho}\right]$ & $G_{\mathcal{C}}(Q_E)$\\
    \hline
    $F_4\left(a_3\right)$ & $\left(10,2\right)$ & $(S_4,1)$ & \raisebox{-0.5\height}{\begin{tikzpicture}
        \node[gauge, label=below:$2$] (1) at (0,0){};
        \node[gauge, label=below:$4$] (2) at (1,0){};
        \draw[-] (1)--(2);
        \draw (2) to[out=-45,in=45,looseness=8](2);
        \end{tikzpicture}} & \raisebox{-0.5\height}{\begin{tikzpicture}
        \node[gauge, label=below:$2$] (1) at (0,0){};
        \node[gauge, label=right:$1$] (2) at (1,0.75){};
        \node[gauge, label=right:$1$] (3) at (1,0.25){};
        \node[gauge, label=right:$1$] (4) at (1,-0.25){};
        \node[gauge, label=right:$1$] (5) at (1,-0.75){};
        \draw[-] (1)--(2) (3)--(1) (4)--(1) (5)--(1);
        \draw[green] (5) circle (0.2cm);
        \draw[green] (2) circle (0.2cm);
        \draw[green] (3) circle (0.2cm);
        \draw[green] (4) circle (0.2cm);
        \end{tikzpicture}} & $S_4$ \\ \hline
    $C_3\left(a_1\right)$ & $\left(8,4\right)$ & $(S_2,S_2)$ & \raisebox{-0.5\height}{\begin{tikzpicture}
        \node[gauge, label=above:$1$] (0) at (0,1){};
        \node[gauge, label=below:$2$] (1) at (0,0){};
        \node[gauge, label=below:$2$] (2) at (1,0){};
        \draw[transform canvas={xshift=1.3pt}] (0)--(1);
        \draw[transform canvas={xshift=-1.3pt}] (0)--(1);
        \draw[-] (-0.2,0.4)--(0,0.6)--(0.2,0.4);
        \draw[-] (1)--(2);
        \draw[red] (0) circle (0.2cm);
        \draw[red] (2) circle (0.2cm);
        \draw (2) to[out=-45,in=45,looseness=8](2);
        \end{tikzpicture}} & \raisebox{-0.5\height}{\begin{tikzpicture}
        \node[gauge, label=below:$2$] (1) at (0,0){};
        \node[gauge, label=right:$1$] (2) at (1,0.5){};
        \node[gauge, label=right:$1$] (3) at (1,-0.5){};
        \node[gauge, label=right:$2$] (4) at (0,1){};
        \draw[-] (1)--(2) (3)--(1) (4)--(1);
        \draw (4) to[out=135,in=45,looseness=8](4);
        \draw[green] (2) circle (0.2cm);
        \draw[green] (3) circle (0.2cm);
        \draw[green] (4) circle (0.2cm);
        \end{tikzpicture}} & $\frac{S_4}{S_2}$ \\ \hline
    $B_2$ & $\left(6,6\right)$ & $(S_2,S_2\times S_2)$ & \raisebox{-0.5\height}{\begin{tikzpicture}
        \node[gauge, label=below:$2$] (0) at (0,0){};
        \node[gauge, label=below:$2$] (1) at (1,0){};
        \draw[transform canvas={yshift=1.3pt}] (0)--(1);
        \draw[transform canvas={yshift=-1.3pt}] (0)--(1);
        \draw[-] (0.4,-0.2)--(0.6,0)--(0.4,0.2);
        \draw (1) to[out=55,in=-45,looseness=8](1);
    \end{tikzpicture}} & \raisebox{-0.5\height}{\begin{tikzpicture}
        \node[gauge, label=below:$2$] (1) at (1,0){};
        \node[gauge, label=below:$2$] (2) at (2,0){};
        \node[gauge, label=right:$2$] (3) at (1,1){};
        \draw[-] (1)--(2) (3)--(1);
        \draw (3) to[out=135,in=45,looseness=8](3);
        \draw (2) to[out=-45,in=45,looseness=8](2);
        \draw[green] (2) circle (0.2cm);
        \draw[green] (3) circle (0.2cm);
        \end{tikzpicture}} & $\frac{S_4}{S_2\times S_2}$ \\ \hline
   $\tilde{A_2}+A_1$ & $\left(6,6\right)$ & $(1,S_3)$ & \raisebox{-0.5\height}{\begin{tikzpicture}
        \node[gauge, label=above:$1$] (0) at (0,1){};
        \node[gauge, label=below:$2$] (1) at (0,0){};
        \node[gauge, label=below:$1$] (2) at (1,0){};
        \draw[transform canvas={xshift=1.3pt}] (0)--(1);
        \draw[transform canvas={xshift=-1.3pt}] (0)--(1);
        \draw[-] (-0.2,0.4)--(0,0.6)--(0.2,0.4);
        \draw[-] (0)--(1)--(2);
        \draw[red] (0) circle (0.2cm);
        \draw[red] (2) circle (0.2cm);
    \end{tikzpicture}} & \raisebox{-0.5\height}{\begin{tikzpicture}
        \node[gauge, label=below:$2$] (1) at (1,0){};
        \node[gauge, label=below:$3$] (2) at (2,0){};
        \node[gauge, label=right:$1$] (3) at (1,1){};
        \draw[-] (1)--(2) (3)--(1);
        \draw (2) to[out=-45,in=45,looseness=8](2);
        \draw[green] (2) circle (0.2cm);
        \draw[green] (3) circle (0.2cm);
        \end{tikzpicture}} & $\frac{S_4}{S_3}$
        \\ \hline
    $A_2+\tilde{A_1}$ & $\left(4,8\right)$ & $(1,S_4)$ & \raisebox{-0.5\height}{\begin{tikzpicture}
        \node[gauge, label=below:$2$] (0) at (0,0){};
        \node[gauge, label=below:$1$] (1) at (1,0){};
        \draw[transform canvas={yshift=1.5pt}] (0)--(1);
        \draw[transform canvas={yshift=0.5pt}] (0)--(1);
        \draw[transform canvas={yshift=-0.5pt}] (0)--(1);
        \draw[transform canvas={yshift=-1.5pt}] (0)--(1);
        \draw[-] (0.4,-0.2)--(0.6,0)--(0.4,0.2);
    \end{tikzpicture}} & \raisebox{-0.5\height}{\begin{tikzpicture}
        \node[gauge, label=below:$2$] (1) at (1,0){};
        \node[gauge, label=below:$4$] (2) at (2,0){};
        \draw[-] (1)--(2);
        \draw (2) to[out=-45,in=45,looseness=8](2);
        \end{tikzpicture}} & $1$ \\ \hline
\end{tabular}
\caption[Electric and Magnetic Quivers for Intersections between $S_4$ Special Piece in ${\mathfrak{f}_4}$ nilcone and $\{B_3,\;A_2 \}$ orbits]{Electric and Magnetic Quivers for Intersections between $S_4$ Special Piece in ${\mathfrak{f}_4}$ nilcone and $\{B_3,\;A_2 \}$ orbits. The orbits $\{C_3(a_1),\;\;B_2,\;\tilde A_2+ A_1\;  \}$ are non-normal and their Coulomb branch constructions give normalisations.}
\label{tab:2-4-Loop}
\end{table}
\end{landscape}

\begin{landscape}
\begin{table}[p]
\centering
\begin{tabular}{|c|c|c|c|c|c|}
\hline
$\rho$ &
      $\left(\left|{\cal C}(Q_M)\right|, \left|{\cal H}(Q_E)\right|\right)$ 
     & $(A(\cal O), C(\cal O))$
& $Q_{\text{M}}\left[\mathcal{S}^{\mathfrak{f}_4}_{\rho,\;A_1+\tilde{A_1}}\right]$ & $Q_{\text{E}}\left[\mathcal{S}^{\mathfrak{f}_4}_{F_4(a_2),\;\rho}\right]$ & $G_{\mathcal{C}}(Q_E)$ \\
\hline
$F_4\left(a_3\right)$ & $\left(12,4\right)$ & $(S_4,1)$ & \raisebox{-0.5\height}{\begin{tikzpicture}
        \node[gauge, label=below:$1$] (0) at (0,0){};
        \node[gauge, label=below:$2$] (1) at (1,0){};
        \node[gauge, label=below:$4$] (2) at (2,0){};
        \draw[-] (0)--(1)--(2);
        \draw (2) to[out=-45,in=45,looseness=8](2);
    \end{tikzpicture}} & \raisebox{-0.5\height}{\begin{tikzpicture}
        \node[gauge, label=below:$1$] (0) at (0,0){};
        \node[gauge, label=below:$2$] (1) at (1,0){};
        \node[gauge, label=right:$1$] (2) at (2,0.75){};
        \node[gauge, label=right:$1$] (3) at (2,0.25){};
        \node[gauge, label=right:$1$] (4) at (2,-0.25){};
        \node[gauge, label=right:$1$] (5) at (2,-0.75){};
        \draw[-] (0)--(1)--(2) (3)--(1) (4)--(1) (5)--(1);
        \draw[green] (5) circle (0.2cm);
        \draw[green] (2) circle (0.2cm);
        \draw[green] (3) circle (0.2cm);
        \draw[green] (4) circle (0.2cm);
        \end{tikzpicture}} & $S_4$\\ \hline
    $C_3\left(a_1\right)$ & $\left(10,6\right)$ & $(S_2,S_2)$ & \raisebox{-0.5\height}{\begin{tikzpicture}
        \node[gauge, label=below:$1$] (0) at (0,0){};
        \node[gauge, label=below:$2$] (1) at (1,0){};
        \node[gauge, label=below:$2$] (2) at (2,0){};
        \node[gauge, label=above:$1$] (3) at (1,1){};
        \draw[transform canvas={xshift=1.3pt}] (3)--(1);
        \draw[transform canvas={xshift=-1.3pt}] (3)--(1);
        \draw[-] (0.8,0.4)--(1,0.6)--(1.2,0.4);
        \draw[red] (2) circle (0.2cm);
        \draw[red] (3) circle (0.2cm);
        \draw[-] (0)--(1)--(2);
        \draw (2) to[out=-45,in=45,looseness=8](2);
    \end{tikzpicture}} & \raisebox{-0.5\height}{\begin{tikzpicture}
        \node[gauge, label=below:$1$] (0) at (0,0){};
        \node[gauge, label=below:$2$] (1) at (1,0){};
        \node[gauge, label=right:$1$] (2) at (2,0.5){};
        \node[gauge, label=right:$1$] (3) at (2,-0.5){};
        \node[gauge, label=right:$2$] (4) at (1,1){};
        \draw[-] (0)--(1)--(2) (3)--(1) (4)--(1);
        \draw (4) to[out=135,in=45,looseness=8](4);
        \draw[green] (2) circle (0.2cm);
        \draw[green] (3) circle (0.2cm);
        \draw[green] (4) circle (0.2cm);
    \end{tikzpicture}} & $\frac{S_4}{S_2}$\\ \hline
     $B_2$ & $\left(8,8\right)$ & $(S_2,S_2\times S_2)$ & \raisebox{-0.5\height}{\begin{tikzpicture}
        \node[gauge, label=below:$1$] (0) at (0,0){};
        \node[gauge, label=below:$2$] (1) at (1,0){};
        \node[gauge, label=below:$2$] (2) at (2,0){};
        \draw[transform canvas={yshift=1.3pt}] (1)--(2);
        \draw[transform canvas={yshift=-1.3pt}] (1)--(2);
        \draw[-] (1.4,0.2)--(1.6,0)--(1.4,-0.2);
        \draw (2) to[out=45,in=-45,looseness=8](2);
        \draw[-] (0)--(1);
    \end{tikzpicture}} & \raisebox{-0.5\height}{\begin{tikzpicture}
        \node[gauge, label=below:$1$] (0) at (0,0){};
        \node[gauge, label=below:$2$] (1) at (1,0){};
        \node[gauge, label=below:$2$] (2) at (2,0){};
        \node[gauge, label=right:$2$] (3) at (1,1){};
        \draw[-] (0)--(1)--(2) (3)--(1);
        \draw (3) to[out=135,in=45,looseness=8](3);
        \draw (2) to[out=-45,in=45,looseness=8](2);
        \draw[green] (2) circle (0.2cm);
        \draw[green] (3) circle (0.2cm);
    \end{tikzpicture}} & $\frac{S_4}{S_2\times S_2}$\\ \hline
    $\tilde{A_2}+A_1$ & $\left(8,8\right)$ & $(1,S_3)$ & \raisebox{-0.5\height}{\begin{tikzpicture}
    \node[gauge, label=below:$1$] (0) at (0,0){};
    \node[gauge, label=below:$2$] (1) at (1,0){};
    \node[gauge, label=below:$1$] (2) at (2,0){};
    \node[gauge, label=above:$1$] (3) at (1,1){};
    \draw[transform canvas={yshift=1.3pt}] (1)--(2);
    \draw[transform canvas={yshift=-1.3pt}] (1)--(2);
    \draw[-] (1.4,-0.2)--(1.6,0)--(1.4,0.2);
    \draw[-] (0)--(1)--(2) (1)--(3);
    \draw[red] (2) circle (0.2cm);
    \draw[red] (3) circle (0.2cm);
    \end{tikzpicture}} & \raisebox{-0.5\height}{\begin{tikzpicture}
        \node[gauge, label=below:$1$] (0) at (0,0){};
        \node[gauge, label=below:$2$] (1) at (1,0){};
        \node[gauge, label=below:$3$] (2) at (2,0){};
        \node[gauge, label=right:$1$] (3) at (1,1){};
        \draw[-] (0)--(1)--(2) (3)--(1);
        \draw (2) to[out=-45,in=45,looseness=8](2);
        \draw[green] (2) circle (0.2cm);
        \draw[green] (3) circle (0.2cm);
        \end{tikzpicture}} & $\frac{S_4}{S_3}$\\\hline
     $A_2+\tilde{A_1}$ & $\left(6,10\right)$ & $(1,S_4)$ & \raisebox{-0.5\height}{\begin{tikzpicture}
        \node[gauge, label=below:$1$] (0) at (0,0){};
        \node[gauge, label=below:$2$] (1) at (1,0){};
        \node[gauge, label=below:$1$] (2) at (2,0){};
        \draw[transform canvas={yshift=1.5pt}] (1)--(2);
        \draw[transform canvas={yshift=0.5pt}] (1)--(2);
        \draw[transform canvas={yshift=-0.5pt}] (1)--(2);
        \draw[transform canvas={yshift=-1.5pt}] (1)--(2);
        \draw[-] (1.4,-0.2)--(1.6,0)--(1.4,0.2);
        \draw[-] (0)--(1);
    \end{tikzpicture}} & \raisebox{-0.5\height}{\begin{tikzpicture}
        \node[gauge, label=below:$1$] (0) at (0,0){};
        \node[gauge, label=below:$2$] (1) at (1,0){};
        \node[gauge, label=below:$4$] (2) at (2,0){};
        \draw[-] (0)--(1)--(2);
        \draw (2) to[out=-45,in=45,looseness=8](2);
        \end{tikzpicture}} & $1$\\ \hline
\end{tabular}
\caption[Electric and Magnetic Quivers for Intersections between $S_4$ Special Piece in ${\mathfrak{f}_4}$ nilcone and $\{F_4(a_2),\; A_1+\tilde{A}_1 \}$ orbits]{Electric and Magnetic Quivers for Intersections between $S_4$ Special Piece in ${\mathfrak{f}_4}$ nilcone and $\{F_4(a_2),\; A_1+\tilde{A}_1 \}$ orbits. The orbits $\{A_2+\tilde A_1, \;B_2,\; C_3(a_1) \}$ are non-normal and their Coulomb branch constructions give normalisations.}
\label{tab:1-2-4-Loop}
\end{table}
\end{landscape}

\subsection{$S_2$ Special Piece}
\label{subsec:f4specialS2}

As explained in section \ref{subsec:SD}, the $S_2$ special piece $\{\tilde A_1, A_1 \}$ of $\mathfrak{f}_4$ does not have a complete image under $d_{SD}$, since both its orbits  are $d_{LS}$ dual to a single special orbit $F_4(a_1)$. While a complete Loop Lace map cannot be presented, it is nonetheless instructive to arrange the known quivers for the orbits and slices in these related $S_2$ and trivial {\it special pieces} as a partial Loop Lace map.

The unitary magnetic quiver \eqref{quiv:12321-S2Wreath}, which was identified in \cite{Bourget:2020bxh}, constructs the next to minimal orbit $\tilde A_1$, which is the parent of the $S_2$ special piece. This quiver manifests an $S_2$ orbifold action via symmetrisation (also termed $S_2$ wreathing). The magnetic quiver for the minimal orbit  ${\mathcal{O}}^{\mathfrak{f}_4}_{A_1}$ and the electric quiver for the subregular slice $\mathcal{S}^{\mathfrak{f}_4}_{F_4(a_1)}$ are both well known, and all three quivers can be arranged as in Table \ref{tab:12321-S2Wreath}.

\begin{equation}
\raisebox{-0.5\height}
{\begin{tikzpicture}
    \node[gauge, label=below:$1$] (1) at (0,0){};
    \node[gauge, label=below:$2$] (2) at (1,0){};
    \node[gauge, label=below:$3$] (3) at (2,0){};
    \node[gauge, label=below:$2$] (4) at (3,0){};
    \node[gauge, label=below:$1$] (5) at (4,0){};
    \draw[-](1)--(2)--(3)--(4)--(5);
    \node[] at (2.5,0){$[$};
    \node[] at (4.3,0){$]$};
    \node[] at (4.6,0){$\; \wr \; S_2$};
    \end{tikzpicture}}
\label{quiv:12321-S2Wreath}
\end{equation}

The $S_2$ symmetrisation in \eqref{quiv:12321-S2Wreath} is achieved by a formal wreathing rather than by a hypermultiplet loop. Indeed, no such loop construction is known for the symmetrisation of a magnetic quiver leg with multiple gauge nodes. As implied by the $S_2$ quotient $G_{\mathcal{C}}(Q_E)$ that acts on the electric quiver for the subregular slice (which takes the form of the $E_6$ Dynkin diagram), this formal wreathing may be viewed as a generalisation of the Loop Lace map between an electric quiver containing a bouquet and its paired magnetic quiver.

The transition from the $S_2$ wreathed to the $S_2$ folded magnetic quiver follows the $A({\cal O})$ and $C({\cal O})$ subgroup structure. The magnetic quiver remains balanced and retains the symmetry of the $F_4$ Dynkin diagram. However, the transition has $\rm{dim}_{\mathbb{C}}=6$, unlike the $\rm{dim}_{\mathbb{C}}=2$ transitions in the canonical cases in Table \ref{tab:AC_Quiver_Map}.

Furthermore, while the Coulomb branches of theories with hypermultiplets match those involving $S_n$ wreathings of $\urm(1)$ or $\sprm(1)$ gauge nodes, no such equivalence exists on the Higgs branch. For such reasons, it is unclear how this Loop Lace map might extend to electric quivers other than the slice to the subregular orbit. Equally, the gap in this Loop Lace map corresponds to the absence of an image of the minimal orbit under $d_{SD}$.

\begin{table}[h!]
\centering
\scalebox{0.8}{
\begin{tabular}{|c|c|c|c|c|c|}
\hline
    $\rho$ & 
     $\left(\left|{\cal C}(Q_M)\right|, \left|{\cal H}(Q_E)\right|\right)$ 
     & $(A(\cal O), C(\cal O))$
    & $Q_{\text{M}}\left[\bar{\mathcal{O}}^{\mathfrak{f}_4}_{\rho}\right]$ 
    & $Q_{\text{E}}\left[\mathcal{S}^{\mathfrak{f}_4}_{d_{SD}(\rho)}\right]$ 
    & $G_{\mathcal{C}}(Q_E)$ \\
    \hline
    $\tilde A_1$ 
    & $\left(\textrm{22},2\right)$ 
    & $(S_2,1)$ &
    \raisebox{-0.5\height}
    {\begin{tikzpicture}
    \node[gauge, label=below:$1$] (1) at (0,0){};
    \node[gauge, label=below:$2$] (2) at (1,0){};
    \node[gauge, label=below:$3$] (3) at (2,0){};
    \node[gauge, label=below:$2$] (4) at (3,0){};
    \node[gauge, label=below:$1$] (5) at (4,0){};
    \draw[-](1)--(2)--(3)--(4)--(5);
    \node[] at (2.5,0){$[$};
    \node[] at (4.3,0){$]$};
    \node[] at (4.6,0){$\; \wr \; S_2$};
    \end{tikzpicture}}  
    & \raisebox{-0.5\height}{\begin{tikzpicture}
    \node[gauge, label=below:$1$] (1) at (0,0){};
    \node[gauge, label=below:$2$] (2) at (1,0){};
    \node[gauge, label=below:$3$] (3) at (2,0){};
    \node[gauge, label=below:$2$] (4) at (3,0){};
    \node[gauge, label=below:$1$] (5) at (4,0){};
    \node[gauge, label=right:$2$] (6) at (2,1){};
    \node[gauge, label=right:$1$] (7) at (2,2){};
    \draw[-] (1)--(2)--(3)--(4)--(5) (3)--(6)--(7);
    \draw[green] \convexpath{6,7} {0.2cm};
    \draw[green] \convexpath{4,5} {0.2cm};
    \end{tikzpicture}} & $S_2$\\ 
    \hline
    $A_1$ & 
    $\left(16,\textrm{N/A}\right)$ 
    \raisebox{-0.5\height}
    & $(1,S_2)$ & \raisebox{-0.5\height}{\begin{tikzpicture}
    \node[gauge, label=below:$1$] (1) at (0,0){};
    \node[gauge, label=below:$2$] (2) at (1,0){};
    \node[gauge, label=below:$3$] (3) at (2,0){};
    \node[gauge, label=below:$2$] (4) at (3,0){};
    \node[gauge, label=below:$1$] (5) at (4,0){};
    \draw[-] (1)--(2)--(3) (4)--(5);
    \draw[transform canvas={yshift=1.3pt}] (3)--(4);
    \draw[transform canvas={yshift=-1.3pt}] (3)--(4);
    \draw[-] (2.4,0.2)--(2.6,0)--(2.4,-0.2);
    \end{tikzpicture}} & \raisebox{-0.5\height}{$\textrm{N/A}$} &
\raisebox{-0.5\height}
{$\textrm{N/A}$} \\ \hline
\end{tabular}}
\caption[Electric and Magnetic Quivers for Orbits and Slices to $S_2$ Special Piece and trivial dual in ${\mathfrak{f}_4}$ nilcone]{Electric and Magnetic Quivers for Orbits and Slices to $S_2$ Special Piece and trivial dual in ${\mathfrak{f}_4}$ nilcone. The Higgs branch of $Q_E$ is the $E_6$ Kleinian singularity.}
\label{tab:12321-S2Wreath}
\end{table}

\FloatBarrier
\section{Special Pieces in the $\mathfrak{e}_6$ Nilcone}
\label{sec:e6special}

The $\mathfrak{e}_6$ nilcone contains 21 orbits and has three non-trivial special pieces; two have Lusztig canonical quotients of $S_2$ and one has $S_3$. The $S_3$ special piece occurs in the middle of the Hasse diagram and has $D_4\left(a_1\right)$ as its defining special orbit. The $S_2$ special pieces are associated with the $\{A_2, E_6(a_3) \}$ special orbits. Under the Lusztig-Spaltenstein map, the $E_6(a_3)$ and $A_2$ orbits are permuted, while $D_4(a_1)$ is mapped to itself. As shown in Table \ref{fig:AC_Map_Data_GFE6}, the $d_{SD}$ map on this nilcone recovers an unambiguous involution under which the $S_3$ special piece is self dual and the two $S_2$ special pieces are dual to each other.

From the gauge theory perspective, there are various quivers known to realise orbits and intersections involving the special pieces inside the $\mathfrak{e}_6$ nilcone and some of these are analysed below.

\subsection{$S_3$ Special Piece}
\label{subsec:e6specialS3}

The $S_3$ special piece is contained within the S\l odowy slice $\mathcal{S}^{\mathfrak{e}_6}_{2A_2}$, which has $\mathfrak{g}_2$ transverse symmetry. The magnetic quiver for the intersection $\mathcal{S}^{\mathfrak{e}_6}_{D_4(a_1),\;2A_2}$ is consequently the same as \ref{quiv:G2_Example}. The $S_3$ special piece is also spanned by intersections of higher dimension, such as $\mathcal{S}^{\mathfrak{e}_6}_{D_4(a_1),\;A_2+A_1}$, for which a magnetic quiver was identified in \cite{Bennett:2024loi}. This quiver is shown in \eqref{quiv:E6_theoriesS3}, along with its alternative form, in which the action of the hypermultiplet is replaced by an $S_3$ wreathing.

\begin{equation}
\raisebox{-0.5\height}
{\begin{tikzpicture}
    \node (a) at (0,0) {$\begin{tikzpicture}
    \node[gauge, label=below:$1$] (1) at (0,0){};
    \node[gauge, label=below:$2$] (2) at (1,0){};
    \node[gauge, label=below:$3$] (3) at (2,0){};
    \node[flavour, label=below:$1$] (4) at (3,0){};
    \draw[-] (1)--(2)--(3)--(4);
    \draw (3) to[out=135,in=45,looseness=8](3);
    \end{tikzpicture}$};
    \node (b) at (6,0) {$\begin{tikzpicture}
     \node[gauge, label=below:$1$] (1) at (0,0){};
    \node[gauge, label=below:$2$] (2) at (1,0){};
    \node[gauge, label=below:$1$] (3) at (2,0){};
    \node[flavour,label=below:$1$] (4) at (3,0){};
    \draw[-] (1)--(2)--(3)--(4);
    \node[] at (1.5,0){$[$};
    \node[] at (3.5,0){$]$};
    \node[] at (4.0,0){$\; \wr \; S_3$};
    \end{tikzpicture}$};
\end{tikzpicture}}
\label{quiv:E6_theoriesS3}
\end{equation}
The relationship of the flavour node to the $S_3$ action involves some subtleties. In moving to the family of quivers under the Loop Lace map in Table \ref{tab:E6_unitary}, the legs of the $S_3$ form of \ref{quiv:E6_theoriesS3} are apparent in the foldings of the magnetic quivers and the bouquet, while the hypermultiplet loops are apparent in the electric quivers.

\begin{table}[h!]
\centering
\scalebox{0.8}{\begin{tabular}{|c|c|c|c|c|c|}
\hline
    $\rho$ & $\left(\left|{\cal C}(Q_M)\right|, \left|{\cal H}(Q_E)\right|\right)$  & $(A(\cal O), C(\cal O))$ & $Q_{\text{M}}\left[\mathcal{S}^{\mathfrak{e}_6}_{\rho,\;A_2+A_1}\right]$ & $Q_{\text{E}}\left[\mathcal{S}^{\mathfrak{e}_6}_{D_5\left(a_1\right),\;\rho}\right]$ & $G_{\mathcal{C}}(Q_E)$ \\
    \hline
    $D_4\left(a_1\right)$ & $\left(12,6\right)$ & $(S_3,1)$ & \raisebox{-0.5\height}{\begin{tikzpicture}
    \node[gauge, label=below:$1$] (1) at (0,0){};
    \node[gauge, label=below:$2$] (2) at (1,0){};
    \node[gauge, label=below:$3$] (3) at (2,0){};
    \node[flavour, label=below:$1$] (4) at (3,0){};
    \draw[-] (1)--(2)--(3)--(4);
    \draw (3) to[out=135,in=45,looseness=8](3);
    \end{tikzpicture}} & \raisebox{-0.5\height}{\begin{tikzpicture}
    \node[gauge, label=below:$1$] (1) at (0,0){};
    \node[gauge, label=below:$2$] (2) at (1,0){};
    \node[gauge, label=below:$1$] (3) at (2,1){};
    \node[gauge, label=below:$1$] (4) at (2,0){};
    \node[gauge, label=below:$1$] (5) at (2,-1){};
    \node[flavour, label=below:$1$] (6) at (3,1){};
    \node[flavour, label=below:$1$] (7) at (3,0){};
    \node[flavour, label=below:$1$] (8) at (3,-1){};
    \draw[-] (1)--(2)--(3)--(6) (2)--(4)--(7) (2)--(5)--(8);
    \draw[green] (3) circle (0.2cm);
    \draw[green] (4) circle (0.2cm);
    \draw[green] (5) circle (0.2cm);
    \end{tikzpicture}} & $S_3$\\ \hline
    $A_3+A_1$ & $\left(10,8\right)$ & $(1,S_2)$ & \raisebox{-0.5\height}{\begin{tikzpicture}
    \node[gauge, label=below:$1$] (1) at (0,0){};
    \node[gauge, label=below:$2$] (2) at (1,0){};
    \node[gauge, label=below:$1$] (3) at (2,0){};
    \node[flavour, label=below:$1$] (4) at (3,0){};
    \node[gauge, label=left:$1$] (5) at (1,1){};
    \node[flavour, label=right:$1$] (6) at (2,1){};
    \draw[-] (1)--(2)--(5)--(6) (3)--(4);
    \draw[transform canvas={yshift=1.3pt}] (2)--(3);
    \draw[transform canvas={yshift=-1.3pt}] (2)--(3);
    \draw[-] (1.4,0.2)--(1.6,0)--(1.4,-0.2);
    \draw[red] (3) circle (0.2cm);
    \draw[red] (5) circle (0.2cm);
    \end{tikzpicture}} & \raisebox{-0.5\height}{\begin{tikzpicture}
    \node[gauge, label=below:$1$] (1) at (0,0){};
    \node[gauge, label=below:$2$] (2) at (1,0){};
    \node[gauge, label=below:$2$] (3) at (2,0.5){};
    \node[gauge, label=below:$1$] (4) at (2,-0.5){};
    \node[flavour, label=below:$1$] (5) at (3,0.5){};
    \node[flavour, label=below:$1$] (6) at (3,-0.5){};
    \draw[-] (1)--(2)--(3)--(5) (2)--(4)--(6);
    \draw[green] (3) circle (0.2cm);
    \draw[green] (4) circle (0.2cm);
    \draw (3) to[out=135,in=45,looseness=8](3);
    \end{tikzpicture}} & $\frac{S_3}{S_2}$\\ \hline
    $2A_2+A_1$ & $\left(8,10\right)$ & $(1,S_3)$ & \raisebox{-0.5\height}{\begin{tikzpicture}
    \node[gauge, label=below:$1$] (1) at (0,0){};
    \node[gauge, label=below:$2$] (2) at (1,0){};
    \node[gauge, label=below:$1$] (3) at (2,0){};
    \node[flavour, label=below:$1$] (4) at (3,0){};
    \draw[-] (1)--(2)--(3)--(4);
    \draw[transform canvas={yshift=1.3pt}] (2)--(3);
    \draw[transform canvas={yshift=-1.3pt}] (2)--(3);
    \draw[-] (1.4,0.2)--(1.6,0)--(1.4,-0.2);
    \end{tikzpicture}} & \raisebox{-0.5\height}{\begin{tikzpicture}
    \node[gauge, label=below:$1$] (1) at (0,0){};
    \node[gauge, label=below:$2$] (2) at (1,0){};
    \node[gauge, label=below:$3$] (3) at (2,0){};
    \node[flavour, label=below:$1$] (4) at (3,0){};
    \draw[-] (1)--(2)--(3)--(4);
    \draw (3) to[out=135,in=45,looseness=8](3);
    \end{tikzpicture}} & $1$\\ \hline
\end{tabular}}

\caption[Electric and Magnetic Quivers for Intersections between $S_3$ Special Piece in ${\mathfrak{e}_6}$ nilcone and $\{D_5\left(a_1\right), \; A_2+A_1 \}$ orbits]{Electric and Magnetic Quivers for Intersections between $S_3$ Special Piece in ${\mathfrak{e}_6}$ nilcone and $\{D_5\left(a_1\right), \; A_2+A_1 \}$ orbits. The special piece is self dual. The orbit $A_3+A_1$ is non-normal and its Coulomb branch construction gives a normalisation. The framing on these unitary quivers has been made explicit.}
\label{tab:E6_unitary}
\end{table}
\FloatBarrier

The magnetic quivers construct S\l odowy intersections between the lower orbit $A_2+A_1$ and orbits $\rho$ within the special piece. The electric quivers construct S\l odowy intersections between the orbits $\rho$ within the special piece and the superior orbit $D_5\left(a_1\right)$. For each orbit $\rho$, the component group and conjugacy class are identifiable from the quiver. 
As with previous examples, the adjoint hypermultiplet on the $\urm(3)$ gauge node sources the $S_3$ component group of the $D_4\left(a_1\right)$ orbit; for the $d_{SD}$ dual quiver, this $S_3$ becomes associated with the Higgs branch decoration.

The magnetic quivers have a $U(3)$ global Coulomb branch symmetry inherited from the $A_2+A_1$ slice. The electric quivers carry Higgs branch symmetries from each $U(1)$ flavour node beyond the first and $A_1$ from each hypermultiplet.

\subsection{$S_2$ Special Pieces}
\label{subsec:e6specialS2}

The $\mathfrak{e}_6$ nilcone presents an interesting example of two $S_2$ special pieces that are dual to each other under $d_{SD}$. They involve the two orbit pairs $\{A_2, \; 3A_1 \}$ and $\{E_6(a_3),\;A_5\}$, such that the orbits of one pair are Special Dual to the S\l odowy slices in the other. The parent magnetic quiver for this family is the orthosymplectic quiver \ref{quiv:E6_theoriesS2}, identified in \cite{Hanany:2022itc}, which constructs the $42\;\rm{dim}_{\mathbb C}$ orbit $\bar {\mathcal{O}}^{\mathfrak{e}_6}_{A_2}$.

\begin{equation}
\raisebox{-0.5\height}
{\begin{tikzpicture}
    \node[gauger, label=below:$D_1$] (0) at (0,0){};
    \node[] (cdots) at (1,0){$\cdots$};
    \node[gaugeb, label=below:$C_3$] (1) at (2,0){};
    \node[gauger, label=below:$D_4$] (2) at (3,0){};
    \node[gaugeb, label=right:$C_2$] (3) at (3,1){};
    \node[gaugeb, label=below:$C_2$] (4) at (4,0){};
    \node[gauger, label=below:$D_1$] (5) at (5,0){};
    \draw[-] (0)--(cdots)--(1)--(2)--(4)--(5) (2)--(3);
    \draw (3) to[out=135,in=45,looseness=8](3);
\end{tikzpicture}}
\label{quiv:E6_theoriesS2}
\end{equation}

 Starting with this magnetic quiver, the construction of a Loop Lace map is straightforward, as shown in Table \ref{tab:E6_orthosym}.

\begin{table}[h!]
\centering
\scalebox{0.75}{\begin{tabular}{|c|c|c|c|c|c|}
\hline
    $\rho$ & $\left(\left|{\cal C}(Q_M)\right|, \left|{\cal H}(Q_E)\right|\right)$  & $(A(\cal O), C(\cal O))$ & $Q_{\text{M}}\left[\mathcal{O}^{\mathfrak{e}_6}_{\rho}\right]$ & $Q_{\text{E}}\left[\mathcal{S}^{\mathfrak{e}_6}_{d_{SD}(\rho)}\right]$ & $G_{\mathcal{C}}(Q_E)$\\
    \hline
    $A_2$ & $\left(42,6\right)$ & $(S_2,1)$ & \raisebox{-0.5\height}{\begin{tikzpicture}
    \node[gauger, label=below:$D_1$] (0) at (0,0){};
    \node[] (cdots) at (1,0){$\cdots$};
    \node[gaugeb, label=below:$C_3$] (1) at (2,0){};
    \node[gauger, label=below:$D_4$] (2) at (3,0){};
    \node[gaugeb, label=right:$C_2$] (3) at (3,1){};
    \node[gaugeb, label=below:$C_2$] (4) at (4,0){};
    \node[gauger, label=below:$D_1$] (5) at (5,0){};
    \draw[-] (0)--(cdots)--(1)--(2)--(4)--(5) (2)--(3);
    \draw (3) to[out=135,in=45,looseness=8](3);
    \end{tikzpicture}} & \raisebox{-0.5\height}{\begin{tikzpicture}
    \node[gauger, label=below:$D_1$] (0) at (0,0){};
    \node[] (cdots) at (1,0){$\cdots$};
    \node[gaugeb, label=below:$C_3$] (1) at (2,0){};
    \node[gauger, label=below:$D_4$] (2) at (3,0){};
    \node[gaugeb, label=below:$C_2$] (3) at (4,0){};
    \node[gauger, label=below:$D_1$] (4) at (5,0){};
    \node[gaugeb, label=left:$C_1$] (5) at (2.25,0.75){};
    \node[gaugeb, label=right:$C_1$] (6) at (3.75,0.75){};
    \draw[-] (0)--(cdots)--(1)--(2)--(3)--(4) (2)--(5) (2)--(6);
    \draw[green] (5) circle (0.2cm);
    \draw[green] (6) circle (0.2cm);
    \end{tikzpicture}} & $S_2$ \\ \hline
    $3A_1$ & $\left(40,8\right)$ & $(1,S_2)$ & \raisebox{-0.5\height}{\begin{tikzpicture}
    \node[gauger, label=below:$D_1$] (0) at (0,0){};
    \node[] (cdots) at (1,0){$\cdots$};
    \node[gaugeb, label=below:$C_3$] (1) at (2,0){};
    \node[gauger, label=below:$D_4$] (2) at (3,0){};
    \node[gaugeb, label=right:$C_1$] (3) at (3,1){};
    \node[gaugeb, label=below:$C_2$] (4) at (4,0){};
    \node[gauger, label=below:$D_1$] (5) at (5,0){};
    \draw[-] (0)--(cdots)--(1)--(2)--(4)--(5);
    \draw[transform canvas={xshift=1.3pt}] (2)--(3);
    \draw[transform canvas={xshift=-1.3pt}] (2)--(3);
    \draw[-] (2.8,0.4)--(3,0.6)--(3.2,0.4);
    \end{tikzpicture}} & \raisebox{-0.5\height}{\begin{tikzpicture}
    \node[gauger, label=below:$D_1$] (0) at (0,0){};
    \node[] (cdots) at (1,0){$\cdots$};
    \node[gaugeb, label=below:$C_3$] (1) at (2,0){};
    \node[gauger, label=below:$D_4$] (2) at (3,0){};
    \node[gaugeb, label=right:$C_2$] (3) at (3,1){};
    \node[gaugeb, label=below:$C_2$] (4) at (4,0){};
    \node[gauger, label=below:$D_1$] (5) at (5,0){};
    \draw[-] (0)--(cdots)--(1)--(2)--(4)--(5) (2)--(3);
    \draw (3) to[out=135,in=45,looseness=8](3);
    \end{tikzpicture}} & $1$\\ \hline
\end{tabular}}
\caption[Electric and Magnetic Quivers for Orbits and Slices to $S_2$ Special Pieces in ${\mathfrak{e}_6}$ nilcone. ]{Electric and Magnetic Quivers for Orbits and Slices to $S_2$ Special Pieces in ${\mathfrak{e}_6}$ nilcone. The special pieces $\{ A_2,\; 3A_1 \}$ and $\{E_6(a_3), A_5\}$ are dual to each other under $d_{SD}$.}
\label{tab:E6_orthosym}
\end{table}

The magnetic quivers construct orbits $\rho$ within the special piece $\{A_2,\; 3A_1 \}$. The electric quivers construct S\l odowy slices from the orbits $d_{SD}(\rho)$ within the dual special piece $\{E_6(a_3),\;A_5\}$. The magnetic quivers naturally have $E_6$ global Coulomb branch symmetry.

\FloatBarrier
\clearpage
\section{Special Pieces in the $\mathfrak{e}_7$ Nilcone}
\label{sec:e7special}

The $\mathfrak{e}_{7}$ nilcone, which has $\rm{dim}_{\mathbb C} $ $126$, contains 45 orbits and eight special pieces. Two special pieces have an $S_3$ Lusztig canonical quotient while the other six have $S_2$. The special orbits inside the $S_3$ special pieces, $E_7\left(a_5\right)$ and $D_4\left(a_1\right)$, are permuted under $d_{LS}$. Four of the six $S_2$ special pieces have complete images under $d_{SD}$; the remaining two pieces, associated with the special orbits $A_2+A_1$ and $D_5\left(a_1\right)$, do not. A full summary of the $d_{SD}$ data for the $\mathfrak{e}_{7}$ nilcone is given in Table \ref{fig:AC_Map_Data_E7E8}.

\subsection{$S_3$ Special Pieces}

The $S_3$ special pieces in the $\mathfrak{e}_7$ nilcone are $\{D_4\left(a_1\right),\;\left(A_3+A_1\right)',\;2A_2+A_1\}$ and $\{E_7\left(a_5\right),\;D_6\left(a_2\right),\;A_5+A_1\}$, with $E_7\left(a_5\right)$ and $D_4\left(a_1\right)$ as the parent special orbits. All orbits in these pieces are normal except $D_6\left(a_2\right)$ and $\left(A_3+A_1\right)'$. These two special pieces are Special Dual to each other under the $d_{SD}$ map and this permits the construction of a complete Loop Lace map.

Various parent magnetic quivers can be given for ${\mathfrak{e}_7}$ intersections ending on $D_4(a_1)$. These include the unitary quiver \eqref{quiv:G2_Example}, which constructs the $10 \; \rm{dim }_{\mathbb C}$ intersection ${\cal S}_{(D_4(a_1),A_2+3A_1)}$, and \eqref{quiv:e7_s3}, which constructs the $12 \; \rm{dim }_{\mathbb C}$ intersection ${\cal S}_{(D_4(a_1),A_2+2A_1)}$. These magnetic quivers are related via the Loop Lace map to electric quivers identified in \cite{Bennett:2024loi} for the Special Dual intersections ${\cal S}_{(A_6, A_5+A_1)}$ and ${\cal S}_{(E_7(a_4), A_5+A_1)}$, respectively.

\begin{equation}
    \raisebox{-0.5\height}{\begin{tikzpicture}
    \node[gauge, label=below:$1$] (0) at (0,-0.5){};
    \node[gauge, label=below:$2$] (1) at (1,0){};
    \node[gauge, label=below:$3$] (2) at (2,0){};
    \node[gauge, label=above:$1$] (3) at (0,0.5){};
    \draw[-] (0)--(1)--(2) (1)--(3);
    \draw (2) to[out=-45,in=45,looseness=8](2);
    \end{tikzpicture}}
\label{quiv:e7_s3}
\end{equation}

The Loop Lace map for the $12 \; \rm{dim }_{\mathbb C}$ example, based on \eqref{quiv:e7_s3} is shown in Table \ref{tab:E7_S3}.

\begin{table}[h!]
\centering
\scalebox{0.8}{\begin{tabular}{|c|c|c|c|c|c|}
\hline
$\rho$ & $\left(\left|{\cal C}(Q_M)\right|, \left|{\cal H}(Q_E)\right|\right)$ & $(A(\cal O), C(\cal O))$ &  $Q_{\text{M}}\left[\mathcal{S}^{\mathfrak{e}_7}_{\rho,\;A_2+2A_1}\right]$ & $Q_{\text{E}}\left[\mathcal{S}^{\mathfrak{e}_7}_{E_7(a_4),\;d_{SD}\left(\rho\right)}\right]$ & $G_{\mathcal{C}}(Q_E)$  \\\hline
$D_4\left(a_1\right)$ & $\left(12,4\right)$ & $(S_3,1)$ & \raisebox{-0.5\height}{\begin{tikzpicture}
    \node[gauge, label=below:$1$] (0) at (0,-0.5){};
    \node[gauge, label=below:$2$] (1) at (1,0){};
    \node[gauge, label=below:$3$] (2) at (2,0){};
    \node[gauge, label=above:$1$] (3) at (0,0.5){};
    \draw[-] (0)--(1)--(2) (1)--(3);
    \draw (2) to[out=-45,in=45,looseness=8](2);
    \end{tikzpicture}} & \raisebox{-0.5\height}{\begin{tikzpicture}
    \node[gauge, label=below:$1$] (0) at (0,-0.5){};
    \node[gauge, label=below:$2$] (1) at (1,0){};
    \node[gauge, label=right:$1$] (2) at (2,0.5){};
    \node[gauge, label=right:$1$] (3) at (2,0){};
    \node[gauge, label=right:$1$] (4) at (2,-0.5){};
    \node[gauge, label=above:$1$] (5) at (0,0.5){};
    \draw[-] (0)--(1)--(2) (1)--(5) (3)--(1) (4)--(1);
    \draw[green] (2) circle (0.2cm);
    \draw[green] (3) circle (0.2cm);
    \draw[green] (4) circle (0.2cm);
    \end{tikzpicture}} & $S_3$ \\ \hline
    $(A_3+A_1)'$ & $\left(10,6\right)$ & $(1,S_2)$ & \raisebox{-0.5\height}{\begin{tikzpicture}
    \node[gauge, label=below:$1$] (0) at (0,-0.5){};
    \node[gauge, label=below:$2$] (1) at (1,0){};
    \node[gauge, label=below:$1$] (2) at (2,0){};
    \node[gauge, label=above:$1$] (3) at (0,0.5){};
    \node[gauge, label=above:$1$] (4) at (1,1){};
    \draw[transform canvas={yshift=1.3pt}] (1)--(2);
    \draw[transform canvas={yshift=-1.3pt}] (1)--(2);
    \draw[-] (1.4,-0.2)--(1.6,0)--(1.4,0.2);
    \draw[-] (0)--(1)--(3) (1)--(4);
    \draw[red] (2) circle (0.2cm);
    \draw[red] (4) circle (0.2cm);
    \end{tikzpicture}} & \raisebox{-0.5\height}{\begin{tikzpicture}
    \node[gauge, label=below:$1$] (0) at (0,-0.5){};
    \node[gauge, label=below:$2$] (1) at (1,0){};
    \node[gauge, label={[xshift=0.55cm, yshift=-0.35cm]$2$}] (2) at (2,0.5){};
    \node[gauge, label=right:$1$] (3) at (2,-0.5){};
    \node[gauge, label=above:$1$] (4) at (0,0.5){};
    \draw[-] (0)--(1)--(2) (3)--(1) (1)--(4);
    \draw[green] (2) circle (0.2cm);
    \draw[green] (3) circle (0.2cm);
    \draw (2) to[out=45,in=-45,looseness=8](2);
    \end{tikzpicture}} & $\frac{S_3}{S_2}$\\ \hline
    $2A_2+A_1$ & $\left(8,8\right)$ & $(1,S_3)$ & \raisebox{-0.5\height}
    {\begin{tikzpicture}
    \node[gauge, label=below:$1$] (0) at (0,-0.5){};
    \node[gauge, label=below:$2$] (1) at (1,0){};
    \node[gauge, label=below:$1$] (2) at (2,0){};
    \node[gauge, label=above:$1$] (3) at (0,0.5){};
    \draw[transform canvas={yshift=1.3pt}] (1)--(2);
    \draw[transform canvas={yshift=-1.3pt}] (1)--(2);
    \draw[-] (0)--(1)--(2) (1)--(3);
    \draw[-] (1.4,-0.2)--(1.6,0)--(1.4,0.2);
    \end{tikzpicture}} & \raisebox{-0.5\height}{\begin{tikzpicture}
    \node[gauge, label=left:$1$] (0) at (0,-0.5){};
    \node[gauge, label=below:$2$] (1) at (1,0){};
    \node[gauge, label=below:$3$] (2) at (2,0){};
    \node[gauge, label=left:$1$] (3) at (0,0.5){};
    \draw[-] (0)--(1)--(2) (1)--(3);
    \draw (2) to[out=-45,in=45,looseness=8](2);
        \end{tikzpicture}} & $1$ \\ \hline
\end{tabular}}
\caption[Electric and Magnetic Quivers for S\l odowy Intersections between $S_3$ Special Pieces in ${\mathfrak{e}_7}$ nilcone and $\{E_7(a_4), \;A_2+2A_1 \}$ orbits]{Electric and Magnetic Quivers for S\l odowy Intersections between $S_3$ Special Pieces in ${\mathfrak{e}_7}$ nilcone and $\{E_7(a_4), \;A_2+2A_1 \}$ orbits. The special piece $\{D_4(a_1), \; (A_3+A_1)',\; 2A_2+A_1 \}$ is dual to $\{E_7(a_5), \; D_6(a_2),\; A_5+A_1 \}$ under $d_{SD}$. The orbit $(A_3+A_1)'$ is non-normal and its Coulomb branch construction gives a normalisation.}
\label{tab:E7_S3}
\end{table}

The magnetic quivers have a $A_1 \times A_1 \times A_1$ global Coulomb branch symmetry inherited from the $A_2+2A_1$ S\l odowy slice. The electric quivers carry Higgs branch symmetries $A_1$ from each hypermultiplet. The sum of the dimensions of the electric and magnetic quivers is constant across the special piece, as summarised in \eqref{eqn:e7_s3_dimsum}.

\begin{equation}
\begin{aligned}
\left|{\cal C}(Q_M)\right|+ \left|{\cal H}(Q_E)\right| &= \rm{dim}_{\mathbb{C}}\left({\cal S}_{D_4\left(a_1\right),\;A_2+2A_1}^{\mathfrak{e}_7}\right)+\rm{dim}_{\mathbb{C}}\left({\cal S}_{E_7\left(a_4\right),\;E_7\left(a_5\right)}^{\mathfrak{e}_7}\right) &= 16\\
\end{aligned}
\label{eqn:e7_s3_dimsum}
\end{equation}

Turning to the $10 \; \rm{dim }_{\mathbb C}$ example, the magnetic and electric quivers from the $\mathfrak{g}_2$ nilcone in Table \ref{Tab:G2_special_piece} also construct intersections in the $\mathfrak{e}_7$ nilcone. This follows from the fact that the two $S_3$ special pieces in the $\mathfrak{e}_7$ nilcone 
are both isomorphic to the $S_3$ special piece
in the $\mathfrak{g}_2$ nilcone. Thus, there are two Loop Lace maps for the $10 \; \rm{dim }_{\mathbb C}$ case, which involve a transposition of the two special pieces related by the $d_{SD}$ map. These are shown in Table \ref{tab:E7_S3B}. 

\begin{table}[h!]
\centering
\scalebox{0.8}{\begin{tabular}{|c|c|c|c|c|c|}
\hline
$\begin{array}{c}
{\rho}\\{\sigma}
\end{array}$
& $\left(\left|{\cal C}(Q_M)\right|, \left|{\cal H}(Q_E)\right|\right)$ & $(A(\cal O), C(\cal O))$ &  
$\begin{array}{c}
Q_{\text{M}}\left[\mathcal{S}^{\mathfrak{e}_7}_{\rho,\;A_2+3A_1}\right]\\Q_{\text{M}}\left[\mathcal{S}^{\mathfrak{e}_7}_{\sigma,\;(A_5)''}\right]
\end{array}$
& 
$\begin{array}{c}
Q_{\text{E}}\left[\mathcal{S}^{\mathfrak{e}_7}_{A_6,\;d_{SD}\left(\rho\right)}\right]\\Q_{\text{E}}\left[\mathcal{S}^{\mathfrak{e}_7}_{D_4,\;d_{SD}\left(\sigma\right)}\right]
\end{array}$
& $G_{\mathcal{C}}(Q_E)$  \\
\hline
$\begin{array}{c}
D_4\left(a_1\right)\\E_7\left(a_5\right)
\end{array}$
& $\left(10,2\right)$ & $(S_3,1)$ & \raisebox{-0.5\height}{\begin{tikzpicture}
    \node[gauge, label=below:$1$] (0) at (0,0){};
    \node[gauge, label=below:$2$] (1) at (1,0){};
    \node[gauge, label=below:$3$] (2) at (2,0){};
    \draw[-] (0)--(1)--(2);
    \draw (2) to[out=-45,in=45,looseness=8](2);
    \end{tikzpicture}} & \raisebox{-0.5\height}{\begin{tikzpicture}
    \node[gauge, label=below:$1$] (0) at (0,0){};
    \node[gauge, label=below:$2$] (1) at (1,0){};
    \node[gauge, label=right:$1$] (2) at (2,0.5){};
    \node[gauge, label=right:$1$] (3) at (2,0){};
    \node[gauge, label=right:$1$] (4) at (2,-0.5){};
    \draw[-] (0)--(1)--(2) (3)--(1) (4)--(1);
    \draw[green] (2) circle (0.2cm);
    \draw[green] (3) circle (0.2cm);
    \draw[green] (4) circle (0.2cm);
    \end{tikzpicture}} 
    & $S_3$ \\
    \hline
$\begin{array}{c}
(A_3+A_1)'\\D_6\left(a_2\right)
\end{array}$
    & $\left(8,4\right)$ & $(1,S_2)$ &
    \raisebox{-0.5\height}{\begin{tikzpicture}
    \node[gauge, label=below:$1$] (0) at (0,0){};
    \node[gauge, label=below:$2$] (1) at (1,0){};
    \node[gauge, label=below:$1$] (2) at (2,0){};
    \node[gauge, label=above:$1$] (4) at (1,1){};
    \draw[transform canvas={yshift=1.3pt}] (1)--(2);
    \draw[transform canvas={yshift=-1.3pt}] (1)--(2);
    \draw[-] (1.4,-0.2)--(1.6,0)--(1.4,0.2);
    \draw[-] (0)--(1)--(4);
    \draw[red] (2) circle (0.2cm);
    \draw[red] (4) circle (0.2cm);
    \end{tikzpicture}} 
    & 
    \raisebox{-0.5\height}{\begin{tikzpicture}
    \node[gauge, label=below:$1$] (0) at (0,0){};
    \node[gauge, label=below:$2$] (1) at (1,0){};
    \node[gauge, label={[xshift=0.55cm, yshift=-0.35cm]$2$}] (2) at (2,0.5){};
    \node[gauge, label=right:$1$] (3) at (2,-0.5){};
    \draw[-] (0)--(1)--(2) (3)--(1);
    \draw[green] (2) circle (0.2cm);
    \draw[green] (3) circle (0.2cm);
    \draw (2) to[out=45,in=-45,looseness=8](2);
    \end{tikzpicture}} 
    & $\frac{S_3}{S_2}$\\ 
    \hline
    $\begin{array}{c}
2A_2+A_1\\A_5+A_1
\end{array}$
    
    & $\left(6,6\right)$ & $(1,S_3)$ &
    \raisebox{-0.5\height}
    {\begin{tikzpicture}
    \node[gauge, label=below:$1$] (0) at (0,0){};
    \node[gauge, label=below:$2$] (1) at (1,0){};
    \node[gauge, label=below:$1$] (2) at (2,0){};
    \draw[transform canvas={yshift=1.3pt}] (1)--(2);
    \draw[transform canvas={yshift=-1.3pt}] (1)--(2);
    \draw[-] (0)--(1)--(2);
    \draw[-] (1.4,-0.2)--(1.6,0)--(1.4,0.2);
    \end{tikzpicture}} 
    & \raisebox{-0.5\height}{\begin{tikzpicture}
    \node[gauge, label=below:$1$] (0) at (0,0){};
    \node[gauge, label=below:$2$] (1) at (1,0){};
    \node[gauge, label=below:$3$] (2) at (2,0){};
    \draw[-] (0)--(1)--(2);
    \draw (2) to[out=-45,in=45,looseness=8](2);
        \end{tikzpicture}} 
    & $1$ \\ \hline
\end{tabular}}
\caption[Electric and Magnetic Quivers for S\l odowy Intersections between $S_3$ Special Pieces in ${\mathfrak{e}_7}$ nilcone and $\{A_6,\; A_2+3A_1 \}$ or $\{D_4,\; (A_5)''\}$ orbits]{Electric and Magnetic Quivers for S\l odowy Intersections between $S_3$ Special Pieces in ${\mathfrak{e}_7}$ nilcone and $\{A_6,\; A_2+3A_1 \}$ or $\{D_4,\; (A_5)''\}$ orbits. The orbits $(A_3+A_1)'$ and $D_6(a_2)$ are non-normal and their Coulomb branch constructions give normalisations.}
\label{tab:E7_S3B}
\end{table}
The magnetic quivers have a $G_2$ global Coulomb branch symmetry inherited from the $A_2+3A_1$ and $(A_5)''$ slices, enhanced to $B_3$ for the normalisation. The electric quivers carry Higgs branch symmetries $A_1$ from each hypermultiplet. The sum of the dimensions of the electric and magnetic quivers is constant across the special piece, as summarised in \eqref{eqn:e7_s3_dimsumB}.

\begin{equation}
\begin{aligned}
\left|{\cal C}(Q_M)\right|+ \left|{\cal H}(Q_E)\right| &= 
 \rm{dim}_{\mathbb{C}}\left({\cal S}_{D_4\left(a_1\right),\;A_2+3A_1}^{\mathfrak{e}_7}\right)+\rm{dim}_{\mathbb{C}}\left({\cal S}_{A_6,\;E_7\left(a_5\right)}^{\mathfrak{e}_7}\right) &= 12\\
&= \rm{dim}_{\mathbb{C}}\left({\cal S}_{E_7\left(a_5\right),\;(A_5)''}^{\mathfrak{e}_7}\right)+\rm{dim}_{\mathbb{C}}\left({\cal S}_{D_4,\;D_4\left(a_1\right)}^{\mathfrak{e}_7}\right) &= 12\\
\end{aligned}
\label{eqn:e7_s3_dimsumB}
\end{equation}
\FloatBarrier

\subsection{$S_2$ Special Pieces}

In the $\mathfrak{e}_7$ algebra there are two pairs of $S_2$ special pieces that are dual to each other. The first pair consists of the $S_2$ special piece $\{A_2,\; (3A_1)'\}$ which is dual to the special piece $\{E_7(a_3),\; D_6\}$. A Loop Lace map constructed from the magnetic quiver for the S\l odowy intersections from the orbit $2A_1$ to the $\{A_2,\; (3A_1)'\}$ special piece is shown in Table \ref{tab:E7_S2}. The magnetic quivers are identifiable from the transitions in the Hasse diagram for $\mathfrak{e}_7$. 

\begin{table}[h!]
\centering
\scalebox{0.8}{\begin{tabular}{|c|c|c|c|c|c|}
\hline
     $\rho$ & $\left(\left|{\cal C}(Q_M)\right|, \left|{\cal H}(Q_E)\right|\right)$ & $(A(\cal O), C(\cal O))$ & $Q_{\text{M}}\left[\mathcal{S}^{\mathfrak{e}_7}_{\rho,\;2A_1}\right]$ & $Q_{\text{E}}\left[\mathcal{S}^{\mathfrak{e}_7}_{E_7\left(a_2\right),\;d_{SD}\left(\rho\right)}\right]$ & $G_{\mathcal{C}}(Q_E)$\\
    \hline
    $A_2$ & $\left(14,2\right)$ & $(S_2,1)$ & \raisebox{-0.5\height}{\begin{tikzpicture}
    \node[gauge, label=below:$1$] (1) at (0,0){};
    \node[gauge, label=below:$2$] (2) at (1,0){};
    \node[gauge, label=below:$2$] (3) at (2,0){};
    \node[gauge, label=below:$2$] (4) at (3,0){};
    \node[gauge, label=above:$1$] (5) at (1,1){};
    \draw[-](1)--(2)--(3)--(4) (2)--(5);
    \draw (4) to[out=-45,in=45,looseness=8](4);
    \end{tikzpicture}} & \raisebox{-0.5\height}{\begin{tikzpicture}
    \node[gauge, label=below:$1$] (1) at (0,0){};
    \node[gauge, label=below:$2$] (2) at (1,0){};
    \node[gauge, label=below:$2$] (3) at (2,0){};
    \node[gauge, label=right:$1$] (4) at (3,0.5){};
    \node[gauge, label=right:$1$] (6) at (3,-0.5){};
    \node[gauge, label=above:$1$] (5) at (1,1){};
    \draw[-](1)--(2)--(3)--(4) (3)--(6) (2)--(5);
    \draw[green] (4) circle (0.2cm);
    \draw[green] (6) circle (0.2cm);
    \end{tikzpicture}} 
    & $S_2$ \\ 
    \hline
    $(3A_1)'$ & $\left(12,4\right)$ & $(1,S_2)$ & \raisebox{-0.5\height}{\begin{tikzpicture}
    \node[gauge, label=below:$1$] (1) at (0,0){};
    \node[gauge, label=below:$2$] (2) at (1,0){};
    \node[gauge, label=below:$2$] (3) at (2,0){};
    \node[gauge, label=below:$1$] (4) at (3,0){};
    \node[gauge, label=above:$1$] (5) at (1,1){};
    \draw[-](1)--(2)--(3) (2)--(5);
    \draw[transform canvas={yshift=1.3pt}] (3)--(4);
    \draw[transform canvas={yshift=-1.3pt}]  (3)--(4);
    \draw[-] (2.4,0.2)--(2.6,0)--(2.4,-0.2);
    \end{tikzpicture}} & \raisebox{-0.5\height}{\begin{tikzpicture}
    \node[gauge, label=below:$1$] (1) at (0,0){};
    \node[gauge, label=below:$2$] (2) at (1,0){};
    \node[gauge, label=below:$2$] (3) at (2,0){};
    \node[gauge, label=below:$2$] (4) at (3,0){};
    \node[gauge, label=above:$1$] (5) at (1,1){};
    \draw[-](1)--(2)--(3)--(4) (2)--(5);
    \draw (4) to[out=-45,in=45,looseness=8](4);
    \end{tikzpicture}} & $1$\\
    \hline
\end{tabular}}
\caption[Electric and Magnetic Quivers for S\l odowy Intersections between $S_2$ Special Pieces in ${\mathfrak{e}_7}$ nilcone and $\{E_7\left(a_2\right),\; 2A_1 \}$ orbits]{Electric and Magnetic Quivers for S\l odowy Intersections between $S_2$ Special Pieces in ${\mathfrak{e}_7}$ nilcone and $\{E_7\left(a_2\right),\; 2A_1 \}$ orbits. The special pieces $\{A_2,\; (3A_1)' \}$ and $\{E_7(a_3), \; D_6 \}$ are dual under $d_{SD}$.}
\label{tab:E7_S2}
\end{table}

In Table \ref{tab:E7_S2} the global symmetry of the Coulomb branch of the magnetic quivers is $B_4$, inherited from the S\l odowy slice to $2A_1$. Indeed, the magnetic quivers construct the minimal and next to minimal orbits of $B_4$. The $2 \;\rm{dim}_{\mathbb{C}}$ Higgs branch is the Kleinian $D_5$ singularity. The electric quivers carry Higgs branch symmetries $A_1$ from each hypermultiplet. 

The second pair consists of the $S_2$ special piece $\{D_4(a_1)+A_1,\; A_3 + 2 A_1 \}$ which is dual to the special piece $\{E_6(a_3),(A_5)'\}$. Loop Lace maps constructed from magnetic quivers for S\l odowy intersections from the orbits $A_2+2A_1$ and $A_2+3A1$ to the $\{D_4(a_1)+A_1,\; A_3 + 2 A_1 \}$ special piece are shown in Tables \ref{tab:E7_S2_A} and \ref{tab:E7_S2_B}, respectively. These are based on electric quivers identified in \cite{Bennett:2024loi}.

\begin{table}[h!]
\centering
\scalebox{0.8}{\begin{tabular}{|c|c|c|c|c|c|}
\hline
     $\rho$ & $\left(\left|{\cal C}(Q_M)\right|, \left|{\cal H}(Q_E)\right|\right)$ & $(A(\cal O), C(\cal O))$ & $Q_{\text{M}}\left[\mathcal{S}^{\mathfrak{e}_7}_{\rho,\;A_2+2A_1}\right]$ & $Q_{\text{E}}\left[\mathcal{S}^{\mathfrak{e}_7}_{E_7\left(a_4\right),\;d_{SD}\left(\rho\right)}\right]$ & $G_{\mathcal{C}}(Q_E)$\\
    \hline
    $D_4(a_1)+A_1$ & $\left(14,6\right)$ & $(S_2,1)$ & \raisebox{-0.5\height}{\begin{tikzpicture}
    \node[gauge, label=below:$1$] (1) at (0,0){};
    \node[gauge, label=below:$2$] (2) at (1,0){};
    \node[gauge, label=below:$2$] (3) at (2,0){};
    \node[gauge, label=below:$2$] (4) at (3,0){};
    \node[gauge, label=above:$1$] (5) at (2,1){};
    \draw[-] (2)--(3)--(4) (3)--(5);
    \draw (4) to[out=-45,in=45,looseness=8](4);
    \draw[transform canvas={yshift=1.3pt}] (1)--(2);
    \draw[transform canvas={yshift=-1.3pt}] (1)--(2);
    \end{tikzpicture}} & \raisebox{-0.5\height}{\begin{tikzpicture}
    \node[gauge, label=below:$1$] (1) at (0,0){};
    \node[gauge, label=below:$2$] (2) at (1,0){};
    \node[gauge, label=below:$2$] (3) at (2,0){};
    \node[gauge, label=right:$1$] (4) at (3,0.5){};
    \node[gauge, label=right:$1$] (6) at (3,-0.5){};
    \node[gauge, label=above:$1$] (5) at (2,1){};
    \draw[-] (2)--(3)--(4) (6)--(3)--(5);
    \draw[transform canvas={yshift=1.3pt}] (1)--(2);
    \draw[transform canvas={yshift=-1.3pt}] (1)--(2);
    \draw[green] (4) circle (0.2cm);
    \draw[green] (6) circle (0.2cm);
    \end{tikzpicture}} & $S_2$ \\ \hline
    $A_3+2A_1$ & $\left(12,8\right)$ & $(1,S_2)$ & \raisebox{-0.5\height}{\begin{tikzpicture}
    \node[gauge, label=below:$1$] (1) at (0,0){};
    \node[gauge, label=below:$2$] (2) at (1,0){};
    \node[gauge, label=below:$2$] (3) at (2,0){};
    \node[gauge, label=below:$1$] (4) at (3,0){};
    \node[gauge, label=above:$1$] (5) at (2,1){};
    \draw[-] (2)--(3)--(5);
    \draw[transform canvas={yshift=1.3pt}] (1)--(2) (3)--(4);
    \draw[transform canvas={yshift=-1.3pt}] (1)--(2) (3)--(4);
    \draw[-] (2.4,0.2)--(2.6,0)--(2.4,-0.2);
    \end{tikzpicture}} & \raisebox{-0.5\height}{\begin{tikzpicture}
    \node[gauge, label=below:$1$] (1) at (0,0){};
    \node[gauge, label=below:$2$] (2) at (1,0){};
    \node[gauge, label=below:$2$] (3) at (2,0){};
    \node[gauge, label=below:$2$] (4) at (3,0){};
    \node[gauge, label=above:$1$] (5) at (2,1){};
    \draw[-] (2)--(3)--(4) (3)--(5);
    \draw (4) to[out=-45,in=45,looseness=8](4);
    \draw[transform canvas={yshift=1.3pt}] (1)--(2);
    \draw[transform canvas={yshift=-1.3pt}] (1)--(2);
    \end{tikzpicture}} & $1$\\ \hline
\end{tabular}}
\caption[Electric and Magnetic Quivers for S\l odowy Intersections between $S_2$ Special Pieces in ${\mathfrak{e}_7}$ nilcone and $\{E_7\left(a_4\right),\; A_2+2A_1 \}$ orbits]{Electric and Magnetic Quivers for S\l odowy Intersections between $S_2$ Special Pieces in ${\mathfrak{e}_7}$ nilcone and $\{E_7\left(a_4\right),\; A_2+2A_1 \}$ orbits. The orbits $D_4(a_1)+A_1$ and $A_3+2A_1$ are non-normal and their Coulomb branch constructions give normalisations. The special pieces $\{D_4(a_1)+A_1,;\ A_3+2A_1 \}$ and $\{E_6(a_3), \; (A_5)' \}$ are dual under $d_{SD}$.}
\label{tab:E7_S2_A}
\end{table}

\begin{table}[h!]
\centering
\scalebox{0.8}{\begin{tabular}{|c|c|c|c|c|c|}
\hline
     $\rho$ & $\left(\left|{\cal C}(Q_M)\right|, \left|{\cal H}(Q_E)\right|\right)$ & $(A(\cal O), C(\cal O))$ & $Q_{\text{M}}\left[\mathcal{S}^{\mathfrak{e}_7}_{\rho,\;A_2+3A_1}\right]$ & $Q_{\text{E}}\left[\mathcal{S}^{\mathfrak{e}_7}_{A_6,\;d_{SD}\left(\rho\right)}\right]$ & $G_{\mathcal{C}}(Q_E)$\\
    \hline
    $D_4(a_1)+A_1$ & $\left(12,4\right)$ & $(S_2,1)$ & \raisebox{-0.5\height}{\begin{tikzpicture}
    \node[gauge, label=below:$1$] (1) at (0,0){};
    \node[gauge, label=below:$2$] (2) at (1,0){};
    \node[gauge, label=below:$2$] (3) at (2,0){};
    \node[gauge, label=below:$2$] (4) at (3,0){};
    \draw[-] (2)--(3)--(4);
    \draw (4) to[out=-45,in=45,looseness=8](4);
    \draw[transform canvas={yshift=1.3pt}] (1)--(2);
    \draw[transform canvas={yshift=-1.3pt}] (1)--(2);
    \end{tikzpicture}} & \raisebox{-0.5\height}{\begin{tikzpicture}
    \node[gauge, label=below:$1$] (1) at (0,0){};
    \node[gauge, label=below:$2$] (2) at (1,0){};
    \node[gauge, label=below:$2$] (3) at (2,0){};
    \node[gauge, label=right:$1$] (4) at (3,0.5){};
    \node[gauge, label=right:$1$] (6) at (3,-0.5){};
    \draw[-] (2)--(3)--(4) (6)--(3);
    \draw[transform canvas={yshift=1.3pt}] (1)--(2);
    \draw[transform canvas={yshift=-1.3pt}] (1)--(2);
    \draw[green] (4) circle (0.2cm);
    \draw[green] (6) circle (0.2cm);
    \end{tikzpicture}} & $S_2$ \\ \hline
    $A_3+2A_1$ & $\left(10,6\right)$ & $(1,S_2)$ & \raisebox{-0.5\height}{\begin{tikzpicture}
    \node[gauge, label=below:$1$] (1) at (0,0){};
    \node[gauge, label=below:$2$] (2) at (1,0){};
    \node[gauge, label=below:$2$] (3) at (2,0){};
    \node[gauge, label=below:$1$] (4) at (3,0){};
    \draw[-] (2)--(3);
    \draw[transform canvas={yshift=1.3pt}] (1)--(2) (3)--(4);
    \draw[transform canvas={yshift=-1.3pt}] (1)--(2) (3)--(4);
    \draw[-] (2.4,0.2)--(2.6,0)--(2.4,-0.2);
    \end{tikzpicture}} & \raisebox{-0.5\height}{\begin{tikzpicture}
    \node[gauge, label=below:$1$] (1) at (0,0){};
    \node[gauge, label=below:$2$] (2) at (1,0){};
    \node[gauge, label=below:$2$] (3) at (2,0){};
    \node[gauge, label=below:$2$] (4) at (3,0){};
    \draw[-] (2)--(3)--(4);
    \draw (4) to[out=-45,in=45,looseness=8](4);
    \draw[transform canvas={yshift=1.3pt}] (1)--(2);
    \draw[transform canvas={yshift=-1.3pt}] (1)--(2);
    \end{tikzpicture}} & $1$\\ \hline
\end{tabular}}
\caption[Electric and Magnetic Quivers for S\l odowy Intersections between $S_2$ Special Pieces in ${\mathfrak{e}_7}$ nilcone and $\{E_7\left(a_4\right),\; A_2+3A_1 \}$ orbits]{Electric and Magnetic Quivers for S\l odowy Intersections between $S_2$ Special Pieces in ${\mathfrak{e}_7}$ nilcone and $\{E_7\left(a_4\right),\; A_2+3A_1 \}$ orbits. The orbits $D_4(a_1)+A_1$ and $A_3+2A_1$ are non-normal and their Coulomb branch constructions give normalisations. The special pieces $\{D_4(a_1)+A_1,;\ A_3+2A_1 \}$ and $\{E_6(a_3), \; (A_5)' \}$ are dual under $d_{SD}$.}
\label{tab:E7_S2_B}
\end{table}

In Table \ref{tab:E7_S2_A} the global symmetry of the Coulomb branch of the magnetic quivers is $A_1 \times A_1 \times A_1$, inherited from the S\l odowy slice to $A_2+2A_1$. In Table \ref{tab:E7_S2_B} the global symmetry of the Coulomb branch of the magnetic quivers is $G_2$, inherited from the S\l odowy slice to $A_2+3A_1$, enhanced by normalisation to $B_3$. The electric quivers carry Higgs branch symmetries $A_1$ from each hypermultiplet.

\FloatBarrier

The final two $S_2$ pieces of the $\mathfrak{e}_7$ nilcone are $d_{LS}$ dual to special orbits with trivial special pieces. In these cases only partial Loop Lace maps can be given. These are shown in Tables \ref{tab:E7S2S1A} and \ref{tab:E7S2S1B}.

\begin{table}[h!]
\centering
\scalebox{0.8}{
\begin{tabular}{|c|c|c|c|c|c|}
\hline
    $\rho$ & 
     $\left(\left|{\cal C}(Q_M)\right|, \left|{\cal H}(Q_E)\right|\right)$ 
     & $(A(\cal O), C(\cal O))$
    & $Q_{\text{M}}\left[\mathcal{S}^{\mathfrak{e}_7}_{\rho,\;(3A_1)''}\right]$ 
    & $Q_{\text{E}}\left[\mathcal{S}^{\mathfrak{e}_7}_{E_6,\;{d_{SD}(\rho)}}\right]$ 
    & $G_{\mathcal{C}}(Q_E)$ \\
    \hline
    $A_2+A_1$ 
    & $\left(\textrm{22},2\right)$ 
    & $(S_2,1)$ &
    \raisebox{-0.5\height}
    {\begin{tikzpicture}
    \node[gauge, label=below:$1$] (1) at (0,0){};
    \node[gauge, label=below:$2$] (2) at (1,0){};
    \node[gauge, label=below:$3$] (3) at (2,0){};
    \node[gauge, label=below:$2$] (4) at (3,0){};
    \node[gauge, label=below:$1$] (5) at (4,0){};
    \draw[-](1)--(2)--(3)--(4)--(5);
    \node[] at (2.5,0){$[$};
    \node[] at (4.3,0){$]$};
    \node[] at (4.6,0){$\; \wr \; S_2$};
    \end{tikzpicture}}  
    & \raisebox{-0.5\height}{\begin{tikzpicture}
    \node[gauge, label=below:$1$] (1) at (0,0){};
    \node[gauge, label=below:$2$] (2) at (1,0){};
    \node[gauge, label=below:$3$] (3) at (2,0){};
    \node[gauge, label=below:$2$] (4) at (3,0){};
    \node[gauge, label=below:$1$] (5) at (4,0){};
    \node[gauge, label=right:$2$] (6) at (2,1){};
    \node[gauge, label=right:$1$] (7) at (2,2){};
    \draw[-] (1)--(2)--(3)--(4)--(5) (3)--(6)--(7);
    \draw[green] \convexpath{6,7} {0.2cm};
    \draw[green] \convexpath{4,5} {0.2cm};
    \end{tikzpicture}} & $S_2$\\ 
    \hline
    $4A_1$ & 
    $\left(16,\textrm{N/A}\right)$ 
    \raisebox{-0.5\height}
    & $(1,S_2)$ & \raisebox{-0.5\height}{\begin{tikzpicture}
    \node[gauge, label=below:$1$] (1) at (0,0){};
    \node[gauge, label=below:$2$] (2) at (1,0){};
    \node[gauge, label=below:$3$] (3) at (2,0){};
    \node[gauge, label=below:$2$] (4) at (3,0){};
    \node[gauge, label=below:$1$] (5) at (4,0){};
    \draw[-] (1)--(2)--(3) (4)--(5);
    \draw[transform canvas={yshift=1.3pt}] (3)--(4);
    \draw[transform canvas={yshift=-1.3pt}] (3)--(4);
    \draw[-] (2.4,0.2)--(2.6,0)--(2.4,-0.2);
    \end{tikzpicture}} & \raisebox{-0.5\height}{$\textrm{N/A}$} & \raisebox{-0.5\height}{$\textrm{N/A}$} \\ \hline
\end{tabular}}
\caption[Electric and Magnetic Quivers for S\l odowy Intersections between $S_2$ Special Piece in ${\mathfrak{e}_7}$ nilcone, its trivial dual and $\{E_6,\; (3A_1)'' \}$ orbits]{Electric and Magnetic Quivers for S\l odowy Intersections between $S_2$ Special Piece in ${\mathfrak{e}_7}$ nilcone, its trivial dual and $\{E_6,\; (3A_1)'' \}$orbits. The $\{A_2+A_1,\; 4A_1 \}$ special piece is $d_{LS}$ dual to the special orbit $E_6(a_1) $ with trivial special piece.}
\label{tab:E7S2S1A}
\end{table}

\begin{table}[h!]
\centering
\scalebox{0.8}{
\begin{tabular}{|c|c|c|c|c|c|}
\hline
    $\rho$ & 
     $\left(\left|{\cal C}(Q_M)\right|, \left|{\cal H}(Q_E)\right|\right)$ 
     & $(A(\cal O), C(\cal O))$
    & $Q_{\text{M}}\left[\mathcal{S}^{\mathfrak{e}_7}_{\rho,\;D_4}\right]$ 
    & $Q_{\text{E}}\left[\mathcal{S}^{\mathfrak{e}_7}_{(A_5)'',\;{d_{SD}(\rho)}}\right]$ 
    & $G_{\mathcal{C}}(Q_E)$ \\
    \hline
    $D_5(a_1)$ 
    & $\left(10,2\right)$ 
    & $(S_2,1)$ &
    \raisebox{-0.5\height}
    {\begin{tikzpicture}
    \node[gauge, label=below:$1$] (1) at (0,0){};
    \node[gauge, label=below:$1$] (2) at (1,0){};
    \node[gauge, label=below:$1$] (3) at (2,0){};
    \node[flavour, label=below:$1$] (4) at (3,0){};
    \draw[-](1)--(2)--(3)--(4);
    \node[] at (0.5,0){$[$};
    \node[] at (3.5,0){$]$};
    \node[] at (4.0,0){$\; \wr \; S_2$};
    \end{tikzpicture}}  
    &
    \raisebox{-0.5\height}{\begin{tikzpicture}
    \node[gauge, label=below:$1$] (1) at (0,0){};
    \node[gauge, label=below:$1$] (2) at (1,0){};
    \node[gauge, label=below:$1$] (3) at (2,0){};
    \node[gauge, label=below:$1$] (4) at (3,0){};
    \node[gauge, label=below:$1$] (5) at (4,0){};
    \node[flavour, label=right:$1$] (6) at (0,-1){};
    \node[flavour, label=right:$1$] (7) at (4,-1){};
    \draw[-] (6)--(1)--(2)--(3)--(4)--(5)--(7);
    \draw[green] \convexpath{6,1,2} {0.2cm};
    \draw[green] \convexpath{4,5,7} {0.2cm};
    \end{tikzpicture}} 
    & $S_2$\\ 
    \hline
    $D_4+A_1$ & 
    $\left(6,\textrm{N/A}\right)$ 
    \raisebox{-0.5\height}
    & $(1,S_2)$ &
    \raisebox{-0.5\height}{\begin{tikzpicture}
    \node[gauge, label=below:$1$] (1) at (0,0){};
    \node[gauge, label=below:$1$] (2) at (1,0){};
    \node[gauge, label=below:$1$] (3) at (2,0){};
    \node[flavour, label=below:$1$] (4) at (3,0){};
    \draw[-] (2)--(3)--(4);
    \draw[transform canvas={yshift=1.3pt}] (1)--(2);
    \draw[transform canvas={yshift=-1.3pt}] (1)--(2);
    \draw[-] (0.4,0.2)--(0.6,0)--(0.4,-0.2);
    \end{tikzpicture}} & \raisebox{-0.5\height}{$\textrm{N/A}$} & \raisebox{-0.5\height}{$\textrm{N/A}$} \\ \hline
\end{tabular}}
\caption[Electric and Magnetic Quivers for S\l odowy Intersections between $S_2$ Special Piece in ${\mathfrak{e}_7}$ nilcone, its trivial dual and $\{(A_5)'',\; D_4 \}$ orbits]{Electric and Magnetic Quivers for S\l odowy Intersections between $S_2$ Special Piece in ${\mathfrak{e}_7}$ nilcone, its trivial dual and $\{(A_5)'',\; D_4 \}$ orbits. The $\{D_5(a_1),\; D_4+A_1 \}$ special piece is $d_{LS}$ dual to the special orbit $A_4$ with trivial special piece. The quivers are presented with explicit framing.}
\label{tab:E7S2S1B}
\end{table}

In Table \ref{tab:E7S2S1A} the global symmetry of the Coulomb branch of the magnetic quivers is $F_4$, inherited from the S\l odowy slice to $(3A_1)''$. In Table \ref{tab:E7S2S1B} the global symmetry of the Coulomb branch of the magnetic quivers is $C_3$, inherited from the S\l odowy slice to $D_4$. These magnetic quivers construct the minimal and next to minimal orbits of $F_4$ and $C_3$, respectively. The transitions between these minimal and next to minimal orbits are $c_3$ and $c_2$ respectively. The electric quivers correspond to the Kleinian $E_6$ (or $F_4$) and $A_5$ (or $B_3$) singularities, respectively.

\section{Special Pieces in the $\mathfrak{e}_8$ Nilcone}
\label{sec:e8special}

The $\mathfrak{e}_{8}$ nilcone, which has $\rm{dim}_{\mathbb C} $ $240$, contains 70 orbits and 16 special pieces. One special piece has an $S_5$ Lusztig canonical quotient, three special pieces have $S_3$ while the other 12 have $S_2$. The $S_5$ special piece is self-dual. The special orbits inside two of the $S_3$ special pieces, $E_8\left(b_5\right)$ and $D_4\left(a_1\right)$, are permuted under $d_{LS}$ and so these two $S_3$ special pieces are images of each other under $d_{SD}$. The 12 $S_2$ special pieces include four pairs that are images of each other under $d_{SD}$.  The remaining pieces, associated with the special orbits $D_4(a_1)+A_1$, $A_2+A_1$, $A_4+2A_1$, $D_5(a_1)$ and $E_7(a_3)$ do not have images under $d_{SD}$. A full summary of the $d_{SD}$ data for the $\mathfrak{e}_{8}$ nilcone is given in Table \ref{fig:AC_Map_Data_E7E8}.

\subsection{$S_5$ Special Piece}

Consider five M5-branes on a $D_5$ singularity with boundary conditions $\rho_{L}=(9,1)$, $\rho_{R}=(1^{10})$ \cite{Hanany:2022itc}. At the origin of the tensor branch the corresponding magnetic quiver is \eqref{quiv:D1--C5_Loop}. The presence of the loop indicates an $S_5$ action on the Coulomb branch, which turns out to construct a S\l odowy intersection in $\bar{\mathcal{O}}_\mathcal{N}^{\mathfrak{e}_8}$.

\begin{equation}
    \raisebox{-0.5\height}{\begin{tikzpicture}
    \node[gauger, label=below:$D_1$] (0) at (0,0){};
    \node[gaugeb, label=below:$C_1$] (1) at (1,0){};
    \node[gauger, label=below:$D_2$] (2) at (2,0){};
    \node[gaugeb, label=below:$C_2$] (3) at (3,0){};
    \node[gauger, label=below:$D_3$] (4) at (4,0){};
    \node[gaugeb, label=below:$C_3$] (5) at (5,0){};
    \node[gauger, label=below:$D_4$] (6) at (6,0){};
    \node[gaugeb, label=below:$C_4$] (7) at (7,0){};
    \node[gauger, label=below:$D_5$] (8) at (8,0){};
    \node[gaugeb, label=below:$C_5$] (9) at (9,0){};
    \draw[-] (0)--(1)--(2)--(3)--(4)--(5)--(6)--(7)--(8)--(9);
    \draw (9) to[out=-45,in=45,looseness=8](9);
    \end{tikzpicture}}
\label{quiv:D1--C5_Loop}
\end{equation}

The orthosymplectic quiver is indeed a magnetic quiver for the $60\;\rm{dim}_{\mathbb C}$ S\l odowy intersection in ${\mathfrak{e}_8}$ between the $208\;\rm{dim}_{\mathbb C}$ orbit $E_8(a_7)$ and the $148\;\rm{dim}_{\mathbb C}$ orbit $A_3$. Under the Loop Lace map, this remarkable orthosymplectic quiver generates a family of magnetic and electric quivers for S\l odowy intersections involving the $S_5$ special piece, of which $E_8(a_7)$ is the parent. These form the Loop Lace map shown in Table \ref{tab:D1-...-C5-Loop}. The global Coulomb branch symmetry of the magnetic quivers is $B_5$, inherited from the S\l odowy slice to the $A_3$ orbit. The sum of the Higgs and Coulomb branch dimensions for each quiver pair in the family is $\rm{dim}_{\mathbb C}= |\bar{\mathcal{O}}_{E_7(a_1)}^{\mathfrak{e}_8}| - |\bar{\mathcal{O}}_{A_3}^{\mathfrak{e}_8}|= 80$.

There are many other $\mathfrak{e}_8$ intersections besides those from $A_3$ (and its dual) to the $S_5$ special piece, and for some of these the quivers are known. Two such examples, which have manifest $S_5$ symmetry, are given by the magnetic quivers in \ref{quiv:1-2-5-loop} and \ref{quiv:2-5-loop}.
\begin{subequations}
\begin{align}
\raisebox{-0.5\height}{\begin{tikzpicture}
        \node[gauge, label=below:$1$] (0) at (-1,0){};
        \node[gauge, label=below:$2$] (1) at (0,0){};
        \node[gauge, label=below:$5$] (2) at (1,0){};
        \draw[-] (0)--(1)--(2);
        \draw (2) to[out=-45,in=45,looseness=8](2);
    \end{tikzpicture}}\label{quiv:1-2-5-loop}\\
\raisebox{-0.5\height}{\begin{tikzpicture}
        \node[gauge, label=below:$2$] (1) at (0,0){};
        \node[gauge, label=below:$5$] (2) at (1,0){};
        \draw[-] (1)--(2);
        \draw (2) to[out=-45,in=45,looseness=8](2);
    \end{tikzpicture}}\label{quiv:2-5-loop}
\end{align}
\end{subequations}

These were identified in \cite{Bennett:2024loi} and are magnetic quivers for the $\mathfrak{e}_8$ intersections between the orbit $E_8(a_7)$ and the orbits $A_4+A_2$ and $A_4+A_2+A_1$, respectively. They form the basis for two further Loop Lace maps, shown in Tables \ref{tab:1-2-5-loop} and \ref{tab:2-5-loop}, respectively.

In Table \ref{tab:1-2-5-loop}, the global symmetry of the magnetic quiver Coulomb branches reduces to $A_1 \times A_1$, inherited from the S\l odowy slice to the $A_4+A_4$ orbit (this symmetry is enhanced to $A_1 \times A_1 \times A_1$ for the normalisations). The sum of the Higgs and Coulomb branch dimensions for each quiver pair is $\rm{dim}_{\mathbb C}= |\bar{\mathcal{O}}_{D_5+A_2}^{\mathfrak{e}_8}| - |\bar{\mathcal{O}}_{A_4+A_2}^{\mathfrak{e}_8}|= 20$.

In Table \ref{tab:2-5-loop}, the global symmetry of the magnetic quiver Coulomb branches reduces to $A_1$, inherited from the S\l odowy slice to the $A_4+A_4+A_1$ orbit (this symmetry is enhanced to $A_1 \times A_1$ for the normalisations). The sum of the Higgs and Coulomb branch dimensions for each quiver pair is $\rm{dim}_{\mathbb C}= |\bar{\mathcal{O}}_{A_6+A_1}^{\mathfrak{e}_8}| - |\bar{\mathcal{O}}_{A_4+A_2+A_1}^{\mathfrak{e}_8}|= 16$.

The Higgs branch symmetry for all these $S_5$ examples is the product of the $A_1$ symmetries on each hypermultiplet loop.

\begin{landscape}
\begin{table}[h!]
\centering
\begin{tabular}{|c|c|c|c|c|c|}
\hline
$\rho$ & $\left(\left|{\cal C}(Q_M)\right|, \left|{\cal H}(Q_E)\right|\right)$  & $(A(\cal O), C(\cal O))$ & $Q_{\text{M}}\left[\mathcal{S}^{\mathfrak{e}_8}_{\rho,\;A_3}\right]$ & $Q_{\text{E}}\left[\mathcal{S}^{\mathfrak{e}_8}_{E_7\left(a_1\right),\;\rho}\right]$ & $G_{\mathcal{C}}(Q_E)$\\
\hline
$E_{8}\left(a_7\right)$ & $\left(60,20\right)$ & $(S_5,1)$ & \raisebox{-0.5\height}
{\begin{tikzpicture}
        \node[gauger, label=below:$D_1$] (0) at (0,0){};
        \node[] (cdots) at (1,0){$\cdots$};
        \node[gaugeb, label=below:$C_4$] (1) at (2,0){};
        \node[gauger, label=below:$D_5$] (2) at (3,0){};
        \node[gaugeb, label=below:$C_5$] (3) at (4,0){};
        \draw[-] (0)--(cdots)--(1)--(2)--(3);
        \draw (3) to[out=-45,in=45,looseness=8](3);
    \end{tikzpicture}} 
& \raisebox{-0.5\height}{\begin{tikzpicture}
        \node[gauger, label=below:$D_1$] (0) at (0,0){};
        \node[] (cdots) at (1,0){$\cdots$};
        \node[gaugeb, label=below:$C_4$] (1) at (2,0){};
        \node[gauger, label=below:$D_5$] (2) at (3,0){};
        \node[gaugeb, label=right:$C_1$] (3) at (4,1){};
        \node[gaugeb, label=right:$C_1$] (4) at (4,0.5){};
        \node[gaugeb, label=right:$C_1$] (5) at (4,0){};
        \node[gaugeb, label=right:$C_1$] (6) at (4,-0.5){};
        \node[gaugeb, label=right:$C_1$] (7) at (4,-1){};
        \draw[-] (0)--(cdots)--(1)--(2)--(3) (2)--(4) (2)--(5) (2)--(6) (2)--(7);
        \draw[green] (3) circle (0.2cm);
        \draw[green] (4) circle (0.2cm);
        \draw[green] (5) circle (0.2cm);
        \draw[green] (6) circle (0.2cm);
        \draw[green] (7) circle (0.2cm);
    \end{tikzpicture}} & $S_5$ \\ \hline
     $E_{7}\left(a_5\right)$ & $\left(58,22\right)$ & $(S_3,S_2)$ & \raisebox{-0.5\height}{\begin{tikzpicture}
        \node[gauger, label=below:$D_1$] (0) at (0,0){};
        \node[] (cdots) at (1,0){$\cdots$};
        \node[gaugeb, label=below:$C_4$] (1) at (2,0){};
        \node[gauger, label=below:$D_5$] (2) at (3,0){};
        \node[gaugeb, label=below:$C_1$] (3) at (4,0){};
        \node[gaugeb, label=right:$C_3$] (4) at (3,1){};
        \draw[-] (0)--(cdots)--(1)--(2)--(4);
        \draw (4) to[out=135,in=45,looseness=8](4);
        \draw[transform canvas={yshift=1.3pt}] (2)--(3);
        \draw[transform canvas={yshift=-1.3pt}] (2)--(3);
        \draw[-] (3.4,0.2)--(3.6,0)--(3.4,-0.2);
        \draw[red] (3) circle (0.2cm);
        \draw[red] (4) circle (0.2cm);
    \end{tikzpicture}} & \raisebox{-0.5\height}{\begin{tikzpicture}
        \node[gauger, label=below:$D_1$] (0) at (0,0){};
        \node[] (cdots) at (1,0){$\cdots$};
        \node[gaugeb, label=below:$C_4$] (1) at (2,0){};
        \node[gauger, label=below:$D_5$] (2) at (3,0){};
        \node[gaugeb, label=right:$C_1$] (3) at (4,0.25){};
        \node[gaugeb, label=right:$C_1$] (4) at (4,0.75){};
        \node[gaugeb, label=right:$C_1$] (5) at (4,-0.25){};
        \node[gaugeb, label={[xshift=0.75cm, yshift=-0.4cm]$C_2$}] (6) at (4,-0.75){};
        \draw[-] (0)--(cdots)--(1)--(2)--(3) (2)--(4) (2)--(5) (2)--(6);
        \draw[green] (3) circle (0.2cm);
        \draw[green] (4) circle (0.2cm);
        \draw[green] (5) circle (0.2cm);
        \draw[green] (6) circle (0.2cm);
        \draw (6) to[out=45,in=-45,looseness=8](6);
    \end{tikzpicture}} & $\frac{S_5}{S_2}$ \\ \hline
     $D_{6}\left(a_2\right)$ & $\left(56,24\right)$ & $(S_2,S_2)$ & \raisebox{-0.5\height}{\begin{tikzpicture}
        \node[gauger, label=below:$D_1$] (0) at (0,0){};
        \node[] (cdots) at (1,0){$\cdots$};
        \node[gaugeb, label=below:$C_4$] (1) at (2,0){};
        \node[gauger, label=below:$D_5$] (2) at (3,0){};
        \node[gaugeb, label=below:$C_2$] (3) at (4,0){};
        \node[gaugeb, label=right:$C_1$] (4) at (3,1){};
        \draw[-] (0)--(cdots)--(1)--(2)--(4);
        \draw (3) to[out=-45,in=45,looseness=8](3);
        \draw[transform canvas={yshift=1.3pt}] (2)--(3);
        \draw[transform canvas={yshift=-1.3pt}] (2)--(3);
        \draw[-] (3.4,0.2)--(3.6,0)--(3.4,-0.2);
        \draw[red] (3) circle (0.2cm);
        \draw[red] (4) circle (0.2cm);
    \end{tikzpicture}} & \raisebox{-0.5\height}{\begin{tikzpicture}
        \node[gauger, label=below:$D_1$] (0) at (0,0){};
        \node[] (cdots) at (1,0){$\cdots$};
        \node[gaugeb, label=below:$C_4$] (1) at (2,0){};
        \node[gauger, label=below:$D_5$] (2) at (3,0){};
        \node[gaugeb, label=right:$C_1$] (3) at (4,0){};
        \node[gaugeb, label={[xshift=0.75cm, yshift=-0.4cm]$C_2$}] (4) at (4,-0.5){};
        \node[gaugeb, label={[xshift=0.75cm, yshift=-0.4cm]$C_2$}] (5) at (4,0.5){};
        \draw[-] (0)--(cdots)--(1)--(2)--(3) (2)--(4) (2)--(5);
        \draw[green] (3) circle (0.2cm);
        \draw[green] (4) circle (0.2cm);
        \draw[green] (5) circle (0.2cm);
        \draw (4) to[out=45,in=-45,looseness=8](4);
        \draw (5) to[out=45,in=-45,looseness=8](5);
    \end{tikzpicture}} & $\frac{S_5}{S_2\times S_2}$ \\ \hline
    $E_{6}\left(a_3\right)+A_1$& $\left(56,24\right)$ & $(S_2,S_3)$ & \raisebox{-0.5\height}{\begin{tikzpicture}
        \node[gauger, label=below:$D_1$] (0) at (0,0){};
        \node[] (cdots) at (1,0){$\cdots$};
        \node[gaugeb, label=below:$C_4$] (1) at (2,0){};
        \node[gauger, label=below:$D_5$] (2) at (3,0){};
        \node[gaugeb, label=below:$C_1$] (3) at (4,0){};
        \node[gaugeb, label=right:$C_2$] (4) at (3,1){};
        \draw[-] (0)--(cdots)--(1)--(2)--(4) (2)--(3);
        \draw (4) to[out=135,in=45,looseness=8](4);
        \draw[transform canvas={yshift=1.3pt}] (2)--(3);
        \draw[transform canvas={yshift=-1.3pt}] (2)--(3);
        \draw[-] (3.4,0.2)--(3.6,0)--(3.4,-0.2);
        \draw[red] (3) circle (0.2cm);
        \draw[red] (4) circle (0.2cm);
    \end{tikzpicture}} & \raisebox{-0.5\height}{\begin{tikzpicture}
        \node[gauger, label=below:$D_1$] (0) at (0,0){};
        \node[] (cdots) at (1,0){$\cdots$};
        \node[gaugeb, label=below:$C_4$] (1) at (2,0){};
        \node[gauger, label=below:$D_5$] (2) at (3,0){};
        \node[gaugeb, label={[xshift=0.75cm, yshift=-0.4cm]$C_3$}] (3) at (4,0){};
        \node[gaugeb, label=right:$C_1$] (4) at (4,-0.5){};
        \node[gaugeb, label=right:$C_1$] (5) at (4,0.5){};
        \draw[-] (0)--(cdots)--(1)--(2)--(3) (2)--(4) (2)--(5);
        \draw[green] (3) circle (0.2cm);
        \draw[green] (4) circle (0.2cm);
        \draw[green] (5) circle (0.2cm);
        \draw (3) to[out=45,in=-45,looseness=8](3);
    \end{tikzpicture}} & $\frac{S_5}{S_3}$\\ \hline
    $A_5+A_1$ & $\left(54,26\right)$ & $(1,S_2\times S_3)$ & \raisebox{-0.5\height}{\begin{tikzpicture}
        \node[gauger, label=below:$D_1$] (0) at (0,0){};
        \node[] (cdots) at (1,0){$\cdots$};
        \node[gaugeb, label=below:$C_4$] (1) at (2,0){};
        \node[gauger, label=below:$D_5$] (2) at (3,0){};
        \node[gaugeb, label=below:$C_1$] (3) at (4,0){};
        \node[gaugeb, label=right:$C_1$] (4) at (3,1){};
        \draw[-] (0)--(cdots)--(1)--(2)--(3);
        \draw[transform canvas={yshift=1.3pt}] (2)--(3);
        \draw[transform canvas={yshift=-1.3pt}] (2)--(3);
        \draw[transform canvas={xshift=1.3pt}] (2)--(4);
        \draw[transform canvas={xshift=-1.3pt}] (2)--(4);
        \draw[-] (3.4,0.2)--(3.6,0)--(3.4,-0.2);
        \draw[-] (2.8,0.4)--(3,0.6)--(3.2,0.4);
        \draw[red] (3) circle (0.2cm);
        \draw[red] (4) circle (0.2cm);
    \end{tikzpicture}} & \raisebox{-0.5\height}{\begin{tikzpicture}
        \node[gauger, label=below:$D_1$] (0) at (0,0){};
        \node[] (cdots) at (1,0){$\cdots$};
        \node[gaugeb, label=below:$C_4$] (1) at (2,0){};
        \node[gauger, label=below:$D_5$] (2) at (3,0){};
        \node[gaugeb, label={[xshift=0.75cm, yshift=-0.4cm]$C_2$}] (3) at (4,0.25){};
        \node[gaugeb, label={[xshift=0.75cm, yshift=-0.4cm]$C_3$}] (4) at (4,-0.25){};
        \draw[-] (0)--(cdots)--(1)--(2)--(3) (2)--(4);
        \draw[green] (3) circle (0.2cm);
        \draw[green] (4) circle (0.2cm);
        \draw (3) to[out=-45,in=45,looseness=8](3);
        \draw (4) to[out=-45,in=45,looseness=8](4);
    \end{tikzpicture}} & $\frac{S_5}{S_2\times S_3}$\\ \hline
    $D_{5}\left(a_1\right)+A_2$ & $\left(54,26\right)$ & $(1,S_4)$ & \raisebox{-0.5\height}{\begin{tikzpicture}
        \node[gauger, label=below:$D_1$] (0) at (0,0){};
        \node[] (cdots) at (1,0){$\cdots$};
        \node[gaugeb, label=below:$C_4$] (1) at (2,0){};
        \node[gauger, label=below:$D_5$] (2) at (3,0){};
        \node[gaugeb, label=below:$C_1$] (3) at (4,0){};
        \node[gaugeb, label=right:$C_1$] (4) at (3,1){};
        \draw[-] (0)--(cdots)--(1)--(2)--(4);
        \draw[transform canvas={yshift=2.25pt}] (2)--(3);
        \draw[transform canvas={yshift=-2.25pt}] (2)--(3);
        \draw[transform canvas={yshift=0.75pt}] (2)--(3);
        \draw[transform canvas={yshift=-0.75pt}] (2)--(3);
        \draw[-] (3.4,0.2)--(3.6,0)--(3.4,-0.2);
        \draw[red] (3) circle (0.2cm);
        \draw[red] (4) circle (0.2cm);
    \end{tikzpicture}} & \raisebox{-0.5\height}{\begin{tikzpicture}
        \node[gauger, label=below:$D_1$] (0) at (0,0){};
        \node[] (cdots) at (1,0){$\cdots$};
        \node[gaugeb, label=below:$C_4$] (1) at (2,0){};
        \node[gauger, label=below:$D_5$] (2) at (3,0){};
        \node[gaugeb, label=right:$C_1$] (3) at (4,0.25){};
        \node[gaugeb, label={[xshift=0.75cm, yshift=-0.4cm]$C_4$}] (4) at (4,-0.25){};
        \draw[-] (0)--(cdots)--(1)--(2)--(3) (2)--(4);
        \draw[green] (3) circle (0.2cm);
        \draw[green] (4) circle (0.2cm);
        \draw (4) to[out=-45,in=45,looseness=8](4);
    \end{tikzpicture}} & $\frac{S_5}{S_4}$\\ \hline
    $A_4+A_3$ & $\left(52,28\right)$ & $(1,S_5)$ & \raisebox{-0.5\height}{\begin{tikzpicture}
        \node[gauger, label=below:$D_1$] (0) at (0,0){};
        \node[] (cdots) at (1,0){$\cdots$};
        \node[gaugeb, label=below:$C_4$] (1) at (2,0){};
        \node[gauger, label=below:$D_5$] (2) at (3,0){};
        \node[gaugeb, label=below:$C_1$] (3) at (4,0){};
        \draw[-] (0)--(cdots)--(1)--(2)--(3);
        \draw[transform canvas={yshift=2pt}] (2)--(3);
        \draw[transform canvas={yshift=-2pt}] (2)--(3);
        \draw[transform canvas={yshift=1pt}] (2)--(3);
        \draw[transform canvas={yshift=-1pt}] (2)--(3);
        \draw[-] (3.4,0.2)--(3.6,0)--(3.4,-0.2);
    \end{tikzpicture}} & \raisebox{-0.5\height}{\begin{tikzpicture}
        \node[gauger, label=below:$D_1$] (0) at (0,0){};
        \node[] (cdots) at (1,0){$\cdots$};
        \node[gaugeb, label=below:$C_4$] (1) at (2,0){};
        \node[gauger, label=below:$D_5$] (2) at (3,0){};
        \node[gaugeb, label=below:$C_5$] (3) at (4,0){};
        \draw[-] (0)--(cdots)--(1)--(2)--(3);
        \draw (3) to[out=-45,in=45,looseness=8](3);
    \end{tikzpicture}} & $1$\\ \hline
\end{tabular}

\caption[Electric and Magnetic Quivers for S\l odowy Intersections between $S_5$ Special Piece in ${\mathfrak{e}_8}$ nilcone and $\{E_7(a_1), \;A_3 \}$ orbits]{Electric and Magnetic Quivers for S\l odowy Intersections between $S_5$ Special Piece in ${\mathfrak{e}_8}$ nilcone and $\{E_7(a_1), \;A_3 \}$ orbits. The special piece is self dual under $d_{SD}$. The orbits $\{E_7(a_5),\; D_6(a_2), \;E_6(a_3)+A_1,\; A_5+A_1,\; D_5(a_1)+A_2  \}$ are non-normal and their Coulomb branch construction gives normalisations. The magnetic quivers have a $B_5$ global symmetry.}

\label{tab:D1-...-C5-Loop}
\end{table}
\end{landscape}

\begin{table}[h!]
\centering
\begin{tabular}{|c|c|c|c|c|c|}
\hline
    $\rho$ & $\left(\left|{\cal C}(Q_M)\right|, \left|{\cal H}(Q_E)\right|\right)$  & $(A(\cal O), C(\cal O))$ & $Q_{\text{M}}\left[\mathcal{S}^{E_8}_{\rho,\;A_4+A_2}\right]$ & $Q_{\text{E}}\left[\mathcal{S}^{E_8}_{D_5+A_2,\;\rho}\right]$ & $G_{\mathcal{C}}(Q_E)$\\\hline
    $E_8\left(a_7\right)$ & $\left(14,6\right)$ & $(S_5,1)$ & \raisebox{-0.5\height}{\begin{tikzpicture}
        \node[gauge, label=below:$1$] (0) at (-1,0){};
        \node[gauge, label=below:$2$] (1) at (0,0){};
        \node[gauge, label=below:$5$] (2) at (1,0){};
        \draw[-] (0)--(1)--(2);
        \draw (2) to[out=-45,in=45,looseness=8](2);
    \end{tikzpicture}} & \raisebox{-0.5\height}{\begin{tikzpicture}
        \node[gauge, label=left:$1$] (0) at (-1,0){};
        \node[gauge, label=below:$2$] (1) at (0,0){};
        \node[gauge, label=right:$1$] (2) at (1,1){};
        \node[gauge, label=right:$1$] (3) at (1,0.5){};
        \node[gauge, label=right:$1$] (4) at (1,-0.5){};
        \node[gauge, label=right:$1$] (5) at (1,-1){};
        \node[gauge, label=right:$1$] (6) at (1,0){};
        \draw[-] (0)--(1)--(2) (3)--(1) (4)--(1) (5)--(1) (6)--(1);
        \draw[green] (5) circle (0.2cm);
        \draw[green] (6) circle (0.2cm);
        \draw[green] (2) circle (0.2cm);
        \draw[green] (3) circle (0.2cm);
        \draw[green] (4) circle (0.2cm);
    \end{tikzpicture}} & $S_5$ \\ \hline
    $E_7(a_5)$ & $\left(12,8\right)$ & $(S_3,S_2)$ & \raisebox{-0.5\height}{\begin{tikzpicture}
        \node[gauge, label=below:$1$] (0) at (0,0){};
        \node[gauge, label=below:$2$] (1) at (1,0){};
        \node[gauge, label=below:$3$] (2) at (2,0){};
        \node[gauge, label=above:$1$] (3) at (1,1){};
        \draw[transform canvas={xshift=1.3pt}] (1)--(3);
        \draw[transform canvas={xshift=-1.3pt}] (1)--(3);
        \draw[-] (0)--(1)--(2);
        \draw (2) to[out=-45,in=45,looseness=8](2);
        \draw[-] (0.8,0.4)--(1,0.6)--(1.2,0.4);
        \draw[red] (3) circle (0.2cm);
        \draw[red] (2) circle (0.2cm);
    \end{tikzpicture}} & \raisebox{-0.5\height}{\begin{tikzpicture}
        \node[gauge, label=below:$1$] (1) at (0,0){};
        \node[gauge, label=below:$2$] (2) at (1,0){};
        \node[gauge, label={[xshift=0.55cm, yshift=-0.35cm]$2$}] (3) at (2,0.75){};
        \draw (3) to[out=45,in=-45,looseness=8](3);
        \node[gauge, label=right:$1$] (4) at (2,0.25){};
        \node[gauge, label=right:$1$] (5) at (2,-0.25){};
        \node[gauge, label=right:$1$] (6) at (2,-0.75){};
        \draw[-] (1)--(2)--(3) (4)--(2) (5)--(2) (6)--(2);
        \draw[green] (3) circle (0.2cm);
        \draw[green] (4) circle (0.2cm);
        \draw[green] (5) circle (0.2cm);
        \draw[green] (6) circle (0.2cm);
    \end{tikzpicture}} & $\frac{S_5}{S_2}$ \\ \hline
    $D_6(a_2)$ & $\left(10,10\right)$ & $(1,S_2\times S_2)$ & \raisebox{-0.5\height}{\begin{tikzpicture}
        \node[gauge, label=below:$1$] (0) at (0,0){};
        \node[gauge, label=below:$2$] (1) at (1,0){};
        \node[gauge, label=below:$2$] (2) at (2,0){};
        \node[gauge, label=above:$1$] (3) at (1,1){};
        \draw[transform canvas={yshift=1.3pt}] (1)--(2);
        \draw[transform canvas={yshift=-1.3pt}] (1)--(2);
        \draw[-] (1.4,-0.2)--(1.6,0)--(1.4,0.2);
        \draw (2) to[out=45,in=-45,looseness=8](2);
        \draw[-] (0)--(1)--(3);
        \draw[red] (2) circle (0.2cm);
        \draw[red] (3) circle (0.2cm);
    \end{tikzpicture}} & \raisebox{-0.5\height}{\begin{tikzpicture}
        \node[gauge, label=below:$1$] (0) at (0,0){};
        \node[gauge, label=below:$2$] (1) at (1,0){};
        \node[gauge, label={[xshift=0.55cm, yshift=-0.35cm]$2$}] (2) at (2,-0.5){};
        \draw (2) to[out=45,in=-45,looseness=8](2);
        \node[gauge, label={[xshift=0.55cm, yshift=-0.35cm]$2$}] (3) at (2,0.5){};
        \draw (3) to[out=45,in=-45,looseness=8](3);
        \node[gauge, label=right:$1$] (4) at (2,0){};
        \draw[-] (0)--(1)--(2) (3)--(1) (1)--(4);
        \draw[green] (2) circle (0.2cm);
        \draw[green] (3) circle (0.2cm);
        \draw[green] (4) circle (0.2cm);
    \end{tikzpicture}} & $\frac{S_5}{S_2\times S_2}$\\ \hline
    $E_6(a_3)+A_1$ & $\left(10,10\right)$ & $(S_2,S_3)$ & \raisebox{-0.5\height}{\begin{tikzpicture}
        \node[gauge, label=below:$1$] (0) at (0,0){};
        \node[gauge, label=below:$2$] (1) at (1,0){};
        \node[gauge, label=below:$2$] (2) at (2,0){};
        \node[gauge, label=above:$1$] (3) at (1,1){};
        \draw[transform canvas={xshift=1.3pt}] (1)--(3);
        \draw[transform canvas={xshift=-1.3pt}] (1)--(3);
        \draw[-] (0.8,0.4)--(1,0.6)--(1.2,0.4);
        \draw[-] (0)--(1)--(2) (1)--(3);
        \draw (2) to[out=-45,in=45,looseness=8](2);
        \draw[red] (2) circle (0.2cm);
        \draw[red] (3) circle (0.2cm);
    \end{tikzpicture}} & \raisebox{-0.5\height}{\begin{tikzpicture}
        \node[gauge, label=below:$1$] (0) at (0,0){};
        \node[gauge, label=below:$2$] (1) at (1,0){};
        \node[gauge, label={[xshift=0.55cm, yshift=-0.35cm]$3$}] (2) at (2,-0.5){};
        \draw (2) to[out=45,in=-45,looseness=8](2);
        \node[gauge, label=right:$1$] (3) at (2,0.5){};
        \node[gauge, label=right:$1$] (4) at (2,0){};
        \draw[-] (0)--(1)--(2) (3)--(1) (1)--(4);
        \draw[green] (2) circle (0.2cm);
        \draw[green] (3) circle (0.2cm);
        \draw[green] (4) circle (0.2cm);
    \end{tikzpicture}} & $\frac{S_5}{S_3}$\\ \hline
    $A_5+A_1$ & $\left(8,12\right)$ & $(1,S_2\times S_3)$ & \raisebox{-0.5\height}{\begin{tikzpicture}
        \node[gauge, label=below:$1$] (0) at (0,0){};
        \node[gauge, label=below:$2$] (1) at (1,0){};
        \node[gauge, label=below:$1$] (2) at (2,0){};
        \node[gauge, label=above:$1$] (3) at (1,1){};
        \draw[transform canvas={xshift=1.3pt}] (3)--(1);
        \draw[transform canvas={xshift=-1.3pt}] (3)--(1);
        \draw[transform canvas={yshift=1.3pt}] (2)--(1);
        \draw[transform canvas={yshift=-1.3pt}] (2)--(1);
        \draw[-] (0.8,0.4)--(1,0.6)--(1.2,0.4);
        \draw[-] (1.4,-0.2)--(1.6,0)--(1.4,0.2);
        \draw[-] (0)--(1)--(3);
        \draw[red] (2) circle (0.2cm);
        \draw[red] (3) circle (0.2cm);
    \end{tikzpicture}} & \raisebox{-0.5\height}{\begin{tikzpicture}
        \node[gauge, label=below:$1$] (0) at (0,0){};
        \node[gauge, label=below:$2$] (1) at (1,0){};
        \node[gauge, label={[xshift=0.55cm, yshift=-0.35cm]$3$}] (2) at (2,-0.25){};
        \draw (2) to[out=45,in=-45,looseness=8](2);
        \node[gauge, label={[xshift=0.55cm, yshift=-0.35cm]$2$}] (3) at (2,0.25){};
        \draw (3) to[out=45,in=-45,looseness=8](3);
        \draw[-] (0)--(1)--(2) (3)--(1);
        \draw[green] (2) circle (0.2cm);
        \draw[green] (3) circle (0.2cm);
    \end{tikzpicture}} & $\frac{S_5}{S_2\times S_3}$ \\ \hline
    $D_5(a_1)+A_2$ & $\left(8,12\right)$  & $(1,S_4)$ & \raisebox{-0.5\height}{\begin{tikzpicture}
        \node[gauge, label=below:$1$] (0) at (0,0){};
        \node[gauge, label=below:$2$] (1) at (1,0){};
        \node[gauge, label=below:$1$] (2) at (2,0){};
        \node[gauge, label=above:$1$] (3) at (1,1){};
        \draw[transform canvas={yshift=1.5pt}] (1)--(2);
        \draw[transform canvas={yshift=-1.5pt}] (1)--(2);
        \draw[transform canvas={yshift=-0.5pt}] (1)--(2);
        \draw[transform canvas={yshift=0.5pt}] (1)--(2);
        \draw[-] (1.4,0.2)--(1.6,0)--(1.4,-0.2);
        \draw[-] (0)--(1)--(3);
        \draw[red] (2) circle (0.2cm);
        \draw[red] (3) circle (0.2cm);
    \end{tikzpicture}} & \raisebox{-0.5\height}{\begin{tikzpicture}
        \node[gauge, label=below:$1$] (0) at (0,0){};
        \node[gauge, label=below:$2$] (1) at (1,0){};
        \node[gauge, label={[xshift=0.55cm, yshift=-0.35cm]$4$}] (2) at (2,-0.25){};
        \draw (2) to[out=45,in=-45,looseness=8](2);
        \node[gauge, label=right:$1$] (3) at (2,0.25){};
        \draw[-] (0)--(1)--(2) (3)--(1);
        \draw[green] (2) circle (0.2cm);
        \draw[green] (3) circle (0.2cm);
    \end{tikzpicture}} & $\frac{S_5}{S_4}$\\ \hline
    $A_4+A_3$ & $\left(6,14\right)$ & $(1,S_5)$ & \raisebox{-0.5\height}{\begin{tikzpicture}
        \node[gauge, label=below:$1$] (0) at (0,0){};
        \node[gauge, label=below:$2$] (1) at (1,0){};
        \node[gauge, label=below:$1$] (2) at (2,0){};
        \draw[transform canvas={yshift=2pt}] (1)--(2);
        \draw[transform canvas={yshift=-2pt}] (1)--(2);
        \draw[transform canvas={yshift=-1pt}] (1)--(2);
        \draw[transform canvas={yshift=1pt}] (1)--(2);
        \draw[-] (1.4,-0.2)--(1.6,0)--(1.4,0.2);
        \draw[-] (0)--(1)--(2);
    \end{tikzpicture}} & \raisebox{-0.5\height}{\begin{tikzpicture}
        \node[gauge, label=below:$1$] (0) at (-1,0){};
        \node[gauge, label=below:$2$] (1) at (0,0){};
        \node[gauge, label=below:$5$] (2) at (1,0){};
        \draw[-] (0)--(1)--(2);
        \draw (2) to[out=-45,in=45,looseness=8](2);
    \end{tikzpicture}} & $1$ \\ \hline
\end{tabular}
\caption[Electric and Magnetic Quivers for S\l odowy Intersections between $S_5$ Special Piece in ${\mathfrak{e}_8}$ nilcone and $\{D_5+A_2, \;A_4+A_2 \}$ orbits]{Electric and Magnetic Quivers for S\l odowy Intersections between $S_5$ Special Piece in ${\mathfrak{e}_8}$ nilcone and $\{D_5+A_2, \;A_4+A_2 \}$ orbits. The special piece is self dual under $d_{SD}$. The orbits $\{E_7(a_5),\; D_6(a_2), \;E_6(a_3)+A_1,\; A_5+A_1,\; D_5(a_1)+A_2  \}$ are non-normal and their Coulomb branch constructions give normalisations. The magnetic quivers have an $A_1 \times A_1 $ global symmetry, enhanced to $A_1 \times A_1 \times A_1 $ by the normalisations.}

\label{tab:1-2-5-loop}
\end{table}

\begin{table}[h!]
\centering
\begin{tabular}{|c|c|c|c|c|c|}
\hline
    $\rho$ & $\left(\left|{\cal C}(Q_M)\right|, \left|{\cal H}(Q_E)\right|\right)$  & $(A(\cal O), C(\cal O))$ & $Q_{\text{M}}\left[\mathcal{S}^{E_8}_{\rho,\;A_4+A_2+A_1}\right]$ & $Q_{\text{E}}\left[\mathcal{S}^{E_8}_{A_6+A_1,\;\rho}\right]$ & $G_{\mathcal{C}}(Q_E)$\\
    \hline
    $E_8\left(a_7\right)$ & $\left(12,4\right)$ & $(S_5,1)$ & \raisebox{-0.5\height}{\begin{tikzpicture}
        \node[gauge, label=below:$2$] (0) at (0,0){};
        \node[gauge, label=below:$5$] (1) at (1,0){};
        \draw[-] (0)--(1)--(2);
        \draw (2) to[out=-45,in=45,looseness=8](2);
    \end{tikzpicture}} & \raisebox{-0.5\height}{\begin{tikzpicture}
        \node[gauge, label=below:$2$] (1) at (0,0){};
        \node[gauge, label=right:$1$] (2) at (1,1){};
        \node[gauge, label=right:$1$] (3) at (1,0.5){};
        \node[gauge, label=right:$1$] (4) at (1,-0.5){};
        \node[gauge, label=right:$1$] (5) at (1,-1){};
        \node[gauge, label=right:$1$] (6) at (1,0){};
        \draw[-] (1)--(2) (3)--(1) (4)--(1) (5)--(1) (6)--(1);
        \draw[green] (5) circle (0.2cm);
        \draw[green] (6) circle (0.2cm);
        \draw[green] (2) circle (0.2cm);
        \draw[green] (3) circle (0.2cm);
        \draw[green] (4) circle (0.2cm);
    \end{tikzpicture}} & $S_5$\\ \hline
    $E_7(a_1)$ & $\left(10,6\right)$ & $(S_3,S_2)$ & \raisebox{-0.5\height}{\begin{tikzpicture}
        \node[gauge, label=above:$1$] (0) at (0,1){};
        \node[gauge, label=below:$2$] (1) at (0,0){};
        \node[gauge, label=below:$3$] (2) at (1,0){};
        \draw[transform canvas={xshift=1.3pt}] (0)--(1);
        \draw[transform canvas={xshift=-1.3pt}] (0)--(1);
        \draw[-] (-0.2,0.4)--(0,0.6)--(0.2,0.4) (1)--(2);
        \draw (2) to[out=-45,in=45,looseness=8](2);
        \draw[red] (0) circle (0.2cm);
        \draw[red] (2) circle (0.2cm);
    \end{tikzpicture}} & \raisebox{-0.5\height}{\begin{tikzpicture}
        \node[gauge, label=below:$2$] (2) at (1,0){};
        \node[gauge, label={[xshift=0.55cm, yshift=-0.35cm]$2$}] (3) at (2,0.75){};
        \draw (3) to[out=45,in=-45,looseness=8](3);
        \node[gauge, label=right:$1$] (4) at (2,0.25){};
        \node[gauge, label=right:$1$] (5) at (2,-0.25){};
        \node[gauge, label=right:$1$] (6) at (2,-0.75){};
        \draw[-] (2)--(3) (4)--(2) (5)--(2) (6)--(2);
        \draw[green] (3) circle (0.2cm);
        \draw[green] (4) circle (0.2cm);
        \draw[green] (5) circle (0.2cm);
        \draw[green] (6) circle (0.2cm);
    \end{tikzpicture}} & $\frac{S_5}{S_2}$\\ \hline
    $D_6(a_2)$ & $\left(8,8\right)$ & $(1,S_2\times S_2)$ & \raisebox{-0.5\height}{\begin{tikzpicture}
        \node[gauge, label=above:$1$] (0) at (0,1){};
        \node[gauge, label=below:$2$] (1) at (0,0){};
        \node[gauge, label=below:$2$] (2) at (1,0){};
        \draw[transform canvas={yshift=1.3pt}] (1)--(2);
        \draw[transform canvas={yshift=-1.3pt}] (1)--(2);
        \draw[-] (0.4,-0.2)--(0.6,0)--(0.4,0.2) (0)--(1);
        \draw (2) to[out=45,in=-45,looseness=8](2);
        \draw[red] (0) circle (0.2cm);
        \draw[red] (2) circle (0.2cm);
    \end{tikzpicture}} & \raisebox{-0.5\height}{\begin{tikzpicture}
        \node[gauge, label=below:$2$] (1) at (1,0){};
        \node[gauge, label={[xshift=0.55cm, yshift=-0.35cm]$2$}] (2) at (2,-0.5){};
        \draw (2) to[out=45,in=-45,looseness=8](2);
        \node[gauge, label={[xshift=0.55cm, yshift=-0.35cm]$2$}] (3) at (2,0.5){};
        \draw (3) to[out=45,in=-45,looseness=8](3);
        \node[gauge, label=right:$1$] (4) at (2,0){};
        \draw[-] (1)--(2) (3)--(1) (1)--(4);
        \draw[green] (2) circle (0.2cm);
        \draw[green] (3) circle (0.2cm);
        \draw[green] (4) circle (0.2cm);
    \end{tikzpicture}} & $\frac{S_5}{S_2\times S_2}$\\ \hline
   $E_6(a_3)+A_1$ & $\left(8,8\right)$ & $(S_2,S_3)$ & \raisebox{-0.5\height}{\begin{tikzpicture}
        \node[gauge, label=above:$1$] (0) at (0,1){};
        \node[gauge, label=below:$2$] (1) at (0,0){};
        \node[gauge, label=below:$2$] (2) at (1,0){};
        \draw[transform canvas={xshift=1.3pt}] (0)--(1);
        \draw[transform canvas={xshift=-1.3pt}] (0)--(1);
        \draw[-] (-0.2,0.4)--(0,0.6)--(0.2,0.4) (0)--(1)--(2);
        \draw (2) to[out=-45,in=45,looseness=8](2);
        \draw[red] (0) circle (0.2cm);
        \draw[red] (2) circle (0.2cm);
    \end{tikzpicture}} & \raisebox{-0.5\height}{\begin{tikzpicture}
        \node[gauge, label=below:$2$] (1) at (1,0){};
        \node[gauge, label={[xshift=0.55cm, yshift=-0.35cm]$3$}] (2) at (2,-0.5){};
        \draw (2) to[out=45,in=-45,looseness=8](2);
        \node[gauge, label=right:$1$] (3) at (2,0.5){};
        \node[gauge, label=right:$1$] (4) at (2,0){};
        \draw[-] (1)--(2) (3)--(1) (1)--(4);
        \draw[green] (2) circle (0.2cm);
        \draw[green] (3) circle (0.2cm);
        \draw[green] (4) circle (0.2cm);
    \end{tikzpicture}} & $\frac{S_5}{S_3}$\\ \hline
    $A_5+A_1$ & $\left(6,10\right)$ & $(1,S_2\times S_3)$ & \raisebox{-0.5\height}{\begin{tikzpicture}
        \node[gauge, label=above:$1$] (0) at (0,1){};
        \node[gauge, label=below:$2$] (1) at (0,0){};
        \node[gauge, label=below:$1$] (2) at (1,0){};
        \draw[transform canvas={xshift=1.3pt}] (0)--(1);
        \draw[transform canvas={xshift=-1.3pt}] (0)--(1);
        \draw[transform canvas={yshift=1.3pt}] (2)--(1);
        \draw[transform canvas={yshift=-1.3pt}] (2)--(1);
        \draw[-] (-0.2,0.4)--(0,0.6)--(0.2,0.4);
        \draw[-] (0.4,-0.2)--(0.6,0)--(0.4,0.2) (0)--(1);
        \draw[red] (0) circle (0.2cm);
        \draw[red] (2) circle (0.2cm);
    \end{tikzpicture}} & \raisebox{-0.5\height}{\begin{tikzpicture}
        \node[gauge, label=below:$2$] (1) at (1,0){};
        \node[gauge, label={[xshift=0.55cm, yshift=-0.35cm]$3$}] (2) at (2,-0.25){};
        \draw (2) to[out=45,in=-45,looseness=8](2);
        \node[gauge, label={[xshift=0.55cm, yshift=-0.35cm]$2$}] (3) at (2,0.25){};
        \draw (3) to[out=45,in=-45,looseness=8](3);
        \draw[-] (1)--(2) (3)--(1);
        \draw[green] (2) circle (0.2cm);
        \draw[green] (3) circle (0.2cm);
    \end{tikzpicture}} & $\frac{S_5}{S_2\times S_3}$\\ \hline
    $D_5(a_1)+A_2$ & $\left(6,10\right)$ & $(1,S_4)$ & \raisebox{-0.5\height}{\begin{tikzpicture}
        \node[gauge, label=above:$1$] (0) at (0,1){};
        \node[gauge, label=below:$2$] (1) at (0,0){};
        \node[gauge, label=below:$1$] (2) at (1,0){};
        \draw[transform canvas={xshift=1.5pt}] (0)--(1);
        \draw[transform canvas={xshift=-1.5pt}] (0)--(1);
        \draw[transform canvas={xshift=-0.5pt}] (0)--(1);
        \draw[transform canvas={xshift=0.5pt}] (0)--(1);
        \draw[-] (-0.2,0.4)--(0,0.6)--(0.2,0.4) (1)--(2);
        \draw[red] (0) circle (0.2cm);
        \draw[red] (2) circle (0.2cm);
    \end{tikzpicture}} & \raisebox{-0.5\height}{\begin{tikzpicture}
        \node[gauge, label=below:$2$] (1) at (1,0){};
        \node[gauge, label={[xshift=0.55cm, yshift=-0.35cm]$4$}] (2) at (2,-0.25){};
        \draw (2) to[out=45,in=-45,looseness=8](2);
        \node[gauge, label=right:$1$] (3) at (2,0.25){};
        \draw[-] (1)--(2) (3)--(1);
        \draw[green] (2) circle (0.2cm);
        \draw[green] (3) circle (0.2cm);
    \end{tikzpicture}} & $\frac{S_5}{S_4}$\\ \hline
    $A_4+A_3$ & $\left(4,12\right)$ & $(1,S_5)$ & \raisebox{-0.5\height}{\begin{tikzpicture}
        \node[gauge, label=below:$2$] (0) at (0,0){};
        \node[gauge, label=below:$1$] (1) at (1,0){};
        \draw[transform canvas={yshift=2pt}] (0)--(1);
        \draw[transform canvas={yshift=-2pt}] (0)--(1);
        \draw[transform canvas={yshift=-1pt}] (0)--(1);
        \draw[transform canvas={yshift=1pt}] (0)--(1);
        \draw[-] (0.4,-0.2)--(0.6,0)--(0.4,0.2) (0)--(1);
    \end{tikzpicture}} & \raisebox{-0.5\height}{\begin{tikzpicture}
        \node[gauge, label=below:$2$] (1) at (0,0){};
        \node[gauge, label=below:$5$] (2) at (1,0){};
        \draw[-] (1)--(2);
        \draw (2) to[out=-45,in=45,looseness=8](2);
    \end{tikzpicture}} & $1$\\ \hline
\end{tabular}

\caption[Electric and Magnetic Quivers for S\l odowy Intersections between $S_5$ Special Piece in ${\mathfrak{e}_8}$ nilcone and $\{A_6+A_1, \;A_4+A_2+A_1 \}$ orbits]{Electric and Magnetic Quivers for S\l odowy Intersections between $S_5$ Special Piece in ${\mathfrak{e}_8}$ nilcone and $\{A_6+A_1, \;A_4+A_2+A_1 \}$ orbits. The special piece is self dual under $d_{SD}$. The orbits $\{E_7(a_5),\; D_6(a_2), \;E_6(a_3)+A_1,\; A_5+A_1,\; D_5(a_1)+A_2  \}$ are non-normal and their Coulomb branch constructions give normalisations. The magnetic quivers have an $A_1$ global symmetry, enhanced to $A_1 \times A_1$ by the normalisations.}

\label{tab:2-5-loop}
\end{table}
\FloatBarrier

\subsection{$S_3$ Special Pieces}

Two of the $S_3$ special pieces in the ${\mathfrak{e}_8}$ nilcone are dual to each other under $d_{SD}$. They are associated with the special orbits $E_8(b_5)$ and $D_4$. It can be seen from a Hasse diagram that the special piece below $E_8(b_5)$ lies within the S\l odowy slice from the orbit $E_6$, which has $G_2$ symmetry and provides an origin for the magnetic quivers for orbits of ${\mathfrak{g}_2}$, as encountered in Table \ref{Tab:G2_special_piece}. These observations guide the construction of a Loop Lace map involving these two $S_3$ pieces as in Table \ref{tab:E8_S3}.

\begin{table}[h!]
\centering
\scalebox{0.8}{\begin{tabular}{|c|c|c|c|c|c|}
\hline
$\rho$ & $\left(\left|{\cal C}(Q_M)\right|, \left|{\cal H}(Q_E)\right|\right)$ & $(A(\cal O), C(\cal O))$ &  $Q_{\text{M}}\left[\mathcal{S}^{\mathfrak{e}_8}_{\rho,\;E_6}\right]$ & $Q_{\text{E}}\left[\mathcal{S}^{\mathfrak{e}_8}_{D_4,\;d_{SD}\left(\rho\right)}\right]$ & $G_{\mathcal{C}}(Q_E)$  \\\hline
$E_8\left(b_5\right)$ & $\left(10,2\right)$ & $(S_3,1)$ & \raisebox{-0.5\height}
{\begin{tikzpicture}
    \node[gauge, label=below:$1$] (0) at (0,0){};
    \node[gauge, label=below:$2$] (1) at (1,0){};
    \node[gauge, label=below:$3$] (2) at (2,0){};
    \draw[-] (0)--(1)--(2);
    \draw (2) to[out=-45,in=45,looseness=8](2);
    \end{tikzpicture}} 
    & \raisebox{-0.5\height}
    {\begin{tikzpicture}
    \node[gauge, label=below:$1$] (0) at (0,0){};
    \node[gauge, label=below:$2$] (1) at (1,0){};
    \node[gauge, label=right:$1$] (2) at (2,0.5){};
    \node[gauge, label=right:$1$] (3) at (2,0){};
    \node[gauge, label=right:$1$] (4) at (2,-0.5){};
    \draw[-] (0)--(1)--(2) (3)--(1) (4)--(1);
    \draw[green] (2) circle (0.2cm);
    \draw[green] (3) circle (0.2cm);
    \draw[green] (4) circle (0.2cm);
    \end{tikzpicture}} 
    & $S_3$ \\ 
    \hline
    $E_7(a_2)$ & $\left(8,4\right)$ & $(1,S_2)$ & \raisebox{-0.5\height}
    {\begin{tikzpicture}
    \node[gauge, label=below:$2$] (1) at (1,0){};
    \node[gauge, label=below:$1$] (2) at (2,0){};
    \node[gauge, label=above:$1$] (3) at (0,0){};
    \node[gauge, label=above:$1$] (4) at (1,1){};
    \draw[transform canvas={yshift=1.3pt}] (1)--(2);
    \draw[transform canvas={yshift=-1.3pt}] (1)--(2);
    \draw[-] (1.4,-0.2)--(1.6,0)--(1.4,0.2);
    \draw[-] (1)--(3) (1)--(4);
    \draw[red] (2) circle (0.2cm);
    \draw[red] (4) circle (0.2cm);
    \end{tikzpicture}} &
    \raisebox{-0.5\height}
    {\begin{tikzpicture}
    \node[gauge, label=below:$1$] (0) at (0,0){};
    \node[gauge, label=below:$2$] (1) at (1,0){};
    \node[gauge, label={[xshift=0.55cm, yshift=-0.35cm]$2$}] (2) at (2,0.5){};
    \node[gauge, label=right:$1$] (3) at (2,-0.5){};
    \draw[-] (0)--(1)--(2) (3)--(1);
    \draw[green] (2) circle (0.2cm);
    \draw[green] (3) circle (0.2cm);
    \draw (2) to[out=45,in=-45,looseness=8](2);
    \end{tikzpicture}} & $\frac{S_3}{S_2}$\\ 
    \hline
    $E_6+A_1$ & $\left(6,6\right)$ & $(1,S_3)$ & \raisebox{-0.5\height}
    {\begin{tikzpicture}
    \node[gauge, label=below:$1$] (0) at (0,0){};
    \node[gauge, label=below:$2$] (1) at (1,0){};
    \node[gauge, label=below:$1$] (2) at (2,0){};
    \draw[transform canvas={yshift=1.3pt}] (1)--(2);
    \draw[transform canvas={yshift=-1.3pt}] (1)--(2);
    \draw[-] (0)--(1)--(2);
    \draw[-] (1.4,-0.2)--(1.6,0)--(1.4,0.2);
    \end{tikzpicture}} &
    \raisebox{-0.5\height}
    {\begin{tikzpicture}
    \node[gauge, label=left:$1$] (0) at (0,0){};
    \node[gauge, label=below:$2$] (1) at (1,0){};
    \node[gauge, label=below:$3$] (2) at (2,0){};
    \draw[-] (0)--(1)--(2);
    \draw (2) to[out=-45,in=45,looseness=8](2);
    \end{tikzpicture}} & $1$ \\
    \hline
\end{tabular}}
\caption[Electric and Magnetic Quivers for S\l odowy Intersections between $S_3$ Special Pieces in ${\mathfrak{e}_8}$ nilcone and $\{E_6, \;D_4 \}$ orbits]{Electric and Magnetic Quivers for S\l odowy Intersections between $S_3$ Special Pieces in ${\mathfrak{e}_8}$ nilcone and $\{E_6, \;D_4 \}$ orbits. The special piece $\{E_8(b_5), \; E_7(a_2),\; E_6+A_1 \}$ is dual to $\{D_4(a_1), \; A_3+A_1,\; 2A_2+A_1 \}$ under $d_{SD}$. The orbit $\{ E_7(a_2) \}$ is non-normal and its Coulomb branch construction gives a normalisation. The Coulomb branch global symmetries of the magnetic quivers are $G_2$ (or $B_3$ for the normalisation).}
\label{tab:E8_S3}
\end{table}

The third $S_3$ special piece in the ${\mathfrak{e}_8}$ nilcone, whose parent is the $D_4(a_1)+A_1$ orbit, maps under $d_{LS}$ to the special orbit $E_8(a_6)$, which has a trivial special piece. Accordingly, only a partial Loop Lace map can be constructed. The example presented in Table \ref{tab:E8_S3trivial} is based on the magnetic quiver with $S_3$ wreathing, \eqref{quiv:E8_dim30}, which constructs the $30\; \rm{dim}_{\mathbb C}$ intersection ${\cal S}_{D_4(a_1)+A_1,\; A_2+2A_1}^{\mathfrak{e}_8}$. By drawing on the isomorphism between $A_3$ and $D_3$ this quiver can be restated as an orthosymplectic quiver, as in \eqref{quiv:E8_dim30osp}.\footnote{These quivers were identified during discussions with Deshuo Liu.} Neither of these wreathed quivers is previously known. The Loop Lace map based on quiver \eqref{quiv:E8_dim30osp} follows straightforwardly as in Table \ref{tab:E8_S3trivialosp}.

\begin{equation}
\raisebox{-0.5\height}
{\begin{tikzpicture}
    \node[gauge, label=below:$1$] (1) at (0,0){};
    \node[gauge, label=below:$2$] (2) at (1,0){};
    \node[gauge, label=below:$3$] (3) at (2,0){};
    \node[gauge, label=below:$4$] (4) at (3,0){};
     \node[gauge, label=below:$2$] (5) at (4,0){};
    \node[] at (3.5,0){$[$};
    \node[] at (4.5,0){$]$};
    \node[] at (5.0,0){$\; \wr \; S_3$};
    \draw[-] (1)--(2)--(3)--(4)--(5);
    \end{tikzpicture}}
\label{quiv:E8_dim30}
\end{equation}

\begin{equation}
\raisebox{-0.5\height}
{\begin{tikzpicture}
        \node[gauger, label=below:$D_1$] (0) at (0,0){};
        \node[] (cdots) at (1,0){$\cdots$};
        \node[gaugeb, label=below:$C_2$] (1) at (2,0){};
        \node[gauger, label=below:$D_3$] (2) at (3,0){};
        \node[gaugeb, label=below:$C_1$] (3) at (4,0){};
        \node[gauger, label=below:$D_1$] (4) at (5,0){};
        \draw[-] (0)--(cdots)--(1)--(2)--(3)--(4);
        \node[] at (3.5,0){$[$};
    \node[] at (5.5,0){$]$};
    \node[] at (6.0,0){$\; \wr \; S_3$};
    \end{tikzpicture}} 
\label{quiv:E8_dim30osp}
\end{equation}

It is notable that the transitions $D_4(a_1)+A_1 \to A_3+2A_1 \to 2A_2+2A_1$ in the Hasse diagram between the orbits in this special piece are each $4\;\rm{dim}_{\mathbb C}$ corresponding, respectively, to the $c_2$ and the (non-normal) $m_2$ singularities \cite{Bennett:2024loi}, which form the singularity structure of $\mathrm{Sym}^{3}\left(\mathbb{C}^{4}\right)$.\footnote{Note that $m_2$ is called called $m'$ in \cite{Generic_singularities}.}

\begin{table}[h!]
\centering
\scalebox{0.75}{\begin{tabular}{|c|c|c|c|c|c|}
\hline
    $\rho$ & $\left(\left|{\cal C}(Q_M)\right|, \left|{\cal H}(Q_E)\right|\right)$  & $(A(\cal O), C(\cal O))$ & $Q_{\text{M}}\left[\mathcal{S}^{\mathfrak{e}_8}_{\rho,\;A_2+2A_1}\right]$ & $Q_{\text{E}}\left[\mathcal{S}^{\mathfrak{e}_8}_{E_8\left(b_4\right),\;d_{SD}(\rho)}\right]$ & $G_{\mathcal{C}}(Q_E)$ \\
    \hline
    $D_4\left(a_1\right)+A_1$ & $\left(30,6\right)$ & $(S_3,1)$ & \raisebox{-0.5\height}
    {\begin{tikzpicture}
    \node[gauge, label=below:$1$] (1) at (0,0){};
    \node[gauge, label=below:$2$] (2) at (1,0){};
    \node[gauge, label=below:$3$] (3) at (2,0){};
    \node[gauge, label=below:$4$] (4) at (3,0){};
     \node[gauge, label=below:$2$] (5) at (4,0){};
    \node[] at (3.5,0){$[$};
    \node[] at (4.5,0){$]$};
    \node[] at (5.0,0){$\; \wr \; S_3$};
    \draw[-] (1)--(2)--(3)--(4)--(5);
    \end{tikzpicture}} 
    & \raisebox{-0.5\height}
    {\begin{tikzpicture}
    \node[gauge, label=below:$1$] (1) at (0,0){};
    \node[gauge, label=below:$2$] (2) at (1,0){};
    \node[gauge, label=below:$3$] (3) at (2,0){};
    \node[gauge, label=below:$4$] (4) at (3,0){};
    \node[gauge, label=below:$2$] (5) at (4,0){};
    \node[gauge, label=left:$2$] (6) at (3.7,0.7){};
      \node[gauge, label=left:$2$] (7) at (3.7,-0.7){};
    \draw[-] (1)--(2)--(3)--(4)--(5) (7)--(4)--(6);
    \draw[green] (5) circle (0.2cm);
    \draw[green] (6) circle (0.2cm);
    \draw[green] (7) circle (0.2cm);
    \end{tikzpicture}} & $S_3$\\ 
    \hline
    $A_3+2A_1$ & $\left(26,\rm{-}\right)$ & $(1,S_2)$ & \raisebox{-0.5\height}{\begin{tikzpicture}
    \node[gauge, label=below:$1$] (1) at (0,0){};
    \node[gauge, label=below:$2$] (2) at (1,0){};
    \node[gauge, label=below:$3$] (3) at (2,0){};
    \node[gauge, label=below:$4$] (4) at (3,0){};
    \node[gauge, label=below:$2$] (5) at (4,0){};
    \node[gauge, label=left:$2$] (6) at (3,1){};
    \draw[-] (1)--(2)--(3)--(4) (4)--(6);
    \draw[transform canvas={yshift=1.3pt}] (4)--(5);
    \draw[transform canvas={yshift=-1.3pt}] (4)--(5);
    \draw[-] (3.4,0.2)--(3.6,0)--(3.4,-0.2);
    \draw[red] (5) circle (0.2cm);
    \draw[red] (6) circle (0.2cm);
    \end{tikzpicture}} & \raisebox{-0.5\height}{$\textrm{N/A}$} &\raisebox{-0.5\height}{$\textrm{N/A}$}\\
    \hline
    $2A_2+2A_1$ & $\left(22,\rm{-}\right)$ & $(1,S_3)$ & \raisebox{-0.5\height}{\begin{tikzpicture}
   \node[gauge, label=below:$1$] (1) at (0,0){};
    \node[gauge, label=below:$2$] (2) at (1,0){};
    \node[gauge, label=below:$3$] (3) at (2,0){};
    \node[gauge, label=below:$4$] (4) at (3,0){};
    \node[gauge, label=below:$2$] (5) at (4,0){};
    \draw[-] (1)--(2)--(3)--(4)--(5);
    \draw[transform canvas={yshift=1.3pt}] (4)--(5);
    \draw[transform canvas={yshift=-1.3pt}] (4)--(5);
    \draw[-] (3.4,0.2)--(3.6,0)--(3.4,-0.2);
    \end{tikzpicture}} & \raisebox{-0.5\height}{$\textrm{N/A}$} & \raisebox{-0.5\height}{$\textrm{N/A}$} \\ 
    \hline
\end{tabular}}

\caption[Electric and Magnetic Unitary Quivers for Intersections between $S_3$ Special Piece in ${\mathfrak{e}_8}$ nilcone, its trivial dual and $\{E_8\left(b_4\right), \; A_2+2A_1 \}$ orbits]{Electric and Magnetic Unitary Quivers for Intersections between $S_3$ Special Piece in ${\mathfrak{e}_8}$ nilcone, its trivial dual and $\{E_8\left(b_4\right), \; A_2+2A_1 \}$ orbits. The $\{D_4(a_1)+A_1,\; A_3+2A_1, \;2A_2 +2A_1 \}$ special piece is $d_{LS}$ dual to the special orbit $E_8(a_6) $ which has a trivial $S_1$ special piece. The orbit $A_3+2A_1$ is non-normal and its Coulomb branch construction gives a normalisation. The Coulomb branch global symmetries of the magnetic quivers are $B_3 \times A_1$ (or $B_3 \times A_1\times A_1$ for the normalisation).}
\label{tab:E8_S3trivial}
\end{table}

\begin{table}[h!]
\centering
\scalebox{0.65}{\begin{tabular}{|c|c|c|c|c|c|}
\hline
    $\rho$ & $\left(\left|{\cal C}(Q_M)\right|, \left|{\cal H}(Q_E)\right|\right)$  & $(A(\cal O), C(\cal O))$ & $Q_{\text{M}}\left[\mathcal{S}^{\mathfrak{e}_8}_{\rho,\;A_2+2A_1}\right]$ & $Q_{\text{E}}\left[\mathcal{S}^{\mathfrak{e}_8}_{E_8\left(b_4\right),\;d_{SD}(\rho)}\right]$ & $G_{\mathcal{C}}(Q_E)$ \\
    \hline
    $D_4\left(a_1\right)+A_1$ & $\left(30,6\right)$ & $(S_3,1)$ & \raisebox{-0.5\height}
  {\begin{tikzpicture}
        \node[gauger, label=below:$D_1$] (0) at (0,0){};
        \node[] (cdots) at (1,0){$\cdots$};
        \node[gaugeb, label=below:$C_2$] (1) at (2,0){};
        \node[gauger, label=below:$D_3$] (2) at (3,0){};
        \node[gaugeb, label=below:$C_1$] (3) at (4,0){};
        \node[gauger, label=below:$D_1$] (4) at (5,0){};
        \draw[-] (0)--(cdots)--(1)--(2)--(3)--(4);
        \node[] at (3.5,0){$[$};
    \node[] at (5.5,0){$]$};
    \node[] at (6.0,0){$\; \wr \; S_3$};
    \end{tikzpicture}} 
    & \raisebox{-0.5\height}
    {\begin{tikzpicture}
    \node[gauger, label=below:$D_1$] (0) at (0,0){};
    \node[] (cdots) at (1,0){$\cdots$};
        \node[gaugeb, label=below:$C_2$] (1) at (2,0){};
        \node[gauger, label=below:$D_3$] (2) at (3,0){};
        \node[gaugeb, label=below:$C_1$] (3) at (4,0){};
        \node[gauger, label=below:$D_1$] (4) at (5,0){};
        \node[gaugeb, label=left:$C_1$] (5) at (3.7,1){};
        \node[gauger, label=right:$D_1$] (6) at (4.7,1){};
         \node[gaugeb, label=left:$C_1$] (7) at (3.7,-1){};
        \node[gauger, label=right:$D_1$] (8) at (4.7,-1){};
    \draw[-](0)--(cdots)--(1)--(2)--(3)--(4) (2)--(5)--(6);
    \draw[-] (2)--(7)--(7);
    \draw[green] \convexpath{5,6} {0.2cm};
    \draw[green] \convexpath{3,4} {0.2cm};
        \draw[green] \convexpath{7,8} {0.2cm};
    \end{tikzpicture}} 
    & $S_3$\\ 
    \hline
    $A_3+2A_1$ & $\left(26,\rm{-}\right)$ & $(1,S_2)$ & \raisebox{-0.5\height}
    {\begin{tikzpicture}
    \node[gauger, label=below:$D_1$] (0) at (0,0){};
        \node[] (cdots) at (1,0){$\cdots$};
        \node[gaugeb, label=below:$C_2$] (1) at (2,0){};
        \node[gauger, label=below:$D_3$] (2) at (3,0){};
        \node[gaugeb, label=below:$C_1$] (3) at (4,0){};
        \node[gauger, label=below:$D_1$] (4) at (5,0){};
        \node[gaugeb, label=left:$C_1$] (5) at (3.7,1){};
        \node[gauger, label=right:$D_1$] (6) at (4.7,1){};
    \draw[-](0)--(cdots)--(1)--(2) (3)--(4) (2)--(5)--(6);
    \draw[transform canvas={yshift=1.3pt}] (2)--(3);
    \draw[transform canvas={yshift=-1.3pt}] (2)--(3);
    \draw[-] (3.4,0.2)--(3.6,0)--(3.4,-0.2);
    \draw[red] \convexpath{5,6} {0.2cm};
    \draw[red] \convexpath{3,4} {0.2cm};
    \end{tikzpicture}} 
    & \raisebox{-0.5\height}{$\textrm{N/A}$} & \raisebox{-0.5\height}{$\textrm{N/A}$}\\
    \hline
    $2A_2+2A_1$ & $\left(22,\rm{-}\right)$ & $(1,S_3)$ & \raisebox{-0.5\height}
    {\begin{tikzpicture}
    \node[gauger, label=below:$D_1$] (0) at (0,0){};
        \node[] (cdots) at (1,0){$\cdots$};
        \node[gaugeb, label=below:$C_2$] (1) at (2,0){};
        \node[gauger, label=below:$D_3$] (2) at (3,0){};
        \node[gaugeb, label=below:$C_1$] (3) at (4,0){};
        \node[gauger, label=below:$D_1$] (4) at (5,0){};
    \draw[-](0)--(cdots)--(1)--(2)--(3)--(4);
    \draw[transform canvas={yshift=1.3pt}] (2)--(3);
    \draw[transform canvas={yshift=-1.3pt}] (2)--(3);
    \draw[-] (3.4,0.2)--(3.6,0)--(3.4,-0.2);
    \end{tikzpicture}}
    & \raisebox{-0.5\height}{$\textrm{N/A}$} & \raisebox{-0.5\height}{$\textrm{N/A}$} \\ \hline
\end{tabular}}

\caption[Electric and Magnetic Orthosymplectic Quivers for Intersections between $S_3$ Special Piece in ${\mathfrak{e}_8}$ nilcone, its trivial dual and $\{E_8\left(b_4\right), \; A_2+2A_1 \}$ orbits]{Electric and Magnetic Orthosymplectic Quivers for Intersections between $S_3$ Special Piece in ${\mathfrak{e}_8}$ nilcone, its trivial dual and $\{E_8\left(b_4\right), \; A_2+2A_1 \}$ orbits. The $\{D_4(a_1)+A_1,\; A_3+2A_1, \;2A_2 +2A_1 \}$ special piece is $d_{LS}$ dual to the special orbit $E_8(a_6) $ which has a trivial $S_1$ special piece. The orbit $A_3+2A_1$ is non-normal and its Coulomb branch construction gives a normalisation. The Coulomb branch global symmetries of the magnetic quivers are $B_3 \times A_1$ (or $B_3 \times A_1\times A_1$ for the normalisation).}
\label{tab:E8_S3trivialosp}
\end{table}

\FloatBarrier

\subsection{$S_2$ Special Pieces}
Amongst the $S_2$ special pieces in the $\mathfrak{e}_8$ nilcone there are four pairs that are dual to each other under $d_{SD}$. Loop Lace maps involving these pairs are shown in Tables \ref{tab:2-2-loop} to \ref{tab:E8-B6n2m}.

\begin{table}[h!]
\centering
\begin{tabular}{|c|c|c|c|c|c|}
\hline
$\rho$ & $\left(\left|{\cal C}(Q_M)\right|, \left|{\cal H}(Q_E)\right|\right)$ & $(A(\cal O), C(\cal O))$ & $Q_{\text{M}}\left[\mathcal{S}^{E_8}_{\rho,\;A_4+A_2}\right]$ & $Q_{\text{E}}\left[\mathcal{S}^{E_8}_{D_5+A_2,\;d_{SD}(\rho)}\right]$ & $G_{\mathcal{C}}(Q_E)$\\
\hline
$E_6\left(a_3\right)$ & $\left(4,4\right)$ & $(S_2,1)$ & \raisebox{-0.5\height}{\begin{tikzpicture}
    \node[flavour, label=below:$2$] (1) at (0,0){};
    \node[gauge, label=below:$2$] (2) at (1,0){};
    \draw (2) to[out=-45,in=45,looseness=8](2);
    \draw[-] (1)--(2);
    \end{tikzpicture}} & \raisebox{-0.5\height}{\begin{tikzpicture}
    \node[flavour, label=below:$2$] (1) at (0,0){};
    \node[gauge, label=right:$1$] (2) at (1,0.5){};
    \node[gauge, label=right:$1$] (3) at (1,-0.5){};
    \draw[-] (1)--(2) (1)--(3);
    \draw[green] (2) circle (0.2cm);
    \draw[green] (3) circle (0.2cm);
    \end{tikzpicture}} & $S_2$ \\ \hline
    $A_5$ & $\left(2,6\right)$ & $(1,S_2)$ & \raisebox{-0.5\height}{\begin{tikzpicture}
    \node[flavour, label=below:$2$] (1) at (0,0){};
    \node[gauge, label=below:$1$] (2) at (1,0){};
    \draw[transform canvas={yshift=1.3pt}] (1)--(2);
    \draw[transform canvas={yshift=-1.3pt}] (1)--(2);
    \draw[-] (0.4,0.2)--(0.6,0)--(0.4,-0.2);
    \end{tikzpicture}} & \raisebox{-0.5\height}{\begin{tikzpicture}
    \node[flavour, label=below:$2$] (1) at (0,0){};
    \node[gauge, label=below:$2$] (2) at (1,0){};
    \draw (2) to[out=-45,in=45,looseness=8](2);
    \draw[-] (1)--(2);
    \end{tikzpicture}} & $1$\\ \hline
\end{tabular}
\caption[Electric and Magnetic Quivers for Intersections between $S_2$ Special Piece $E_6(a_3)$ in ${\mathfrak{e}_8}$ nilcone and $\{D_5+A_2, \; A_4+A_2 \}$ orbits]{Electric and Magnetic Quivers for Intersections between $S_2$ Special Piece $E_6(a_3)$ in ${\mathfrak{e}_8}$ nilcone and $\{D_5+A_2, \; A_4+A_2 \}$ orbits. The $\{E_6(a_3),\; A_5 \}$ special piece is $d_{SD}$ dual to the $\{D_6(a_1),\; D_5+A_1 \}$ special piece. The Coulomb branch global symmetry of the magnetic quivers is $A_1$. The Higgs branch global symmetry of the electric quivers is $D_2$. The quivers are presented with explicit framing.}
\label{tab:2-2-loop}
\end{table}

\begin{table}[h!]
\centering
\begin{tabular}{|c|c|c|c|c|c|}
\hline
$\rho$ & $\left(\left|{\cal C}(Q_M)\right|, \left|{\cal H}(Q_E)\right|\right)$ & $(A(\cal O), C(\cal O))$ & $Q_{\text{M}}\left[\mathcal{S}^{E_8}_{\rho,\;A_3+A_2}\right]$ & $Q_{\text{E}}\left[\mathcal{S}^{E_8}_{D_7(a_1),\;d_{SD}(\rho)}\right]$ & $G_{\mathcal{C}}(Q_E)$\\
\hline
$D_4\left(a_1\right)+A_2$ & $\left(6,2\right)$ & $(S_2,1)$ & \raisebox{-0.5\height}
{\begin{tikzpicture}
    \node[flavour, label=below:$1$] (0) at (2,0){};
    \node[gauge, label=below:$1$] (1) at (0,0){};
    \node[gauge, label=below:$2$] (2) at (1,0){};
    \draw (2) to[out=45,in=135,looseness=8](2);
    \draw[-] (1)--(2)--(0);
    \end{tikzpicture}} 
& \raisebox{-0.5\height}
{\begin{tikzpicture}
\node[flavour, label=below:$1$] (0) at (1.7,.7){};
    \node[gauge, label=below:$1$] (1) at (0.7,0.7){};
    \node[gauge, label=below:$1$] (2) at (0,0){};
    \node[gauge, label=below:$1$] (3) at (.7,-.7){};
\node[flavour, label=below:$1$] (4) at (1.7,-.7){};
    \draw[-] (0)--(1)--(2)--(3)--(4);
    \draw[green] (1) circle (0.2cm);
    \draw[green] (3) circle (0.2cm);
    \end{tikzpicture}} 
    & $S_2$ \\ 
\hline
    $A_3+A_2+A_1$ & $\left(4,4\right)$ & $(1,S_2)$ & \raisebox{-0.5\height}
    {\begin{tikzpicture}
    \node[gauge, label=below:$1$] (1) at (0,0){};
    \node[gauge, label=below:$1$] (2) at (1,0){};
    \node[flavour, label=below:$1$] (3) at (2,0){};
    \draw[-] (2)--(3);
    \draw[transform canvas={yshift=1.3pt}] (1)--(2);
    \draw[transform canvas={yshift=-1.3pt}] (1)--(2);
    \draw[-] (0.4,0.2)--(0.6,0)--(0.4,-0.2);
    \end{tikzpicture}} 
    & \raisebox{-0.5\height}
   {\begin{tikzpicture}
    \node[flavour, label=below:$1$] (0) at (2,0){};
    \node[gauge, label=below:$1$] (1) at (0,0){};
    \node[gauge, label=below:$2$] (2) at (1,0){};
    \draw (2) to[out=45,in=135,looseness=8](2);
    \draw[-] (1)--(2)--(0);
    \end{tikzpicture}} 
    & $1$\\ \hline
\end{tabular}
\caption[Electric and Magnetic Quivers for Intersections between $S_2$ Special Piece $D_4(a_1)+A_2$ in ${\mathfrak{e}_8}$ nilcone and $\{D_7(a_1), \; A_3+A_2 \}$ orbits]{Electric and Magnetic Quivers for Intersections between $S_2$ Special Piece $D_4(a_1)+A_2$ in ${\mathfrak{e}_8}$ nilcone and $\{D_7(a_1), \; A_3+A_2 \}$ orbits. The $\{D_4(a_1)+A_2,\; A_3+A_2+A_1 \}$ special piece is $d_{SD}$ dual to the $\{E_8(b_6),\; A_7\}$ special piece. The Coulomb branch global symmetry of the magnetic quivers is $B_2$. The Higgs branch global symmetry of the electric quivers is $A_1$. The orbit $D_7(a_1)$ is non-normal and the Higgs branch construction gives a normalisation. The quivers are presented with explicit framing.}
\label{tab:1-2-loop-1}
\end{table}

\begin{table}[h!]
\centering
\scalebox{1}{
\begin{tabular}{|c|c|c|c|c|c|}
\hline
$\rho$ & $\left(\left|{\cal C}(Q_M)\right|, \left|{\cal H}(Q_E)\right|\right)$ & $(A(\cal O), C(\cal O))$ & $Q_{\text{M}}\left[\mathcal{S}^{E_8}_{\rho,\;A_2+2A_1}\right]$ & $Q_{\text{E}}\left[\mathcal{S}^{E_8}_{E_8(b_4),\;d_{SD}(\rho)}\right]$ & $G_{\mathcal{C}}(Q_E)$\\
\hline
$2A_2$ & $\left(10,2\right)$ & $(S_2,1)$ & \raisebox{-0.5\height}
{\begin{tikzpicture}
    \node[gauge, label=left:$1$] (0) at (1,1){};
    \node[gauge, label=below:$1$] (1) at (0,0){};
    \node[gauge, label=below:$2$] (2) at (1,0){};
       \node[gauge, label=below:$2$] (3) at (2,0){};
    \draw (3) to[out=45,in=-45,looseness=8](3);
    \draw[-] (1)--(2)--(3) (0)--(2);
    \end{tikzpicture}} 
& \raisebox{-0.5\height}
{\begin{tikzpicture}
 \node[gauge, label=left:$1$] (0) at (-0.7,0.7){};
    \node[gauge, label=below:$1$] (1) at (-0.7,-0.7){};
    \node[gauge, label=below:$2$] (2) at (0,0){};
       \node[gauge, label=below:$1$] (3) at (0.7,0.7){};
        \node[gauge, label=below:$1$] (4) at (0.7,-0.7){};
    \draw[-] (1)--(2)--(3) (0)--(2) (2)--(4);
    \draw[green] (3) circle (0.2cm);
    \draw[green] (4) circle (0.2cm);
    \end{tikzpicture}} 
    & $S_2$ \\ 
\hline
    $A_2+3A_1$ & $\left(8,4\right)$ & $(1,S_2)$ & \raisebox{-0.5\height}
    {\begin{tikzpicture}
    \node[gauge, label=left:$1$] (0) at (1,1){};
    \node[gauge, label=below:$1$] (1) at (0,0){};
    \node[gauge, label=below:$2$] (2) at (1,0){};
       \node[gauge, label=below:$1$] (3) at (2,0){};
    \draw[-] (1)--(2) (0)--(2);
    \draw[transform canvas={yshift=1.3pt}] (2)--(3);
    \draw[transform canvas={yshift=-1.3pt}] (2)--(3);
    \draw[-] (1.4,0.2)--(1.6,0)--(1.4,-0.2);
    \end{tikzpicture}} 
    & \raisebox{-0.5\height}
  {\begin{tikzpicture}
    \node[gauge, label=left:$1$] (0) at (1,1){};
    \node[gauge, label=below:$1$] (1) at (0,0){};
    \node[gauge, label=below:$2$] (2) at (1,0){};
       \node[gauge, label=below:$2$] (3) at (2,0){};
    \draw (3) to[out=45,in=-45,looseness=8](3);
    \draw[-] (1)--(2)--(3) (0)--(2);
    \end{tikzpicture}} 
    & $1$\\ \hline
\end{tabular}
}
\caption[Electric and Magnetic Quivers for Intersections between $S_2$ Special Piece $2A_2$ in ${\mathfrak{e}_8}$ nilcone and $\{E_8(b_4), \; A_2+2A_1 \}$ orbits]{Electric and Magnetic Quivers for Intersections between $S_2$ Special Piece $2A_2$ in ${\mathfrak{e}_8}$ nilcone and $\{E_8(b_4), \; A_2+2A_1 \}$ orbits. The $\{2A_2,\;A_2+ 3A_1 \}$ special piece is $d_{SD}$ dual to the $\{E_8(a_5),\; D_7\}$ special piece. The Coulomb branch global symmetry of the magnetic quivers is $B_3$.}
\label{tab:E8-B3n2m}
\end{table}

\begin{table}[h!]
\centering
\scalebox{0.75}{
\begin{tabular}{|c|c|c|c|c|c|}
\hline
$\rho$ & $\left(\left|{\cal C}(Q_M)\right|, \left|{\cal H}(Q_E)\right|\right)$ & $(A(\cal O), C(\cal O))$ & $Q_{\text{M}}\left[\mathcal{S}^{E_8}_{\rho,\;2A_1}\right]$ & $Q_{\text{E}}\left[\mathcal{S}^{E_8}_{E_8(a_2),\;d_{SD}(\rho)}\right]$ & $G_{\mathcal{C}}(Q_E)$\\
\hline
$A_2$ & $\left(22,2\right)$ & $(S_2,1)$ & \raisebox{-0.5\height}
{\begin{tikzpicture}
    \node[gauge, label=left:$1$] (0) at (1,1){};
    \node[gauge, label=below:$1$] (1) at (0,0){};
    \node[gauge, label=below:$2$] (2) at (1,0){};
    \node[gauge, label=below:$2$] (3) at (2,0){};
    \node[gauge, label=below:$2$] (4) at (3,0){};
    \node[gauge, label=below:$2$] (5) at (4,0){};
    \node[gauge, label=below:$2$] (6) at (5,0){};
    \draw (6) to[out=45,in=-45,looseness=8](6);
    \draw[-] (1)--(2)--(3)--(4)--(5)--(6) (0)--(2);
    \end{tikzpicture}} 
& \raisebox{-0.5\height}
{\begin{tikzpicture}
 \node[gauge, label=left:$1$] (0) at (1,1){};
    \node[gauge, label=below:$1$] (1) at (0,0){};
    \node[gauge, label=below:$2$] (2) at (1,0){};
    \node[gauge, label=below:$2$] (3) at (2,0){};
    \node[gauge, label=below:$2$] (4) at (3,0){};
    \node[gauge, label=below:$2$] (5) at (4,0){};
    \node[gauge, label=below:$1$] (6) at (4.7,0.7){};
    \node[gauge, label=below:$1$] (7) at (4.7,-0.7){};
    \draw[-] (1)--(2)--(3)--(4)--(5)--(6) (0)--(2) (5)--(7);
    \draw[green] (6) circle (0.2cm);
    \draw[green] (7) circle (0.2cm);
    \end{tikzpicture}} 
    & $S_2$ \\ 
\hline
    $3A_1$ & $\left(20,4\right)$ & $(1,S_2)$ & \raisebox{-0.5\height}
    {\begin{tikzpicture}
    \node[gauge, label=left:$1$] (0) at (1,1){};
    \node[gauge, label=below:$1$] (1) at (0,0){};
    \node[gauge, label=below:$2$] (2) at (1,0){};
    \node[gauge, label=below:$2$] (3) at (2,0){};
    \node[gauge, label=below:$2$] (4) at (3,0){};
    \node[gauge, label=below:$2$] (5) at (4,0){};
    \node[gauge, label=below:$1$] (6) at (5,0){};
    \draw[-] (1)--(2)--(3)--(4)--(5) (0)--(2);
    \draw[transform canvas={yshift=1.3pt}] (5)--(6);
    \draw[transform canvas={yshift=-1.3pt}] (5)--(6);
    \draw[-] (4.4,0.2)--(4.6,0)--(4.4,-0.2);
    \end{tikzpicture}} 
    & \raisebox{-0.5\height}
   {\begin{tikzpicture}
    \node[gauge, label=left:$1$] (0) at (1,1){};
    \node[gauge, label=below:$1$] (1) at (0,0){};
    \node[gauge, label=below:$2$] (2) at (1,0){};
    \node[gauge, label=below:$2$] (3) at (2,0){};
    \node[gauge, label=below:$2$] (4) at (3,0){};
    \node[gauge, label=below:$2$] (5) at (4,0){};
    \node[gauge, label=below:$2$] (6) at (5,0){};
    \draw (6) to[out=45,in=-45,looseness=8](6);
    \draw[-] (1)--(2)--(3)--(4)--(5)--(6) (0)--(2);
    \end{tikzpicture}} 
    & $1$\\ \hline
\end{tabular}
}
\caption[Electric and Magnetic Quivers for Intersections between $S_2$ Special Piece $A_2$ in ${\mathfrak{e}_8}$ nilcone and $\{E_8(a_2), \; 2A_1 \}$ orbits]{Electric and Magnetic Quivers for Intersections between $S_2$ Special Piece $A_2$ in ${\mathfrak{e}_8}$ nilcone and $\{E_8(a_2), \; 2A_1 \}$ orbits. The $\{A_2,\; 3A_1 \}$ special piece is $d_{SD}$ dual to the $\{E_8(a_3),\; E_7\}$ special piece. The Coulomb branch global symmetry of the magnetic quivers is $B_6$.}
\label{tab:E8-B6n2m}
\end{table}

In Table \ref{tab:2-2-loop}, the magnetic quivers construct the $a_1 \equiv {\mathbb C}^2/{\mathbb Z}_2$ and $\rm{Sym}^2({\mathbb C}^2/{\mathbb Z}_2)$ singularities.\footnote{The $\rm{Sym}^2({\mathbb C}^2/{\mathbb Z}_2)$ singularity also corresponds to a Dihedral group gauging of a pair of $SU(2)$ fundamental fields transforming in the 2 dimensional irrep of $Dih_4$.} The top electric quiver constructs the $\mathfrak{d}_2$ maximal orbit. The lower electric quiver constructs the $\rm{dim}_{\mathbb C} = 6$ space of  $2-SU(2)$ instantons on $\mathbb{R}^4$.

In Table \ref{tab:1-2-loop-1}, the magnetic quivers construct closures of the $\mathfrak{b}_2$ minimal and next to minimal orbits. The top electric quiver constructs the Kleinian $A_3 \equiv D_3$ singularity, which is a normalisation of the $\mu$ transition, as discussed in \cite{Generic_singularities}. The lower electric quiver constructs the $\rm{dim}_{\mathbb C} = 4$ S\l odowy slice of $\mathfrak{b}_2 \equiv \mathfrak{c}_2$.

In Table \ref{tab:E8-B3n2m}, the magnetic quivers construct closures of the $\mathfrak{b}_3$ minimal and next to minimal orbits. The top electric quiver constructs the Kleinian $D_4$ singularity and the lower electric quiver constructs the $\rm{dim}_{\mathbb C} = 4$ S\l odowy slice of $\mathfrak{c}_3$.

In Table \ref{tab:E8-B6n2m}, the magnetic quivers construct closures of the $\mathfrak{b}_6$ minimal and next to minimal orbits. The top electric quiver constructs the Kleinian $D_7$ singularity.

The transitions between the magnetic quivers for these pieces correspond in all cases to $a_1$ Kraft Procesi transitions \cite{kraft1982geometry} between the minimal and next to minimal orbits of $\mathfrak{b}_k$. The transitions between the electric quivers correspond to the $A_1$ transitions between the sub-regular and sub-sub-regular orbits of the GNO dual algebra $\mathfrak{c}_k$. 

\FloatBarrier

The final four $S_2$ special pieces in $\mathfrak{e}_8$ are somewhat more enigmatic. Their $d_{LS}$ dual special orbits have trivial special pieces, so partial Loop Lace maps can be constructed at best. Nonetheless, all these $S_2$ special pieces fit into the general schema in Table \ref{tab:SpecialTrans}, in which the transition across a $S_2$ special piece of $\rm{dim}_{\mathbb H} = m$ is always the singularity $c_m$.

The $S_2$ special piece with the $D_5(a_1)$ parent contains a $c_3$ transition and lies above a $f_4$ transition in the Hasse diagram. These correspond to the lowest two transitions in the $\mathfrak{f}_4$ nilcone. So, magnetic quivers can be identified, and a partial Loop Lace map can be constructed, as in Table \ref{tab:E8-S2Wreath}. The quivers are those associated with the $S_2$ special piece of $\mathfrak{f}_4$ examined in section \ref{subsec:f4specialS2}.

\begin{table}[h!]
\centering
\scalebox{0.78}{
\begin{tabular}{|c|c|c|c|c|c|}
\hline
    $\rho$ & 
     $\left(\left|{\cal C}(Q_M)\right|, \left|{\cal H}(Q_E)\right|\right)$ 
     & $(A(\cal O), C(\cal O))$
    & $Q_{\text{M}}\left[\mathcal{S}^{\mathfrak{e}_8}_{\rho), \; D_4}\right]$ 
    & $Q_{\text{E}}\left[\mathcal{S}^{\mathfrak{e}_8}_{E_6, \; d_{SD}(\rho)}\right]$ 
    & $G_{\mathcal{C}}(Q_E)$ \\
    \hline
    $D_5(a_1)$ 
    & $\left(\textrm{22},2\right)$ 
    & $(S_2,1)$ &
    \raisebox{-0.5\height}
    {\begin{tikzpicture}
    \node[gauge, label=below:$1$] (1) at (0,0){};
    \node[gauge, label=below:$2$] (2) at (1,0){};
    \node[gauge, label=below:$3$] (3) at (2,0){};
    \node[gauge, label=below:$2$] (4) at (3,0){};
    \node[gauge, label=below:$1$] (5) at (4,0){};
    \draw[-](1)--(2)--(3)--(4)--(5);
    \node[] at (2.5,0){$[$};
    \node[] at (4.3,0){$]$};
    \node[] at (4.6,0){$\; \wr \; S_2$};
    \end{tikzpicture}}  
    & \raisebox{-0.5\height}{\begin{tikzpicture}
    \node[gauge, label=below:$1$] (1) at (0,0){};
    \node[gauge, label=below:$2$] (2) at (1,0){};
    \node[gauge, label=below:$3$] (3) at (2,0){};
    \node[gauge, label=below:$2$] (4) at (3,0){};
    \node[gauge, label=below:$1$] (5) at (4,0){};
    \node[gauge, label=right:$2$] (6) at (2,1){};
    \node[gauge, label=right:$1$] (7) at (2,2){};
    \draw[-] (1)--(2)--(3)--(4)--(5) (3)--(6)--(7);
    \draw[green] \convexpath{6,7} {0.2cm};
    \draw[green] \convexpath{4,5} {0.2cm};
    \end{tikzpicture}} & $S_2$\\ 
    \hline
    $D_4+A_1$ & 
    $\left(16,\textrm{N/A}\right)$ 
    \raisebox{-0.5\height}
    & $(1,S_2)$ & \raisebox{-0.5\height}{\begin{tikzpicture}
    \node[gauge, label=below:$1$] (1) at (0,0){};
    \node[gauge, label=below:$2$] (2) at (1,0){};
    \node[gauge, label=below:$3$] (3) at (2,0){};
    \node[gauge, label=below:$2$] (4) at (3,0){};
    \node[gauge, label=below:$1$] (5) at (4,0){};
    \draw[-] (1)--(2)--(3) (4)--(5);
    \draw[transform canvas={yshift=1.3pt}] (3)--(4);
    \draw[transform canvas={yshift=-1.3pt}] (3)--(4);
    \draw[-] (2.4,0.2)--(2.6,0)--(2.4,-0.2);
    \end{tikzpicture}} & \raisebox{-0.5\height}{$\textrm{N/A}$} & \textrm{N/A} \\ \hline
\end{tabular}}
\caption[Electric and Magnetic Quivers for Intersections between $S_2$ Special Piece $D_5(a_1)$ in ${\mathfrak{e}_8}$ nilcone and $\{E_6, \; D_4 \}$ orbits]{Electric and Magnetic Quivers for Intersections between $S_2$ Special Piece $D_5(a_1)$ in ${\mathfrak{e}_8}$ nilcone and $\{E_6, \; D_4 \}$ orbits. The $\{D_5(a_1),\; D_4+A_1 \}$ special piece is $d_{LS}$ dual to the special orbit $\{E_6(a_1)\}$ which has a trivial special piece. The Coulomb branch global symmetry of the magnetic quivers is $F_4$. The Higgs branch of $Q_E$ is the $E_6$ Kleinian singularity.}
\label{tab:E8-S2Wreath}
\end{table}

The $S_2$ special piece, $A_4+2A_1$ contains a $b_2 \equiv c_2$ transition, with an $a_2$ transition immediately below in the Hasse diagram. This $c_2$ transition is the $S_2$ wreathing of the $a_2$ transition. A partial Loop lace map building on a magnetic quiver for $a_2$ follows as in Table \ref{tab:E8-S2Wreatha2}. $Q_E$ has been verified by the localisation formula and the bouquet feature does not appear, being replaced by an $S_2$ wreathing of the Kleinian $A_2$ singularity. The reasons for this feature are unclear – perhaps related to the absence of a central node in the quiver or to some offsetting $S_2$ factor.

\begin{table}[h!]
\centering
\scalebox{0.8}{
\begin{tabular}{|c|c|c|c|c|c|}
\hline
    $\rho$ & 
     $\left(\left|{\cal C}(Q_M)\right|, \left|{\cal H}(Q_E)\right|\right)$ 
     & $(A(\cal O), C(\cal O))$
    & $Q_{\text{M}}\left[\mathcal{S}^{\mathfrak{e}_8}_{\rho, \; D_4(a_1)+A_2}\right]$ 
    & $Q_{\text{E}}\left[\mathcal{S}^{\mathfrak{e}_8}_{E_8(b_6), \; d_{SD}(\rho)}\right]$ 
    & $G_{\mathcal{C}}(Q_E)$ \\
    \hline
    $A_4+2A_1$ 
    & $\left(\textrm{8},4\right)$ 
    & $(S_2,1)$ &
    \raisebox{-0.5\height}
    {\begin{tikzpicture}
    \node[gauge, label=below:$1$] (1) at (0,0){};
    \node[gauge, label=right:$1$] (2) at (1,1){};
    \node[gauge, label=right:$1$] (3) at (1,-1){};
    \draw[-](2)--(1)--(3)--(2);
    \node[] at (-0.5,0){$[$};
    \node[] at (1.5,0){$]$};
    \node[] at (-1,0){$S_2 \; \wr $};
    \end{tikzpicture}}  
    & 
    \raisebox{-0.5\height}
    {\begin{tikzpicture}
    \node[gauge, label=below:$1$] (1) at (0,0){};
    \node[gauge, label=right:$1$] (2) at (1,1){};
    \node[gauge, label=right:$1$] (3) at (1,-1){};
    \draw[-](2)--(1)--(3)--(2);
    \node[] at (-0.5,0){$[$};
    \node[] at (1.5,0){$]$};
    \node[] at (-1,0){$S_2 \; \wr $};
    \end{tikzpicture}}  
    & 1 \\ 
    \hline
    $2A_3$ & 
    $\left(4,\textrm{N/A}\right)$ 
    \raisebox{-0.5\height}
    & $(1,S_2)$ & \raisebox{-0.5\height}    {\begin{tikzpicture}
    \node[gauge, label=below:$1$] (1) at (0,0){};
    \node[gauge, label=right:$1$] (2) at (1,1){};
    \node[gauge, label=right:$1$] (3) at (1,-1){};
    \draw[-](2)--(1)--(3)--(2);
    \end{tikzpicture}} & \raisebox{-0.5\height}{$\textrm{N/A}$} & \textrm{N/A} \\ \hline
\end{tabular}}
\caption[Electric and Magnetic Quivers for Intersections between $S_2$ Special Piece $A_4+2A_1$ in ${\mathfrak{e}_8}$ nilcone and $\{E_8(b_6), \; D_4(a_1)+A_2 \}$ orbits]{Electric and Magnetic Quivers for Intersections between $S_2$ Special Piece $A_4+2A_1$ in ${\mathfrak{e}_8}$ nilcone and $\{E_8(b_6), \; D_4(a_1)+A_2 \}$ orbits. The $\{A_4+2A_1,\; 2A_3 \}$ special piece is $d_{LS}$ dual to the special orbit $D_7(a_2)$ which has a trivial special piece. The Coulomb branch global symmetry of the magnetic quivers is $A_2$. The Higgs branch of $Q_E$ is the $\rm{Sym}^2(A_2)$ symmetric product of the Kleinian singularity which carries a $U(1)$ symmetry.}
\label{tab:E8-S2Wreatha2}
\end{table}

The special piece $A_2+A_1$ contains a $c_4$ transition,  with an $f_4$ transition immediately below in the Hasse diagram. While the magnetic quivers for this compound transition can be constructed, the $S_2$ folding action is not apparent, so not even a partial Loop Lace map is known.

The remaining $S_2$ special piece, $\{E_7(a_3), \; D_6\}$ has the $d_{LS}$ dual orbit $A_4$. The piece contains a $b_2 \equiv c_2$ transition. A set of quivers that gives a partial Loop Lace map for this special piece has yet to be identified.

\FloatBarrier

\section{Discussion and Conclusions}
\label{sec:discussion}

This study extends the notion of {\it special duality} between nilpotent orbits and S\l odowy slices, or more generally S\l odowy intersections, to resolve some of the complications that follow from the non-involutive nature of the $d_{LS}/d_{BV}$ maps, when applied to orbits within the special pieces of nilcones.

In particular, in all the Classical and Exceptional algebras studied, it is shown that a modification to the $d_{LS}/d_{BV}$ maps, termed the $d_{SD}$ map, combined with a defined Loop Lace map between magnetic and electric quivers, is consistent with {\it special duality} between intersections whenever these involve orbits that lie within special pieces that satisfy certain conditions. The proviso is that these special pieces should be dual in the sense of (a) having the same Lusztig's canonical quotient and (b) having parent orbits which are $d_{LS}/d_{BV}$ dual to each other.

In the cases where the special pieces of an orbit and its $d_{LS}/d_{BV}$ dual are both trivial, the $d_{SD}$ map reverts to the $d_{LS}/d_{BV}$ map. 

In essence, the $d_{SD}$ map extends the $d_{LS}/d_{BV}$ maps to incorporate Lusztig's canonical quotient data, which persists through dualisation, thereby providing a 1:1 map between intersections and their duals. This appears related to the coupling of finite group data to Hitchin poles encountered in \cite{chacaltana2013tinkertoys, chacaltana2015gaiotto, chacaltana2015tinkertoys, chacaltana2018tinkertoys, chacaltana2018tinkertoyse8theory}.

Correspondingly, the Loop Lace map between magnetic and electric quivers relates different encodings of Lusztig's canonical quotient group and subgroup data. In particular, the map draws upon the complementary nature of wreathings and foldings. The resulting related pairs of magnetic and electric quivers construct S\l odowy intersections that are dual to each other within the nilcone containing the special piece(s).

The examples herein have been drawn from both unitary and (unframed) orthosymplectic quivers. Interestingly the Loop Lace map is similar across these two quiver types.

Particularly clear examples of the action of the Loop Lace map on quivers are provided by the self-dual $S_2, S_3, S_4 \text{ and } S_5$ {\it special pieces} that arise in the $\mathfrak{b}_2 \cong \mathfrak{c}_2, \mathfrak{g}_2, \mathfrak{f}_4, \text{ and }\mathfrak{e}_8$ algebras. In all these cases, the $d_{SD}$ map reduces to the identity, and the duality between orbits and slices is seen to correspond to the reciprocity of the Coulomb and Higgs branches of quiver theories carrying different diagrammatic encodings of the same symmetric group factors.

In the cases where only one of the special pieces is trivial, it is sometimes, but not always, possible to construct a partial Loop Lace map. It is noteworthy (a) that trivial special pieces appear as duals to all the $\textrm{Sym}^k(\mathbb{H}^n)/{\mathbb{H}^n}$ special pieces with $n>1$ (see Table \ref{tab:SpecialTrans}) and (b) that the missing images in the $d_{SD}$ maps (see Tables \ref{fig:AC_Map_Data_BC} to \ref{fig:AC_Map_Data_E7E8}) correspond to the missing images in the various partial Loop Lace maps (see Tables \ref{tab:ckn2m} to \ref{tab:E8-S2Wreatha2}).

The problematic cases where it is not possible to construct Loop Lace maps highlight that there are different ways of encoding symmetric group factors in quivers besides foldings and symmetrisations or wreathings \cite{Bourget:2020bxh}. Furthermore, since there is no known algorithmic way to construct magnetic quivers for general S\l odowy intersections (beyond $A$ type nilcones), the identification of magnetic quivers amenable to a Loop Lace map treatment remains an open research problem.

This study touches on the relationship between quiver wreathings, foldings and Lusztig's special pieces conjecture. In particular, in Section \ref{subsec:LL}, it is verified using the localisation formula that the transitions between $S_{2,3,4,5}$ wreathed magnetic quivers (at the top of a special piece) and $S_{2,3,4,5}$ folded magnetic quivers (at the bottom of a special piece) are of the form $\text{Sym}^{2,3,4,5}\left[\mathbb{H}^{m}\right]/\mathbb{H}^{m}$, where the transition across the special piece has dimension $\rm{dim}_{\mathbb H} =$ $m$, in accordance with \cite{fu2024localgeometryspecialpieces}. The extent to which this pattern extends beyond the special pieces of nilcones remains an open question.

\paragraph{Additional open questions}

The examples studied here have drawn on a limited selection of intersections that involve the special pieces of exceptional nilcones. Furthermore, the study of classical nilcones is inherently open-ended and the analysis presented in this work extends only to $BC$-type algebras of rank 4. It may be interesting to extend the analysis of {\it special duality} to other Classical algebras.

The component group factors $A(\cal O)$ that appear in special pieces are a subset of the finite group factors encoded in the nilcones of Classical and Exceptional algebras and their Weyl groups. It could be interesting to explore how these other finite group factors relate to quiver theories.

The Loop Lace map, coupled with developments in Higgs and Coulomb branch quiver subtractions (discussed elsewhere), may provide a means for extending the families of known magnetic and electric quivers for S\l odowy intersections. Thus, it may be interesting to develop a subtraction algorithm on the Higgs branch of orthosymplectic theories to complement the subtraction algorithm on unitary theories in \cite{Bennett:2024loi}. Also, the Higgs branch decorations identified in Table \ref{tab:AC_Quiver_Map} carry over to magnetic quivers and may play a role in refining the existing Coulomb branch methods to determine the Hilbert series for non-normal orbits exactly (rather than their normalisations).

The orthosymplectic quivers in the examples herein are unframed. Framed orthosymplectic quivers are also known to construct intersections in the nilcones of Classical algebras \cite{Hanany:2019tji}. It would be interesting to understand whether an analogue of the Loop Lace map exists for such framed theories.

\acknowledgments
We thank Eric Sommers, Nicholas Proudfoot, Paul Levy, Travis Schedler, Deshuo Liu, Guhesh Kumaran and Ben Webster for discussions. The work of SB, AH and RK is partially supported by STFC Consolidated Grant ST/X000575/1. The work of SB is supported by the STFC DTP research studentship grant ST/Y509231/1.

\appendix

\FloatBarrier

\section{Nilpotent Orbit Data}
\label{app:OrbitData}
This note makes extensive use of various sources of group theoretic data associated with the nilpotent orbits of Lie algebras.

Tables \ref{tab:orbitsbc2} to \ref{tab:orbitsbc4} contain a compilation of key data for nilpotent orbits of $\mathfrak{b}$ and $\mathfrak{c}$ algebras of rank up to 4. The data includes, for each orbit, the partition, the Characteristic, the complex dimension, whether the orbit is normal or {\it special}, the Barbasch-Vogan dual orbit, any $A(\cal O)$ and $C(\cal O)$ subgroups of Lusztig's canonical quotient group, the S\l odowy slice Lie symmetry group and the CSA weight map for its S\l odowy slice.

Tables \ref{tab:orbitsg2} to \ref{tab:orbitse8c} contain a compilation of key data for nilpotent orbits of Exceptional algebras. Data includes, for each orbit, the Bala-Carter label, the Characteristic, the complex dimension, whether the orbit is normal or {\it special}, the Lusztig-Spaltenstein dual orbit, any $A(\cal O)$ and $C(\cal O)$ subgroups of Lusztig's canonical quotient group, the S\l odowy slice Lie symmetry group and the CSA weight map for its S\l odowy slice. The CSA weight maps for $E_6$, $E_7$ and $E_8$ draw upon those presented in \cite{chacaltana2015tinkertoys, chacaltana2018tinkertoys, chacaltana2018tinkertoyse8theory}. The $C(\cal O)$ values for non-special orbits correspond to the finite group actions of Hitchin poles noted in \cite{chacaltana2013tinkertoys, chacaltana2015gaiotto,chacaltana2015tinkertoys, chacaltana2018tinkertoys, chacaltana2018tinkertoyse8theory}.

It may be helpful to read the tables in these Appendices in conjunction with the relevant Hasse diagrams, as shown for example in \cite{Generic_singularities,collingwood1993nilpotent, kraft1982geometry}. 

\begin{table}{}
\includegraphics[scale=1]{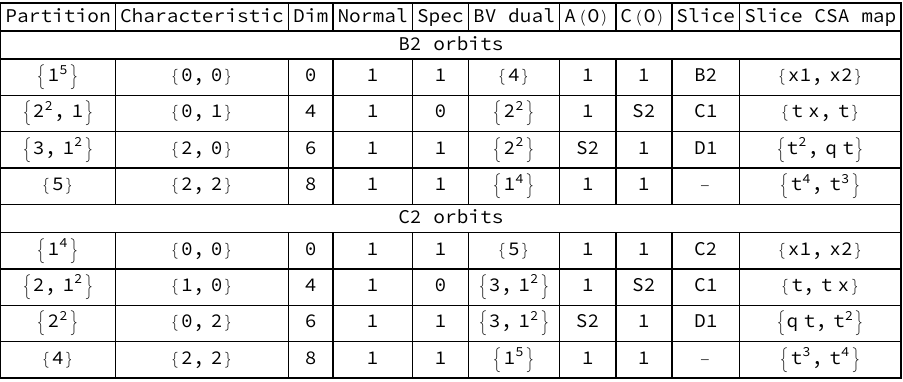}
\caption[Data for Nilpotent Orbits of $\mathfrak{b}_2 \cong \mathfrak{c}_2$]{Data for Nilpotent Orbits of $\mathfrak{b}_2 \cong \mathfrak{c}_2$.}
\label{tab:orbitsbc2}
\end{table}

\begin{table}{}
\includegraphics[scale=1]{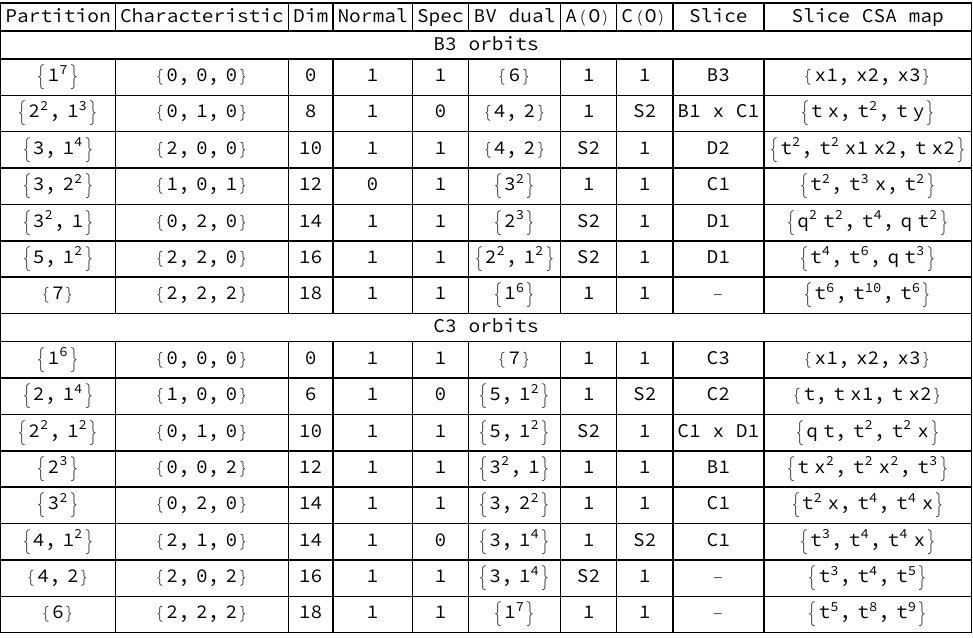}
\caption[Data for Nilpotent Orbits of $\mathfrak{b}_3$ and  $\mathfrak{c}_3$]{Data for Nilpotent Orbits of $\mathfrak{b}_3$ and  $\mathfrak{c}_3$.}
\label{tab:orbitsbc3}
\end{table}

\begin{table}{}
\includegraphics[scale=0.8]{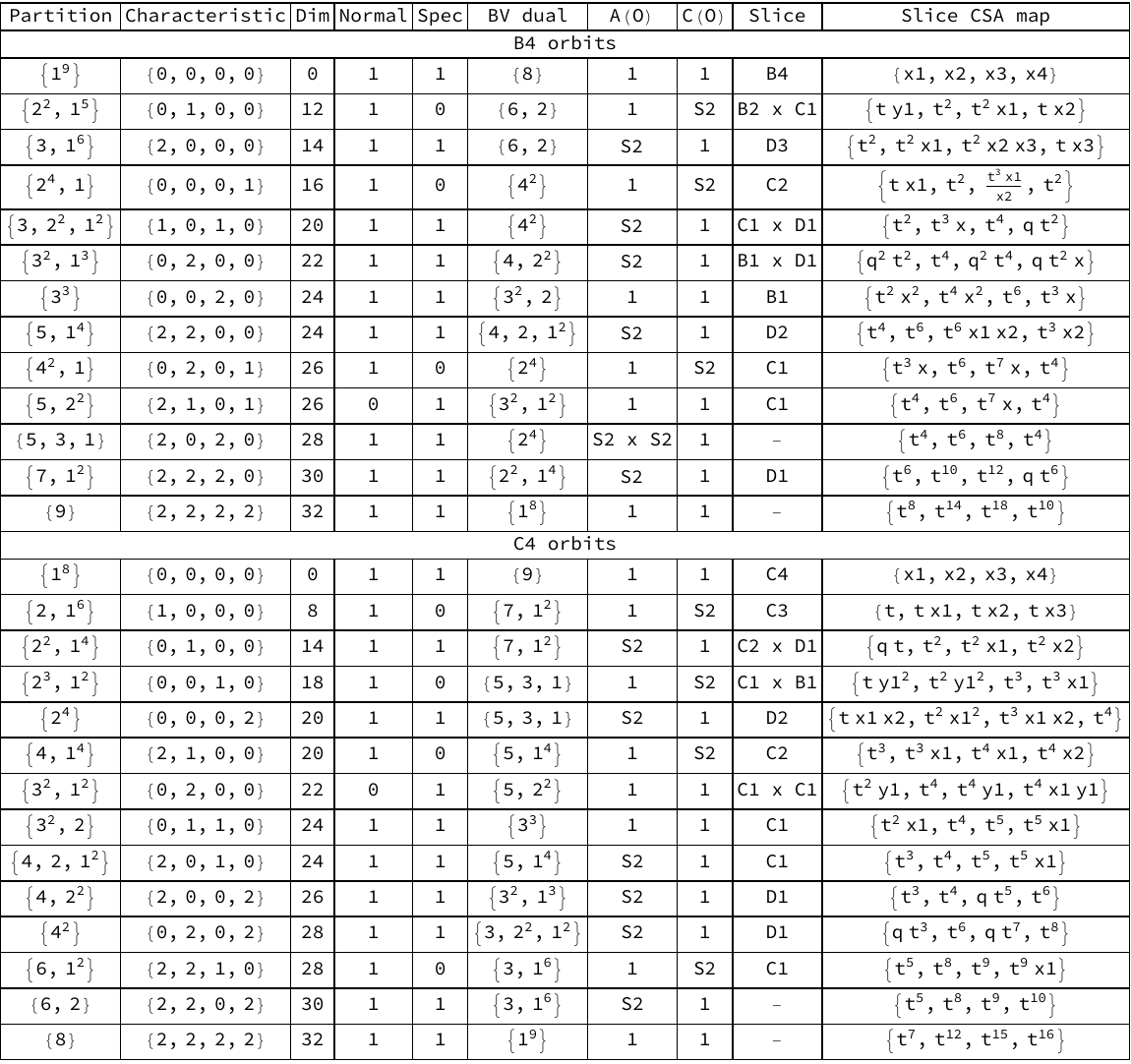}
\caption[Data for Nilpotent Orbits of $\mathfrak{b}_4$ and  $\mathfrak{c}_4$]{Data for Nilpotent Orbits of $\mathfrak{b}_4$ and  $\mathfrak{c}_4$.}
\label{tab:orbitsbc4}
\end{table}

\begin{table}{}
\includegraphics[scale=1]{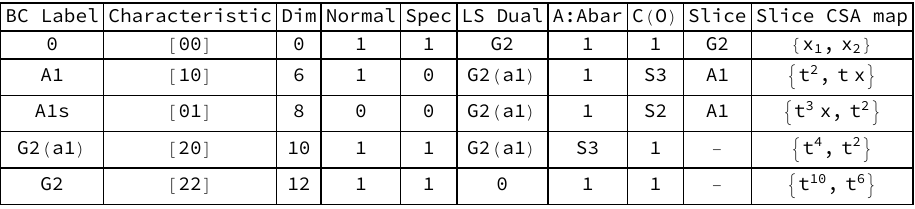}
\caption[Data for Nilpotent Orbits of $\mathfrak{g}_2$]{Data for Nilpotent Orbits of $\mathfrak{g}_2$.}
\label{tab:orbitsg2}
\end{table}

\begin{sidewaystable}{
\includegraphics[scale=1]{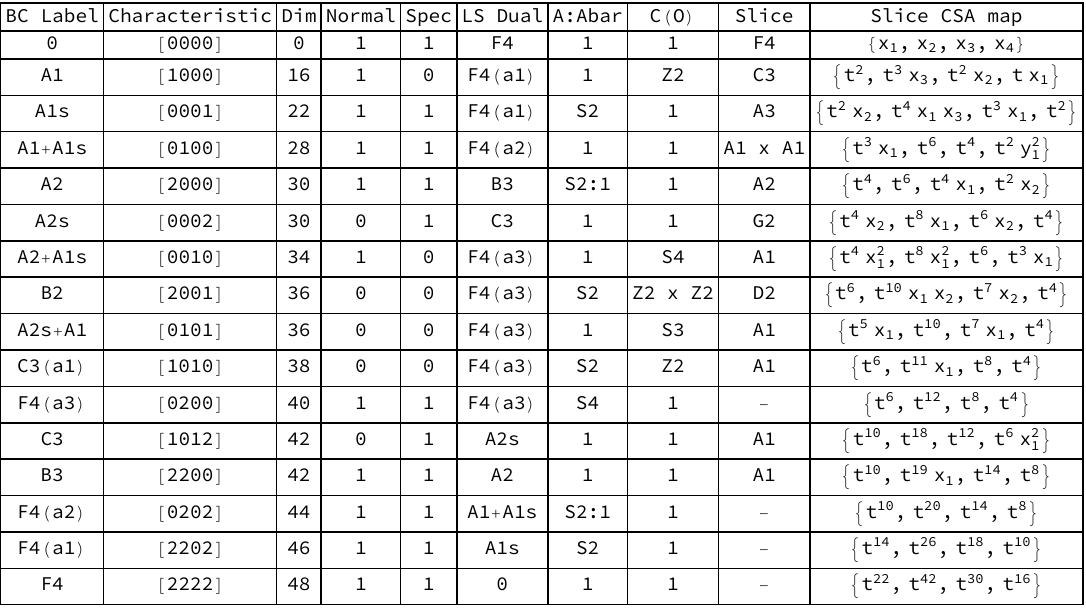}
}
\caption[Data for Nilpotent Orbits of $\mathfrak{f}_4$]{Data for Nilpotent Orbits of $\mathfrak{f}_4$.}
\label{tab:orbitsf4}
\end{sidewaystable}

\begin{sidewaystable}{
\includegraphics[scale=1]{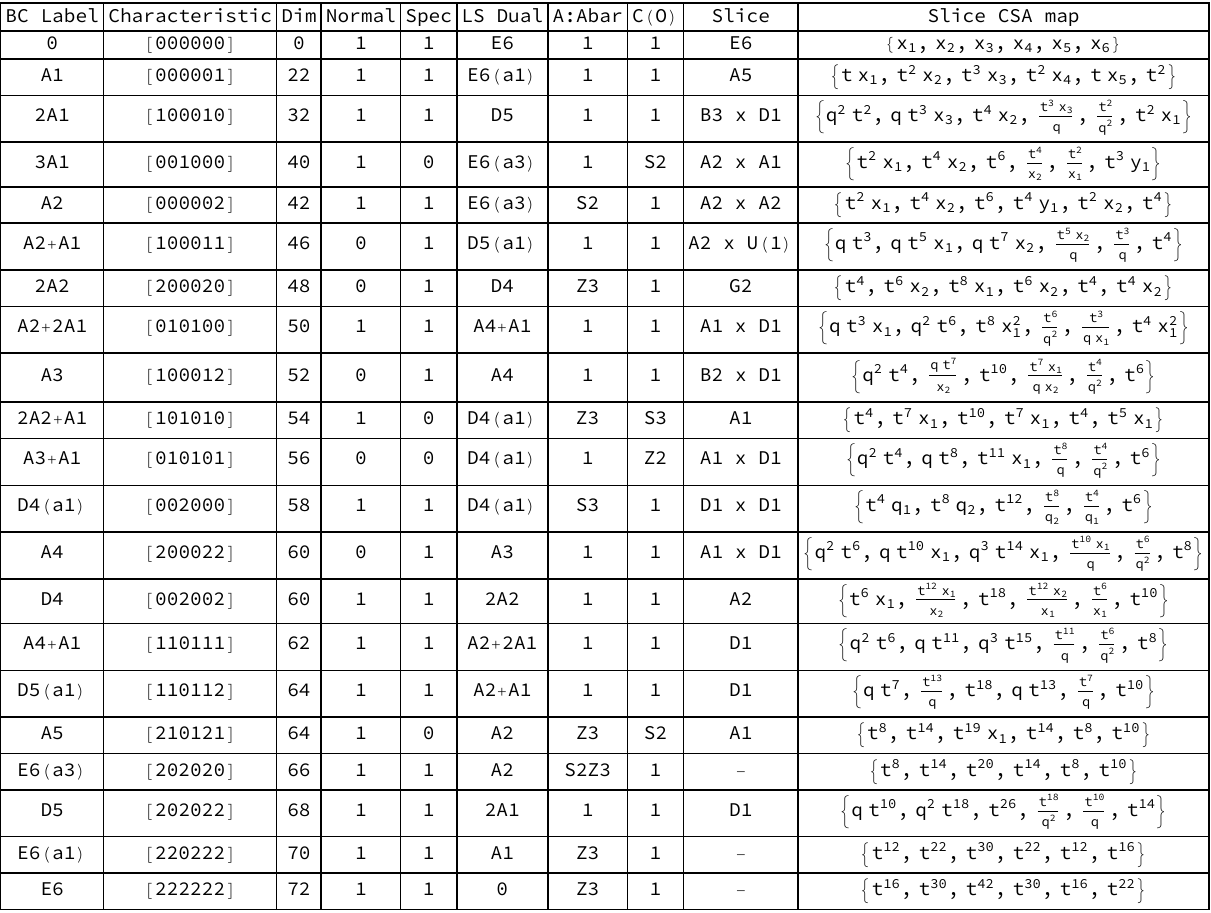}
}
\caption[Data for Nilpotent Orbits of $\mathfrak{e}_6$]{Data for Nilpotent Orbits of $E_6$.}
\label{tab:orbitse6}
\end{sidewaystable}

\begin{sidewaystable}{
\includegraphics[scale=.9]{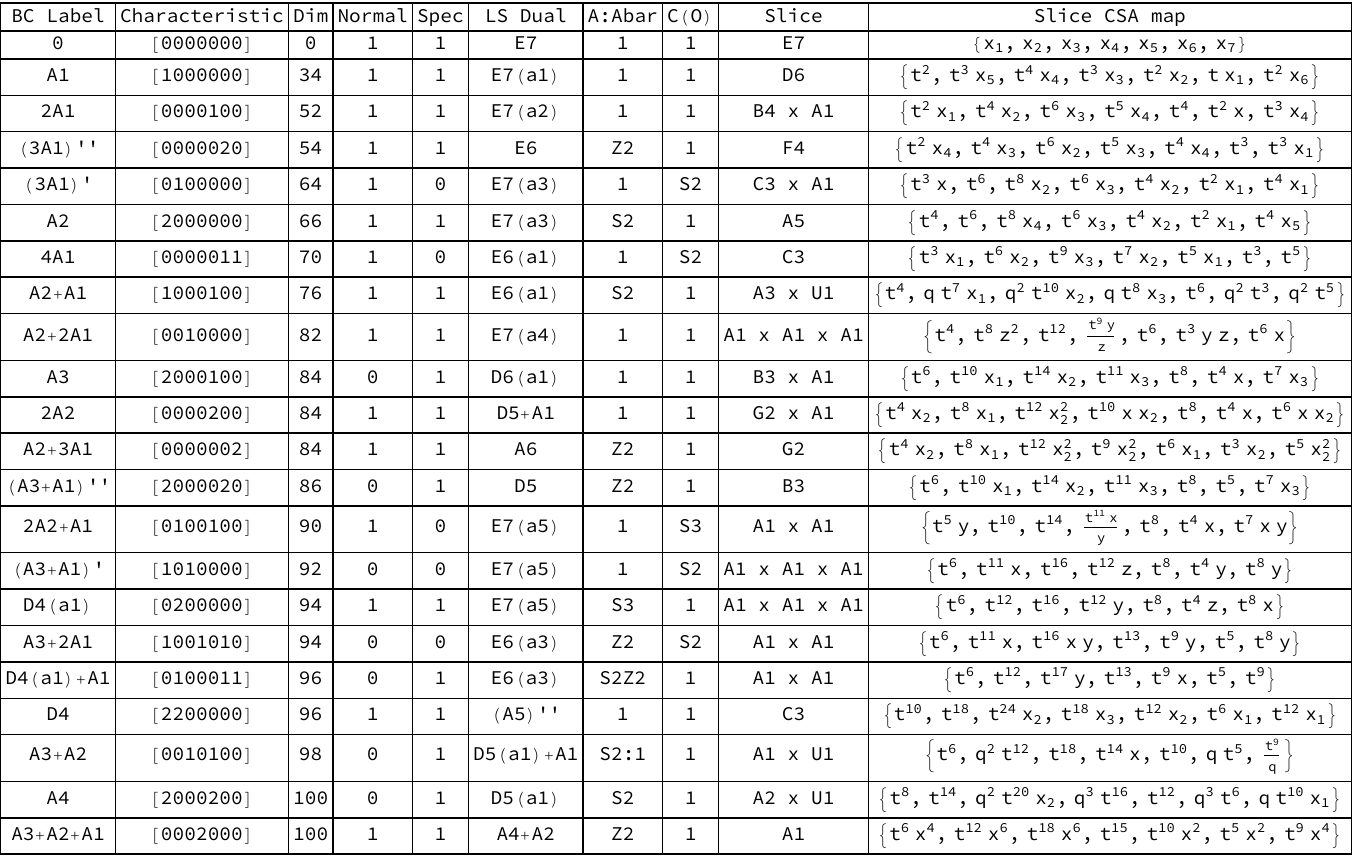}
}
\caption[Data for Nilpotent Orbits of $\mathfrak{e}_7 \;(1)$]{Data for Nilpotent Orbits of $E_7$ (1 of 2).}
\label{tab:orbitse7a}
\end{sidewaystable}

\begin{sidewaystable}{
\includegraphics[scale=.9]{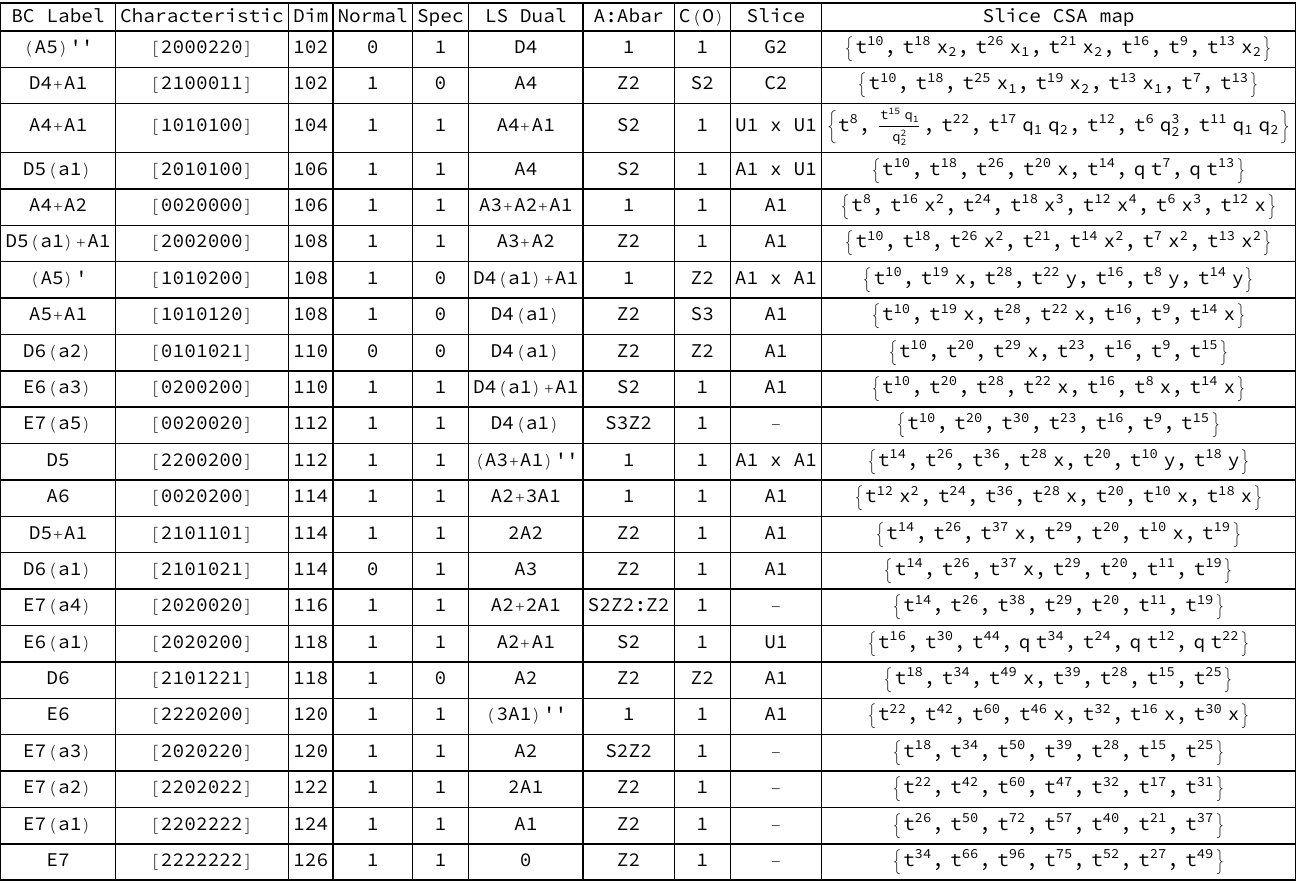}
}
\caption[Data for Nilpotent Orbits of $\mathfrak{e}_7 \;(2)$]{Data for Nilpotent Orbits of $E_7$ (2 of 2).}
\label{tab:orbitse7b}
\end{sidewaystable}

\begin{sidewaystable}{
\includegraphics[scale=.8]{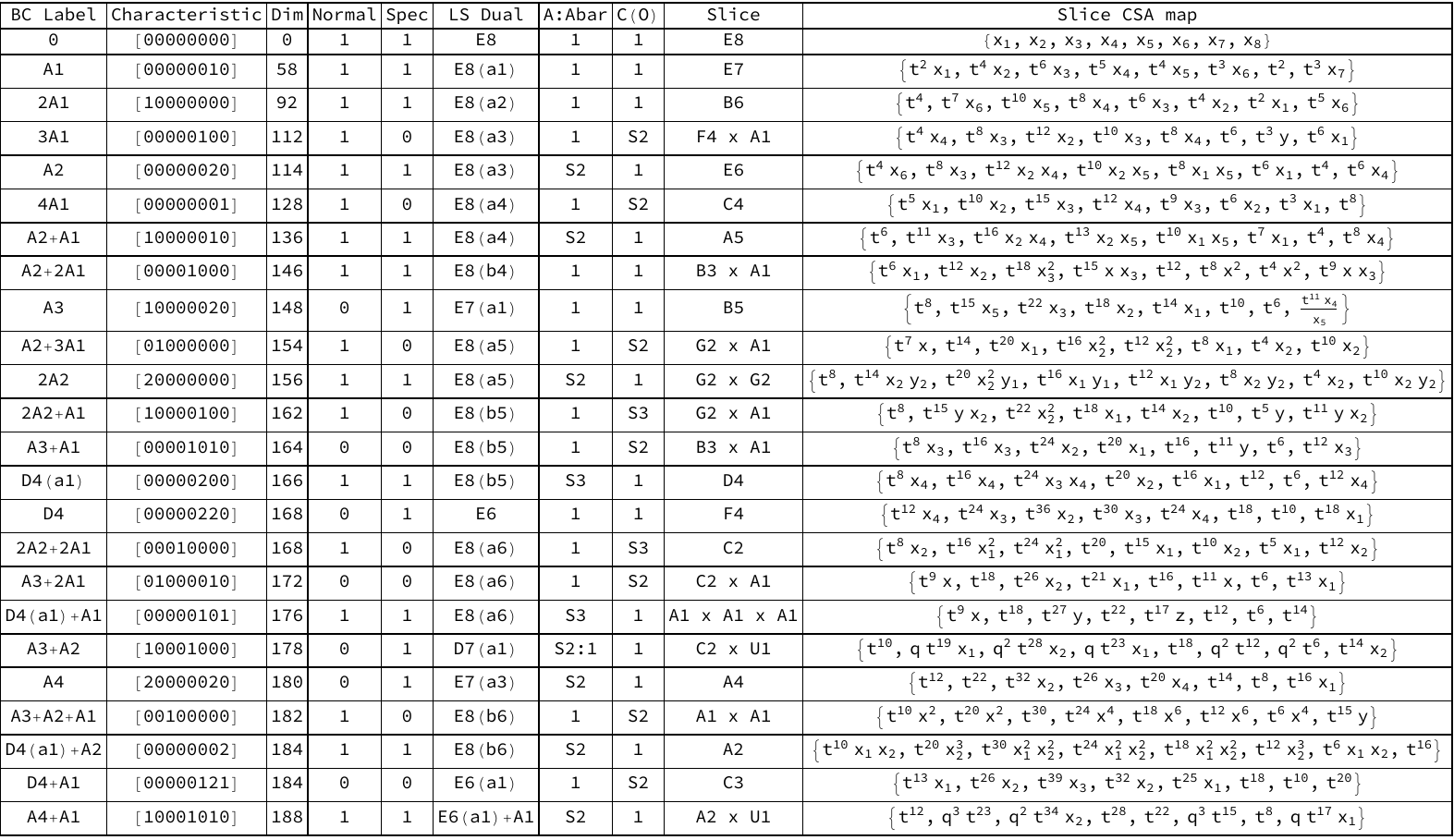}
}
\caption[Data for Nilpotent Orbits of $\mathfrak{e}_8 \;(1)$]{Data for Nilpotent Orbits of $E_8$ (1 of 3).}
\label{tab:orbitse8a}
\end{sidewaystable}

\begin{sidewaystable}{
\includegraphics[scale=.8]{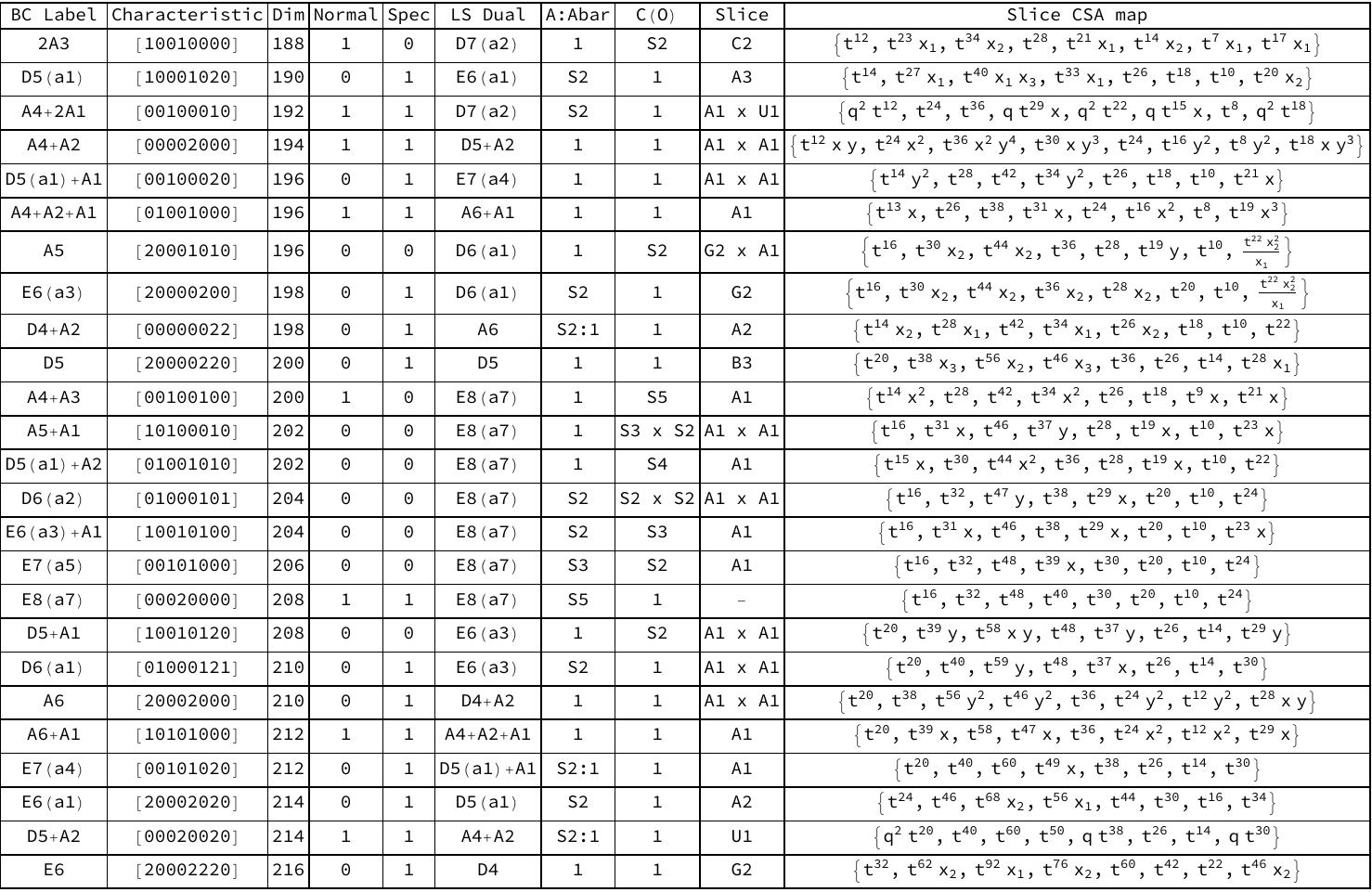}
}
\caption[Data for Nilpotent Orbits of $\mathfrak{e} _8 \;(2)$]{Data for Nilpotent Orbits of $E_8$ (2 of 3).}
\label{tab:orbitse8b}
\end{sidewaystable}

\begin{sidewaystable}{
\includegraphics[scale=.9]{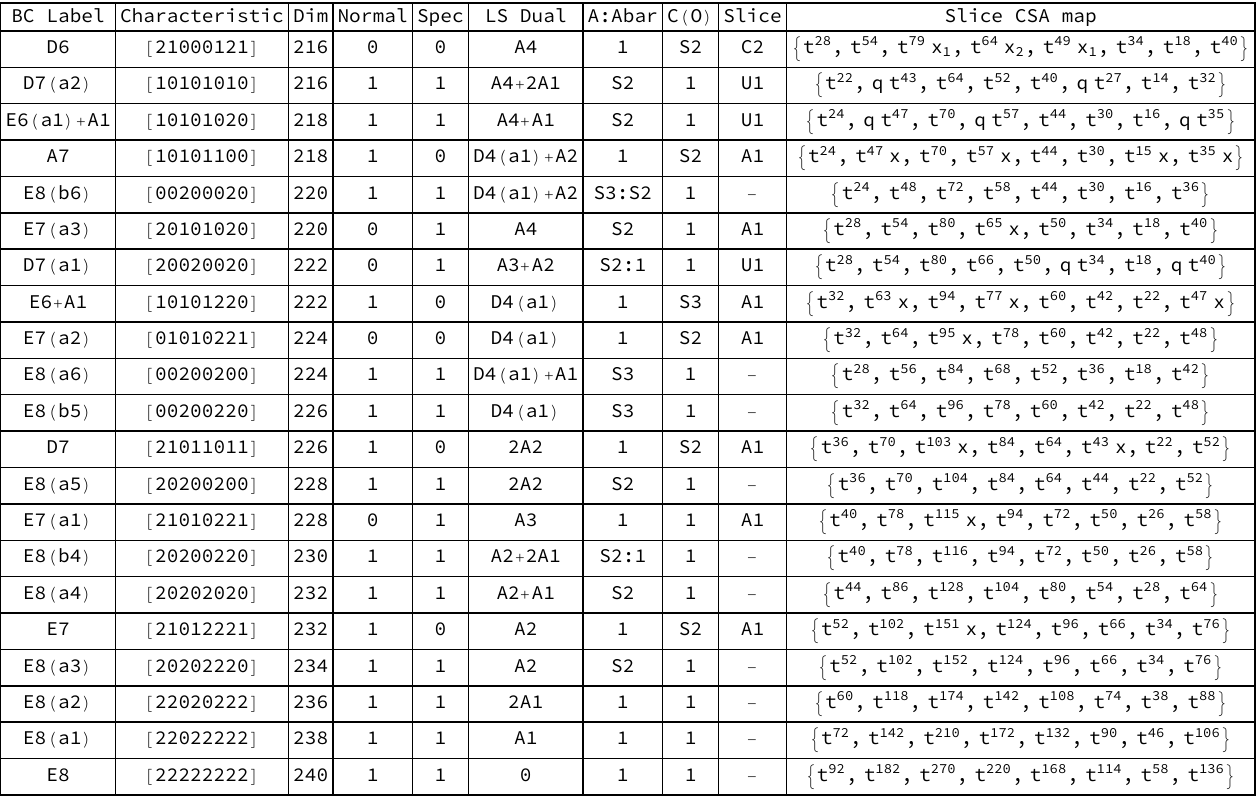}
}
\caption[Data for Nilpotent Orbits of $\mathfrak{e} _8 \;(3)$]{Data for Nilpotent Orbits of $E_8$ (3 of 3).}
\label{tab:orbitse8c}
\end{sidewaystable}

\FloatBarrier
\clearpage

\bibliographystyle{JHEP}

\bibliography{bibli.bib}
\end{document}